\def\slashchar#1{\setbox0=\hbox{$#1$}           % set a box for #1
   \dimen0=\wd0                                 % and get its size
   \setbox1=\hbox{/} \dimen1=\wd1               % get size of /
   \ifdim\dimen0>\dimen1                        % #1 is bigger
      \rlap{\hbox to \dimen0{\hfil/\hfil}}      % so center / in box
      #1                                        % and print #1
   \else                                        % / is bigger
      \rlap{\hbox to \dimen1{\hfil$#1$\hfil}}   % so center #1
      /                                         % and print /
   \fi}                                         %
 
\input epsf
\magnification=\magstep1
\baselineskip=16pt
 
% for equation, section, and figure numbering
\newcount \myeq
\myeq=0
\newcount \myfig
\myfig=0
\newcount \mysec
\mysec=0
% the \global in the following was frustrating to figure out
\def\myeqno {\global\advance\myeq  by 1 \eqno(\the\myeq)}
\def\myfigno{\global\advance\myfig by 1 \noindent Fig.~{\the\myfig. }}
\def\mysecno{\ifnum \the\mysec>0 {\vfill\eject}\fi
\noindent\global\advance\mysec by 1 \uppercase
\expandafter{\romannumeral\the\mysec}. }

\def\myeqname#1{ \myeqno  \xdef#1{Eq.~(\the\myeq)~}}
\def\myfigname#1{\myfigno \xdef#1{Fig.~(\the\myfig)~}}

\def\Tr{\rm Tr}
\def\half{{1\over 2}}

\centerline{{\bf Aspects of Chiral Symmetry and the Lattice}\footnote*
{For {\bf Reviews of Modern Physics.}}}
\medskip
\centerline{Michael Creutz}
\centerline{Physics Department}
\centerline{Brookhaven National Laboratory}
\centerline{Upton, NY 11973}
\centerline{creutz@bnl.gov}
\bigskip
{\narrower
\centerline {Abstract}

I explore the non-perturbative issues entwining lattice gauge theory,
anomalies, and chiral symmetry.  After briefly reviewing the
importance of chiral symmetry in particle physics, I discuss how
anomalies complicate lattice formulations.  Considerable information
can be deduced from effective chiral Lagrangians, helping interpret
the expectations for lattice models and elucidating the role of the CP
violating parameter $\Theta$.  I then turn to a particularly elegant
scheme for exploring this physics on the lattice.  This uses an
auxiliary extra space-time dimension, with the physical world being a
four dimensional interface.

}
\bigskip
{\bf \mysecno Introduction}
\medskip

Difficulties with chiral symmetry have plagued lattice gauge theory
from the earliest days.  Sometimes regarded as a technical problem
associated with Fermion doublers, the issues delve deeply into quantum
anomalies of field theory.  My goal here is a pedagogical discussion
of the interplay between chiral symmetry and the lattice.  The
discussion is biased towards my own, sometimes unconventional, point
of view, and is not an attempt to cover the extensive and rapidly
growing literature on the topic.  Rather I hope to convince the reader
of the deep consequences for the understanding of quantum field
theory.  I will be mainly qualitative, concentrating on
non-perturbative issues.  I will loosely intermingle considerations on
the behavior of gauge theories as functions of Fermion masses with
speculations on the obstacles to coupling gauge fields with chiral
currents.  While many profound topological issues arise in these
connections, I will concentrate more on the physical interpretations.

The lattice plays a dual role in particle physics.  On a practical
level it is a tool for calculating non-perturbative phenomena.  Here
it has taught much about hadronic physics, both in isolation and in
terms of corrections to electro-weak phenomena.  On a more conceptual
level the lattice is a defining regulator for quantum field theory.
The continuum theory is the zero lattice spacing limit of the lattice
formulation.  If this limit does not exist, can the continuum theory
make sense?  If unexpected phenomena occur as the limit is taken,
these must be understood in the context of potential manifestations in
the final theory.  From this point of view, it is essential to
understand all properties of the continuum theory in lattice language.
Chiral symmetry has played a major role in the history of theoretical
particle physics, but has a long and tortured history in lattice
models.  These difficulties with chiral issues on the lattice may well
be revealing something deep.

I assume that the reader is familiar with the basics (Creutz, 1983) of
lattice gauge theory, i.e. concepts such as gauge fields as elements
of the gauge group on the links of a four dimensional lattice, a gauge
action formed out of plaquette variables, and the exact local gauge
invariance of the lattice theory.  I also assume a basic familiarity
with chiral symmetry in the continuum and the $SU(N)\times SU(N)$
symmetry of the strong interactions with $N$ massless quarks.  I
concentrate on the aspects that make formulating chiral symmetry on
the lattice tricky, and will bypass several fascinating topics on the
interplay of chiral symmetry breaking and confinement.  These include
anomaly matching conditions (Coleman and Grossman, 1982), the
structure of low lying Fermionic states (Banks and Casher, 1980;
Verbaarschot, 1994), and various models of the {\ae}ther (van Baal,
1998).  I will assume from the outset that confinement of quarks into
mesons and baryons is a fundamental property of the strong
interactions.  I also will not discuss the rich interplay of chiral
symmetry and finite temperature.

The next section provides an overview of the importance of chiral
symmetry to particle physics.  The following two sections discuss how
quantum mechanical anomalies eliminate some symmetries from classical
field theory, and show that these effects have analogues in simple
band theory.  Section V shows how ignoring these phenomena in lattice
models can lead to multiplication of species.  Here I mention some of
the older traditional schemes for eliminating these difficulties.
Section VI lists some of the generic schemes to restore the
non-anomalous chiral symmetries through the use of large numbers of
auxiliary fields.

To stress some of the physics that must arise in any correct
formulation of chiral symmetry, several sections, VII through X,
explore how anomalies introduce a possible CP violating parameter into
the strong interactions.  These sections emphasize how these effects
are strongly dependent on the number of Fermion species.  The physics
of the anomaly results in a rather intriguing phase structure as a
function of complex quark masses.  Throughout this treatment effective
chiral Lagrangians, another powerful non-perturbative tool, play a
major role.

Returning to the lattice, section XI speculates on how these flavor
dependent phenomenon may appear with standard Wilson lattice Fermions.
Section XII shows that the picture of section XI is not quite
complete, and explores the Aoki phase, a rather fascinating lattice
artifact expected whenever a lattice action has explicit chiral
breaking terms.

Section XIII reviews how anomalies are elegantly incorporated into
effective Lagrangians through the use of an extra dimension.  This is
a partial motivation for the domain-wall Fermion approach introduced
in Sections XIV through XVI.  Section XVII gives a physical
interpretation for anomalous processes in in this approach via a flow
of Fermion states in the extra dimension.  Sections XIII and XIX treat
two contrasting schemes for incorporating the parity violation of the
weak interactions into lattice gauge theory.  Section XX derives an
exact chiral identity for domain-wall Fermions, casting light on the
distinction between singlet and non-singlet chiral currents.

Section XI briefly discusses recent efforts to eliminate the extra
dimension and implement exact chiral symmetries directly on four
dimensional systems.  The final two sections contain speculations
about the domain-wall Fermion idea and some concluding remarks.

\bigskip
{\bf \mysecno Why chirality?}
\medskip

If the lattice is the definition of a quantum field theory, why bother
understanding chiral symmetry at anything but a superficial level?
Just put the model on the computer, calculate the particle properties,
and go home.  This attitude, however, ignores a long history of
fascinating developments tied to chirality in theoretical particle
physics.

Lord Kelvin (1904) introduced the word ``chiral,'' coming from the
Greek word for hand, to refer to objects, such as hands, that are
distinguishable under reflection.  The most common use of the term is
in chemistry, where a chiral molecule has a distinct mirror image.  A
carbon atom tetrahedrally bound to four different groups is the
simplest example; tartaric acid with its two chiral centers is
historically the most famous.  In particle physics the term is adapted
to distinguish massless particles by their helicity, i.e. their spin
along the direction of motion.

For the particle physicist, chirality is deeply entwined with the
Lorentz group.  Indeed, massless representations are qualitatively
different from their massive counterparts.  Without mass, the helicity
of a particle is invariant under boosts and rotations.  The helicity
becomes a Lorentz invariant concept.  When a spin 1/2 Fermion is
coupled minimally to a gauge theory, this helicity conservation
survives interactions, and currents associated with left and right
handed particles are naively separately conserved.  The Fermion fields
naturally break into two independent parts, $\psi_R\equiv{1\over
2}(1+\gamma_5)\psi$ and $\psi_L\equiv{1\over 2}(1-\gamma_5)\psi$.
With only one spatial dimension the concept is even simpler.  A
massless excitation traveling at the speed of light can never be
overtaken.  Thus a particle moving to the right does so in all frames,
and the fields break up into left and right moving parts.

Infamous anomalies complicate this simple picture.  Indeed, with gauge
interactions present, not all axial currents can be conserved.  This
lies at the root of much fascinating physics, entwined with such
issues as strong $CP$ violation, the presence of an unanticipated
parameter $\Theta$, the so called $U(1)$ problem, the mass of the
$\eta$ meson, etc.  The lattice, by removing all infinities from the
problem, forces the study of these effects at a deep and fundamental
level.

There are at least three easily identifiable reasons for particle
physicists to study chiral symmetry.  First, the observed world is
chiral.  Only left handed neutrinos emerge from beta decays.  The
underlying theory must incorporate parity violation.  In the standard
model this appears through chiral couplings of the gauge fields to
Fermions.  In the massless limit, the intermediate vector Bosons of
the weak interactions couple only to left handed helicities.  It is
essential that any fundamental formulation incorporate this asymmetric
coupling.  Perturbatively this has not proven to be a major problem,
at least if anomalies are appropriately canceled.  However for the
lattice, a direct non-perturbative scheme for defining field theories,
this anomaly cancellation enters at a rather deep level.  Indeed,
there still is not a universal agreement on how to formulate the full
standard model on the lattice.  Minefields of unproven technical
details lie scattered along the paths of all existing approaches.

A lattice formulation of the standard model is perhaps a rather
technical issue, not particularly relevant for observable phenomena.
The weak interactions are weak, and perturbation theory works
extremely well for any experimental observations.  The main exception
lies in attempts to understand the early universe, when the
temperature was of the order of the weak scale.  At this time
non-perturbative phenomena may have played a major role in the
generation of the baryon number asymmetry in the universe.

The second reason to explore chiral symmetry in a lattice context is
more phenomenological, and is increasingly showing its import to
ongoing experimental results.  This has its roots in the strong
interactions and the low mass of the pion relative to other hadrons.
The underlying classical Lagrangian for the quark confining theory of
the strong interactions has extra symmetry in the massless limit.  All
terms in this limiting Lagrangian independently conserve the number of
left handed and right handed Fermions.  With massless quarks this is
an exact classical symmetry of the theory.  There exists a vast
repository of literature on how this symmetry is spontaneously broken
by the vacuum, with the pions being approximate Goldstone Bosons.  The
pion mass is small because the two lightest quarks have small masses.
Indeed, the phenomenology of chiral symmetry-breaking works remarkably
well even when the strange quark is also treated as light.

The immense success of the phenomenology based on spontaneous chiral
symmetry breaking compels an effort to understand its role on the
lattice.  To ignore the chiral issues, simply calculating hadronic
phenomenology as functions of the quark masses and adjusting to fit
experiment, deprives one of much understanding of how the particles
derive their masses and the vast successes of the effective chiral
Lagrangian approach to low energy meson interactions.

Finally, chiral symmetry plays a crucial role in most proposed
extensions of the standard model.  These often involve unifications at
scales much larger than those of ordinary hadronic physics.  In the
process of renormalizing and projecting out the low energy theory, it
appears peculiar that the physical particles in the laboratory are so
much lighter than these possibly more fundamental scales.  To argue
around this unnaturalness, chiral symmetry serves as a protector of
light masses, which only receive multiplicative renormalizations.  A
massless particle should remain so under renormalization.  This is
also one of the main motivations for super-symmetry; by relating
fundamental scalars such as the Higgs meson to Fermionic partners,
chiral symmetry can restrict the renormalization of their masses as
well.

\bigskip
{\bf \mysecno Anomalies}
\medskip

It is well known that not all classical symmetries survive
quantization.  The most basic example, the scale anomaly, has been so
fully absorbed into the lattice lore that it is rarely mentioned.
When the quark masses vanish, the classical Lagrangian for the strong
interactions contains no dimensional parameters.  But the quantum
theory is supposed to describe baryons and mesons, and the lightest
baryon, the proton, definitely has mass.  This is understood through
the phenomenon of ``dimensional transmutation,'' wherein the classical
coupling constant of the theory is traded, through the process of
renormalization, for an overall scale parameter (Coleman and Weinberg,
1973).

This scale anomaly is central to understanding asymptotically free
gauge theories; so, let me briefly repeat the basic words.  On
quantization of the initial theory, the usual divergences of loop
diagrams are encountered.  These are removed by an ultraviolet cutoff,
but the latter inherently involves a length scale.  To remove the
cutoff one must adjust, or ``renormalize'' the bare parameters of the
theory so that physical parameters remain finite.  For the strong
interactions with massless quarks, the only bare parameter is the
gauge coupling $g$.  This quantity becomes a function of the cutoff
$\Lambda$.  The coupling ``runs,'' following the behavior dictated by
the ``renormalization group equation.''  Because of asymptotic
freedom, the bare coupling runs to zero as the cutoff is removed.

In this way of thinking, any physical mass parameter, say the proton
mass, takes the form of a coupling dependent factor times the only
dimensional parameter around, the cutoff
$$
m_p=f_p(g) \Lambda.
\myeqno
$$
The factor $f_p$ goes to zero as the cutoff $\Lambda$ goes to
infinity, leaving the product finite.  Given two different physical
masses, say $m_\rho$ and $m_p$ for the rho and the nucleon, the scale
drops out of their ratio.  Thus
$$
{m_\rho \over m_p} = {f_\rho(g) \over f_p(g)}
\myeqno
$$
is a function only of the bare coupling.  The physical value is this
ratio evaluated at zero bare coupling.  If the quantum field theory is
well defined, this final result must be independent of the scheme used
for regularization.  At this point there are no more parameters to
play with.  Any dimensionless ratio is completely determined, with no
free parameters.  If a calculation gets some ratio of masses wrong,
then either the calculation or the theory is incorrect.  The net
outcome is really quite amazing --- a fully parameter free quantum
field theory.  It is no longer scale invariant, but has an overall
scale fixed by the physical particle masses.  The ratios of any of
these masses should be exactly determined, and calculable via lattice
simulations.

This discussion applies to the massless theory.  When the quark masses
are turned on, these, measured in units of the above generated scale,
provide additional parameters to adjust.  What determines their values
is one of the outstanding questions of particle physics.  Conceivably
they also come from some sort of ``dimensional transmutation'' in a
higher level theory, but this goes well beyond the context of this
review.

\midinsert
\epsfxsize .3\hsize
\centerline{\epsfbox {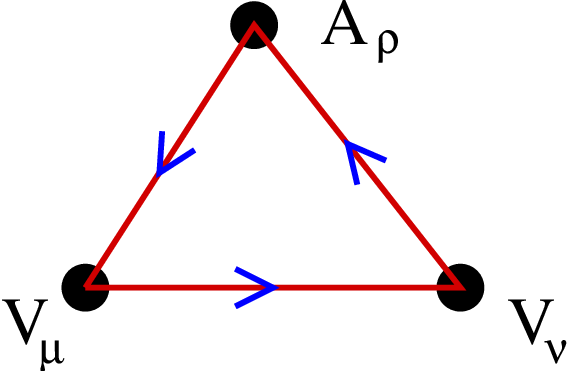}}
\narrower{\myfigname\triangle In four dimensions, regularization of
triangle diagrams involving both axial and vector currents allows only
one of the two to be conserved, even as the Fermion mass goes to
zero.\par}
\endinsert

The scale anomaly is not the only symmetry of the strong interactions
of massless quarks that is lost upon quantization.  The most famous,
and central to this review, are the anomalies in the axial vector
Fermion currents (Bell and Jackiw, 1969; Adler, 1969).  Working in a
helicity basis, the classical Lagrangian has no terms to change the
number of left or right handed Fermions.  On quantization, however,
these numbers can not be separately conserved.  Technically this comes
about because of the famous triangle diagram of \triangle.  This
introduces a divergence to the theory which requires regularization
via a dimensional cutoff.  For the strong interactions this cutoff is
removed so that the vector current, representing total Fermion number,
is conserved.  But if this choice is taken, then the axial current,
representing the difference of right and left handed Fermion numbers,
cannot be.  There is a freedom in choosing which currents are
conserved; however, in a gauge theory, consistency requires that gauge
fields couple only to conserved currents.

In the standard model, anomalies require some time honored
conservation laws to be violated.  The most famous example ('t Hooft,
1976a,b) is baryon number, which in the standard model is sacrificed
so that the chiral currents that couple to the vector Bosons are
conserved.  Baryon violating semi-classical processes have been
identified and must be present, although at a very low rate.  While
not of observable strength, at a conceptual level any scheme for
non-perturbatively regulating the standard model must either contain
baryon violating terms (Eichten and Preskill, 1986) or extend the
model to cancel these anomalies with, say, mirror species.
Furthermore, this consistency has non-trivial implications for the
allowed species of Fermions.  To conserve all the gauged currents of
the standard model requires the cancellation of all potential
anomalies in currents coupled to gauge fields.  In particular, the
standard model is not consistent if either the leptons or the quarks
are left out.  This connection between quarks and leptons is a deep
subtlety of the theory and must play a key role in placing the theory
on a lattice.

For the strong interactions, the chiral anomaly has dramatic
consequences for the low lying hadronic spectrum.  As mentioned
earlier, the light nature of the quark masses gives rise to the
interpretation of pions as approximate Goldstone Bosons associated
with a spontaneous breaking of chiral symmetry.  But mesons associated
with broken chiral currents need not be as light.  Considering only
the up and down quarks as light, the eta meson is heavier than the
pion even though it is constructed from the same quarks.  Under the
eightfold way involving the strange quark, it is the $\eta^\prime$
meson which is anomalously heavy.

\bigskip
{\bf \mysecno Band theory and anomalies}
\medskip

The necessity for chiral anomalies is easily understood in lower
dimensions.  In a world with only one time and one space dimension,
there is no spin, and the concept of chirality reduces to a
distinction between particles moving in one direction relative to the
other.  If a particle is massless, the inability to boost by more than
the speed of light indicates that a right moving particle is so in all
Lorentz frames.  For free particles the number of left movers and the
number of right movers are separately conserved quantities.

A gauge interaction would at first sight seem to preserve this
symmetry since the classical form of the current
$\overline\psi\gamma_\mu\psi=\overline\psi_R\gamma_\mu
\psi_R+\overline\psi_L\gamma_\mu\psi_L$
separates into two independent terms.  However, in analogy with the
triangle diagram in four dimensions, there is a bubble diagram which
cannot be regulated so that both the vector and axial currents are
conserved.

\midinsert
\hfill
\epsfxsize=.45\hsize
{\epsfbox {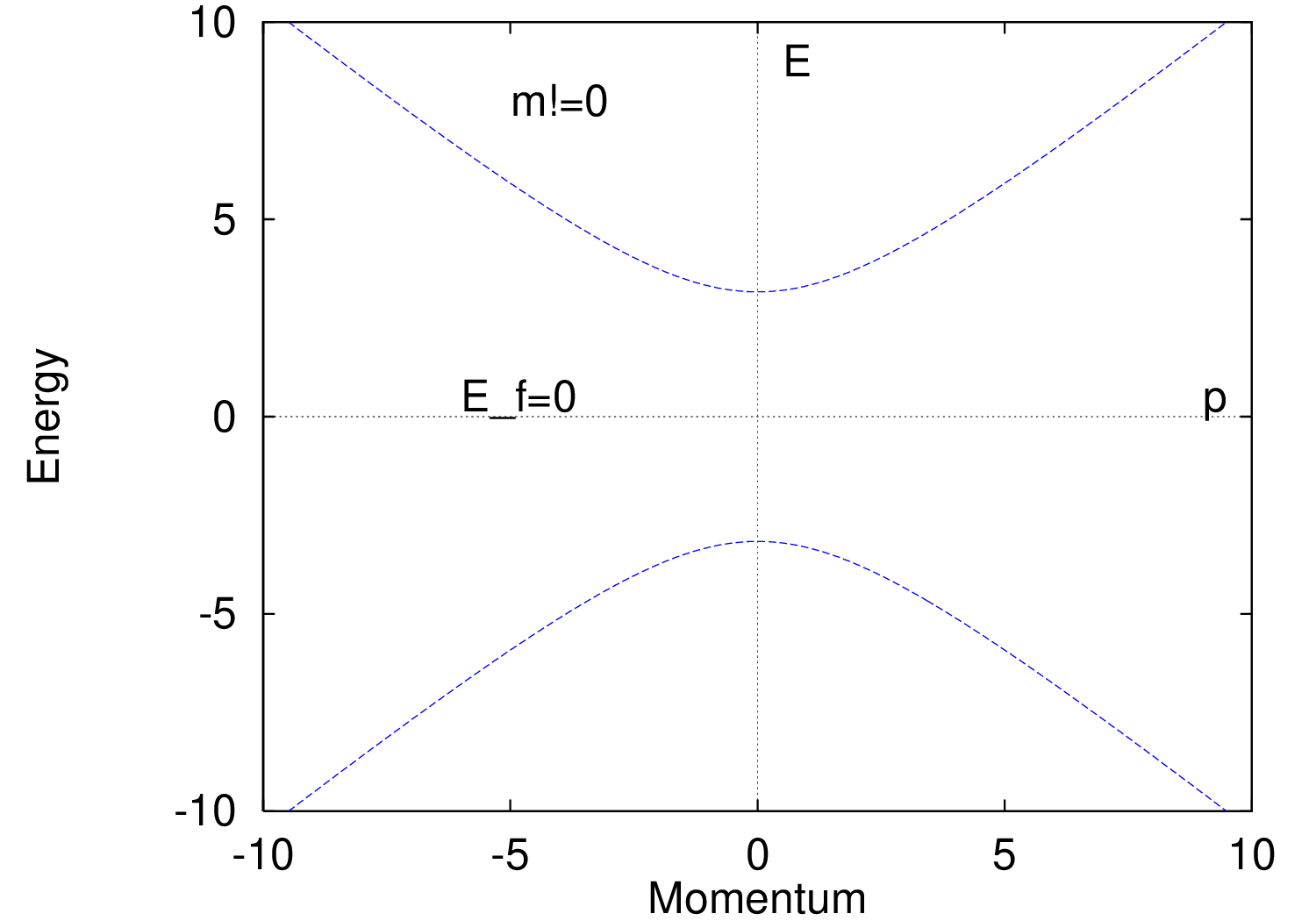}} 
\hfill \epsfxsize=.45\hsize
{\epsfbox {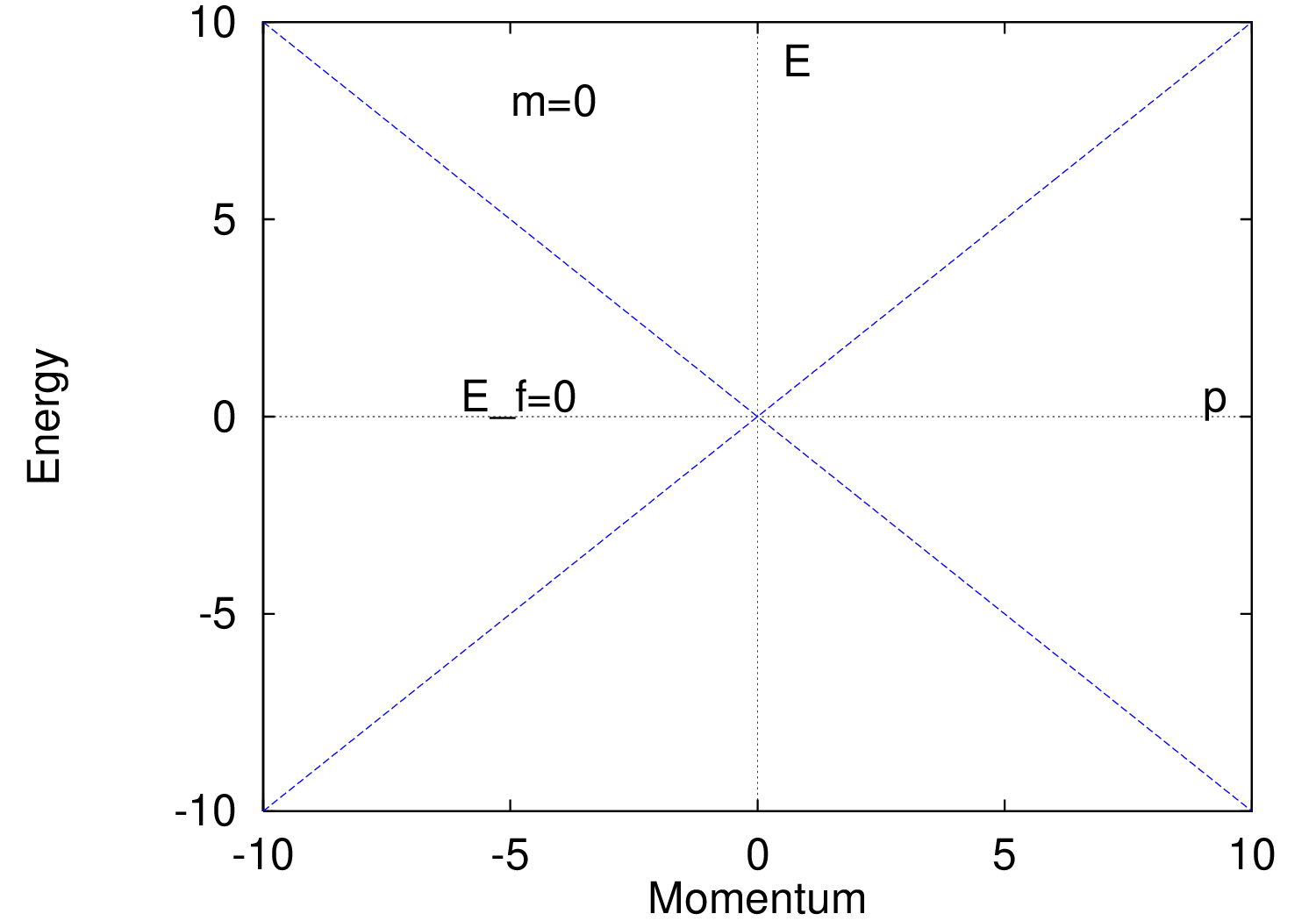}} \hfill

\narrower \myfigname\conductor A massive Fermion corresponds to
an insulating vacuum.  A massless Fermion gives a conductor.  The
induction of currents in this conductor corresponds to the chiral
anomaly.

\endinsert

In a Hamiltonian language, the anomaly arises from a sliding of states
in and out of the Dirac sea (Ambjorn, Greensite, and Peterson 1983;
Holstein, 1993).  The mechanism can be seen intuitively from simple
band theory.  Consider a massive Fermion with energy spectrum
$E=\sqrt{p^2+m^2}$, as sketched in
\conductor.  In the Dirac picture, there is a filled sea of
negative energy states, in which holes represent anti-particles.  The
Fermi level of the vacuum lies at $E=0$, between these two bands, and
this is the classic picture of an insulator.

Now for the massless case the energy spectrum reduces to $E=\pm p$.
The right moving particles are represented by the increasing branch,
$E=p$ and the left movers occupy the decreasing branch.  I note in
passing that an anti-particle of a right-mover also moves to the
right.  It corresponds to a hole in the negative energy, negative
momentum spectrum.  Thus it has an effective positive momentum.  This
is in contrast to the three dimensional case where an anti-particle
has opposite helicity.

In the band theory language, the theory no longer has a gap.  It is a
conductor.  The current is proportional to the number of right movers
minus the number of left movers.  Here is where the anomaly comes in:
electric fields can induce currents in a conductor.  Thus an electric
field will change the relative number of right and left movers.
Without the anomaly, wires would not conduct and transformers would
not work.

The mechanism of the anomaly is particularly clear in this Dirac sea
picture. An electric field pushes on the filled levels, increasing
their momentum.  As the momentum changes, a negative energy state can
acquire positive energy.  The anomaly is not due to a left moving
state jumping to a right mover, but rather a shifting of the negative
energy states up or down depending on the branch.

This picture can be generalized to three dimensions.  First the
Fermions are subjected to a magnetic field that confines them
transversely into Landau levels.  The lowest of these Landau levels
has precisely the spectrum of a one dimensional model, where that
dimension is the distance along the magnetic field.  Then an applied
electric field parallel to that direction can induce currents in that
level.  The anomaly is proportional to $\vec E\cdot \vec B\sim \tilde
F_{\mu\nu}F_{\mu\nu}$.

Note that this picture of level flow relies on an infinite number of
states on each branch.  A new right mover arises from a shift out of
an infinite negative energy sea, while an anti left-mover is a new
hole arising from a depression of the Dirac sea.  This discussion of
how the anomaly works also hints at why it is difficult in a lattice
approach.  With such a regulator, the momentum is bounded by the
cutoff and the spectral branches cannot go to infinity; the maximum
allowed momentum on a lattice of spacing $a$ is $\pi/a$.  When the
states slide up under the influence of a field, a mechanism must be
provided to absorb one state from the top.  Generally this requires
either an explicit breaking of the chiral symmetry or another
infinity.

\bigskip
{\bf \mysecno Lattice Fermions}
\medskip
That there is a difficulty with lattice Fermions was realized quite
early.  Take the simple continuum Lagrangian for a free massless Dirac
field
$$
S_{\rm cont}=\int d^4x\ \overline\psi i\slashchar\partial\psi \hfill \myeqno
$$
and use a simple difference scheme for the derivative on an $L^4$ lattice
$$
S_{\rm lattice}={1\over 2}\sum_j \sum_{\mu=0}^4 
\ i\overline \psi_j \gamma_\mu (\psi_{j+e_\mu}-\psi_{j-e_\mu}) \myeqno
$$
where $e_\mu$ represents a unit step in the $\mu$ direction.  To
remove explicit factors of the lattice spacing, I have rescaled the
Fermionic fields by a factor of $\sqrt {a^3}$; thus, one would relate
the continuum field and the lattice one by $\psi_j \leftrightarrow
\sqrt {a^3}\psi(x=ja)$.  Going to momentum space on a periodic lattice,
$$\eqalign{ &\psi_j={1\over L^2}\sum_q e^{-2\pi i j\cdot q/L}
\tilde\psi_q \cr &\tilde\psi_q={1\over L^2}\sum_j e^{2\pi i j\cdot
q/L} \psi_j, \cr }\myeqno
$$
the lattice action becomes diagonal
$$
S_{\rm lattice}=\sum_q\ i \overline{\tilde\psi}_q 
\gamma_\mu \sin(2\pi q_\mu/L) \tilde\psi_q \myeqno
$$
and I can explicitly evaluate the propagator
$$
\langle \overline\psi_j \psi_k \rangle =
\sum_q e^{-2\pi iq\cdot(j-k)/L} D(q)
\myeqno
$$
where
$$
D(q)=
{1\over i\sum_\mu \gamma_\mu \sin(2\pi q_\mu/L)}.
\myeqname\propagator
$$
For small momentum, this reduces to the conventional propagator
${1 \over i\slashchar p}$ with the identification
$p=2\pi q/La$
and the continuum position $x=a j$.

Because of the properties of the sine function, the free quark
propagator in \propagator has poles not only at $p_\mu=0$ but also when
any component of the momentum is of magnitude $\pi/a$.  Since the
natural range of momentum can be taken as $0\le p_\mu < 2\pi/a$, in four
dimensions there are sixteen poles representing 16 Fermionic species.

These degenerate species have different chiralities (Karsten and Smit,
1981).  Since the slope of the sine function is negative at $\pi$,
when a component of momentum is near that value, the Fermion
effectively uses a gamma matrix that differs by a sign.  For the
matrix $\gamma_5=\gamma_0\gamma_1\gamma_2\gamma_3$, half the species
use the opposite sign.  The naive axial current, $\overline\psi
\gamma_\mu
\gamma_5 \psi$ represents a flavor non-singlet current in the
resulting 16 flavor theory.  Such currents are not required to be
anomalous, and this theory has good chiral properties, albeit with
more flavors than naively expected.

A similar doubling problem should be expected in any fully regulated
local formulation that ignores anomalies but attempts to keep the
classical chiral symmetries.  Something must go wrong in the path
integral, and in this case it is the appearance of multiple species in
the quantum theory.  Doubling is more than simply an artifact of the
lattice action.  This phenomenon has a topological interpretation as
emphasized in early papers on the topic (Nielsen and Ninomiya, 1981a,b,c),
and recently generalized in (Shamir, 1993a).

I note in passing that a single generation in the standard model has
eight Fermionic species: three colors of quark doublet and one lepton
doublet (allowing for a sterile right-handed neutrino).  It is
tempting to relate this power of two with some sort of doubling, and
indeed this full structure of quarks and leptons is involved in the
electroweak anomaly cancellation.  While this view of Fermion
multiplicities within a Fermion generation is unconventional, it is
sufficiently amusing that I will explore it a bit more in a later
section.

With the goal of eliminating doubling as a lattice artifact, much
effort has gone into more complex Fermionic actions.  One popular
approach (Kogut and Susskind, 1975) staggers the four components of
the Fermions over separate lattice sites.  This removes one factor of
four from the doubling, leaving four species.  The resulting theory
has one remaining exact chiral symmetry.  The approach has been
extremely popular for simulations because 1.)  the Fermion matrices
are smaller, and 2.) the remnant chiral symmetry protects the Fermion
mass from additive renormalization.  The main penalty of this approach
is that flavor symmetry is sacrificed; with four flavors there is only
one exact Goldstone Boson in the chiral limit and the other pions
remain massive.  In later sections I will argue for a strong physical
dependence on the number of light quark flavors.  Much of this physics
may be hidden by this sacrifice of flavor symmetry.

The other traditional way (Wilson, 1977) to deal with this ``problem''
destroys the chiral symmetry at the outset.  The doublers occur at
non-zero momentum, and thus can be given a mass via a term that can
be thought of as a momentum dependent mass.  Having this vanish at
zero momentum keeps the desired Fermion light, while allowing it to be
large at the doubling points with $p_\mu=\pi/a$ removes the doublers.
This can be accomplished with only nearest neighbor hoppings via the
free lattice action
$$
S=
\sum_{q,\mu} 
{2K}
\overline{\tilde\psi} 
(i\gamma_\mu \sin(2\pi q_\mu/L) + r(1-\cos(2\pi q_\mu/L)) 
\tilde\psi 
+m\overline{\tilde\psi}\tilde\psi.
\myeqno
$$
Here I have explicitly added a mass term and made the nearest neighbor
coupling proportional to a ``hopping parameter'' $K$.  In position
space this reads
$$
S=K\sum_{j,\mu} 
\overline\psi_j (r+\gamma_\mu)\psi_{j+e_\mu}
+\overline\psi_j (r-\gamma_\mu)\psi_{j-e_\mu}
+M\sum_j
\overline\psi_j \psi_j \myeqname{\wilsonferm}
$$
where the diagonal terms have been combined to give $M=8Kr+m$.  This
``Wilson Fermion'' action indeed solves the doubling problem, leaving
only one light Fermion species.  Note that when the ``Wilson
parameter'' $r$ is unity the hopping terms involve projection
operators $1\pm\gamma_\mu$.  This simplifies many algebraic
manipulations and is the most popular choice.  The two parameters $M$
and $K$ are the natural ones in lattice language, and relate to the
physical mass and lattice spacing.

In this scheme the price paid for removing the doublers is any exact
chiral symmetry in the cutoff theory.  One consequence is that when
interactions are turned on, renormalizations may be required which
involve operators that explicitly break chiral symmetry.  In
particular, the bare mass can receive an additive renormalization.
There is nothing to protect a massless Fermion, so preserving a light
pion mass will require finely tuning a mass counter-term.  For the
strong interactions there is nothing in principle known to be wrong
with such an approach, and indeed one philosophy is to just do this
renormalization, fitting particle masses and seeing that the final
spectrum comes out acceptably.  If one does not care about chiral
issues, the Wilson approach is a perfectly acceptable regularization
for Fermionic fields.

For a chiral gauge theory, such as the standard model, this approach
is less clear.  The breaking of chiral symmetry inherent in the Wilson
term will break the gauge symmetry, introducing an infinite number of
possible gauge variant counter-terms.  This is a rather unpleasant
situation which would be nice to avoid.  Whether that is possible is
still unresolved, although extensive studies using gauge fixing in
conjunction with the lattice are ongoing (Borelli, Maiani, Sisto,
Rossi, and Testa, 1990; Alonso, Boucaud, Cor\'es, and Rivas, 1991;
Bock, Golterman, and Shamir, 1998a,b).

\bigskip
{\bf \mysecno An infinity of Fermions}
\medskip

The last few years have seen an intense revitalisation of interest in
chiral symmetry in the lattice context.  Much of this was sparked by
Kaplan (Kaplan, 1992) but the ideas have spread and joined with
several other schemes.  Indeed, (Narayanan and Neuberger, 1993a,b,
1994, 1995; Randjbar-Daemi and Strathdee, 1996a,b) emphasized the
common feature that for the anomaly to work in a lattice theory, there
must be some mechanism to absorb the Fermionic modes involved in
anomalous processes.  In the band theory language discussed earlier,
for a state upwardly moving in energy not to be accompanied by a
lowering one of the same chirality, it must be absorbed somewhere.  In
the continuum it comes and goes from the infinite Dirac sea.  On a
lattice with a finite number of degrees of freedom this is impossible.
This suggests extending the theory with an infinite reservoir of
Fermionic states.  Indeed, one point of view is that any chiral gauge
formulation on the lattice that preserves all the chiral symmetries
must have an infinite number of auxiliary states in the Fermionic
sector.  In this sense all the recent schemes do something like this,
differing only in where these states appear in the formalism.

Kaplan's suggestion has become known as the domain-wall approach.
This involves allowing the Fermions to move in an extra space-time
dimension.  The low energy states are surface modes bound to a four
dimensional surface of the underlying five dimensional system.  As
energies increase, the states penetrate further into the extra
dimension, which absorbs the upwardly rising states of anomalous
processes.  But as far as low energies are concerned, the system
appears four dimensional.  I will discuss this mechanism in
considerably more detail in later sections, but for now I note that an
important part of the construction is that the gauge fields themselves
do not observe this new coordinate, it is only for Fermions.  In some
sense the extra dimension is not really physical, and might be thought
of as an ``internal symmetry'' or ``flavor'' space.  As the size of
this extra space goes to infinity, the consequences of chiral symmetry
should become exact.

Frolov and Slavnov (1994) present a more abstract description building
directly on an infinite tower of auxiliary states.  In an extensive
sequence of papers Neuberger and Narayanan (1993a,b, 1994, 1995)
unified this approach with that of Kaplan, emphasizing how the effect
of the bottom of the tower can be compactly represented as an overlap
of two states in an auxiliary Hilbert space, effectively the space of
a transfer matrix in the extra dimension.

Another scheme, a bit less explored, involves a finer lattice for the
Fermions and a double limit, with the Fermion lattice spacing going to
zero first (Hernandez and Sundrum, 1995; Hsu, 195; Bodwin, 1995;
Hernandez and Boucaud, 1998).  A variation on this treats the Fermions
directly as continuum fields (Gockeler, Kronfeld, Schierholz and
Wiese, 1993; Kronfeld, 1995; 't Hooft, 1995), acting under the
influence of the gauge fields interpolated from the courser gauge
lattice.  A closely related procedure involves taking continuum
Fermion fields and expanding in a complete set of basis functions
motivated by a lattice holding the gauge fields (Friedberg, Lee, and
Pang, 1994).  On truncation the doubling returns, but as this basis
goes to infinity one again has an infinite tower of Fermionic states.
These approaches are a sharpening of a much older scheme based on
directly using the continuum Fermion propagator (Drell, Weinstein, and
Yankielowicz, 1976; Svetitsky, Drell, Quinn, and Weinstein, 1980).
Whether these approaches are practical remains controversial; however
they do share the property of introducing a large number of additional
degrees of freedom over naive Fermion approaches.

My later presentation will concentrate on the domain wall approach.
Nevertheless, I become more general in one section where I briefly
discuss a promising class of schemes based on the Ginsparg-Wilson
(1982) relation to generalize the continuum symmetries to an exact
lattice symmetry.  This approach parallels continuum discussions in
diverting the complications of anomalies appear to the Fermionic
measure.

\bigskip
{\bf \mysecno Strong CP violation}
\medskip

As discussed above, scale anomalies reduce the number of physically
independent dimensionless parameters in the quark confining dynamics
of quarks and gluons.  For massless quarks, dimensional transmutation
leaves no undetermined continuous parameters; the theory only depends
the gauge group and the number of Fermion flavors.

With masses for the quarks, the number of parameters increases.
Naively there is one additional mass parameter for each quark.
However, due to chiral anomalies, one more parameter, usually called
$\Theta$, is hidden in the phases in these quark masses.  If
non-zero, this parameter gives rise to CP violating processes in the
strong interactions.  Such appear to be extremely small in nature,
suggesting that this parameter may vanish.  In a grand unified context
this raises puzzles since CP violation is present in the weak
interactions and has no particular reason to be small.  Possible
resolutions to this problem go beyond the scope of this paper; for
reviews see Turner (1990) and Raffelt (1990).

For the lattice it is crucial that any Fermion formulation be able to
account for this parameter.  To define a phase for a single Fermion
mass, break the naive mass term into two parts
$$
\overline\psi \psi = \overline\psi_L \psi_R +\overline\psi_R \psi_L \myeqno
$$
where the left and right parts are eigenstates of $\gamma_5$
$$
\psi_{R,L}={1\over 2} (1\pm \gamma_5) \psi_{R,L}. \myeqno
$$
Then a generic complex mass term has the form 
$$
\overline\psi_L M\psi_R +\overline\psi_R M^*\psi_L \myeqno
$$
with $M$ a complex number, i.e. of form $M=me^{i\theta}$.
Another convenient form for the mass term is
$$
m_1 \overline\psi\psi + im_2\overline\psi\gamma_5\psi
\myeqname\monemtwo
$$
so that $M=m_1\cos(\theta)+im_2\sin(\theta)$.

At the classical level the phase of the quark mass is easily removed
by a chiral rotation.  The change of variables
$$\eqalign{
&\psi \rightarrow e^{i\theta\gamma_5/2} \psi \cr
&\overline\psi \rightarrow \overline\psi e^{i\theta\gamma_5/2} \cr
} \myeqname\chiralrot
$$
takes
$$
M\rightarrow e^{i\theta} M. \myeqno
$$
Since this is just a redefinition, if the path integral measure is
invariant under the same phase change, theta can be rotated away.
However, here the chiral anomalies come into play.  Flavor non-singlet
axial symmetries are believed to be preserved; so, one can eliminate
any non-singlet phases in the mass.  However one overall phase always
remains.  Using the anomaly, this can be moved from the quark masses
into pure gauge terms of the action; indeed, the latter form is the
starting point for most conventional discussions of $\Theta$ in terms
of topological structures in the gauge sector ('t Hooft, 1986).

How this physics manifests itself in a theoretical formalism depends
on approach.  With a Pauli-Villars regulator, the heavy auxiliary
field has a phase in its mass; the theta parameter is the relative
phase between the regulator mass and that of the physical Fermion.  In
other continuum schemes the phase is often pushed into the path
integral measure (Fujikawa, 1979).

In usual lattice approaches, the Fermion measure is a direct product
of discrete Grassmann integrals, leaving no room in the measure for
forbidding the change of variables alluded to above.  However with
Wilson Fermions we have the added Wilson term, which is itself of a
mass like form.  Theta then becomes a relative phase between the
Wilson term and the explicit mass (Seiler and Stamatescu, 1982).

For another angle on the meaning of theta, it is useful to think in
terms of effective chiral Lagrangians ('t Hooft, 1986; Creutz, 1995a;
Evans, Hsu, Nyffeler, and Schwetz, 1997; Smilga, 1999; Tytgat, 1999).
This temporarily leads away from the lattice, but gives insight into
the expected structure of Wilson Fermions with complex mass.  Also,
the chiral Lagrangian approach will be useful in later sections for
understanding other phenomena of direct relevance to the lattice, such
as the Aoki phase and the Wess-Zumino motivation for exploring higher
dimensions.

\midinsert
\epsfxsize .4\hsize
\centerline {\epsfbox{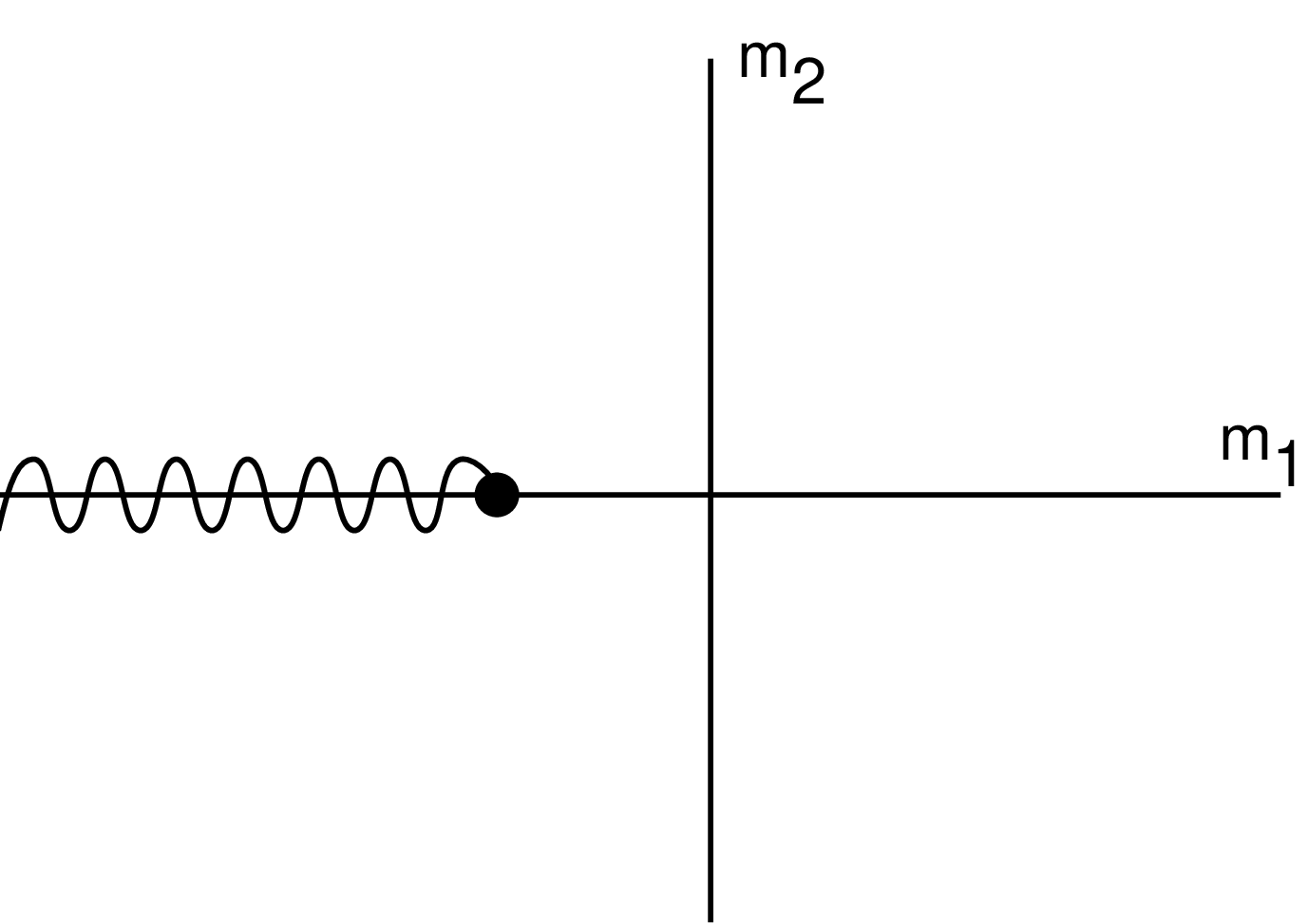}}\bigskip
\narrower{\myfigname\oneflavor The phase diagram for the one flavor case.
The wavy line represents a first-order phase transition, along
which $i\overline\psi\gamma_5\psi$ acquires an expectation value.  The
end point of this transition line is renormalized away from the origin
towards negative $m_1$.}
\endinsert

\midinsert
\epsfxsize .4\hsize
\centerline {\epsfbox{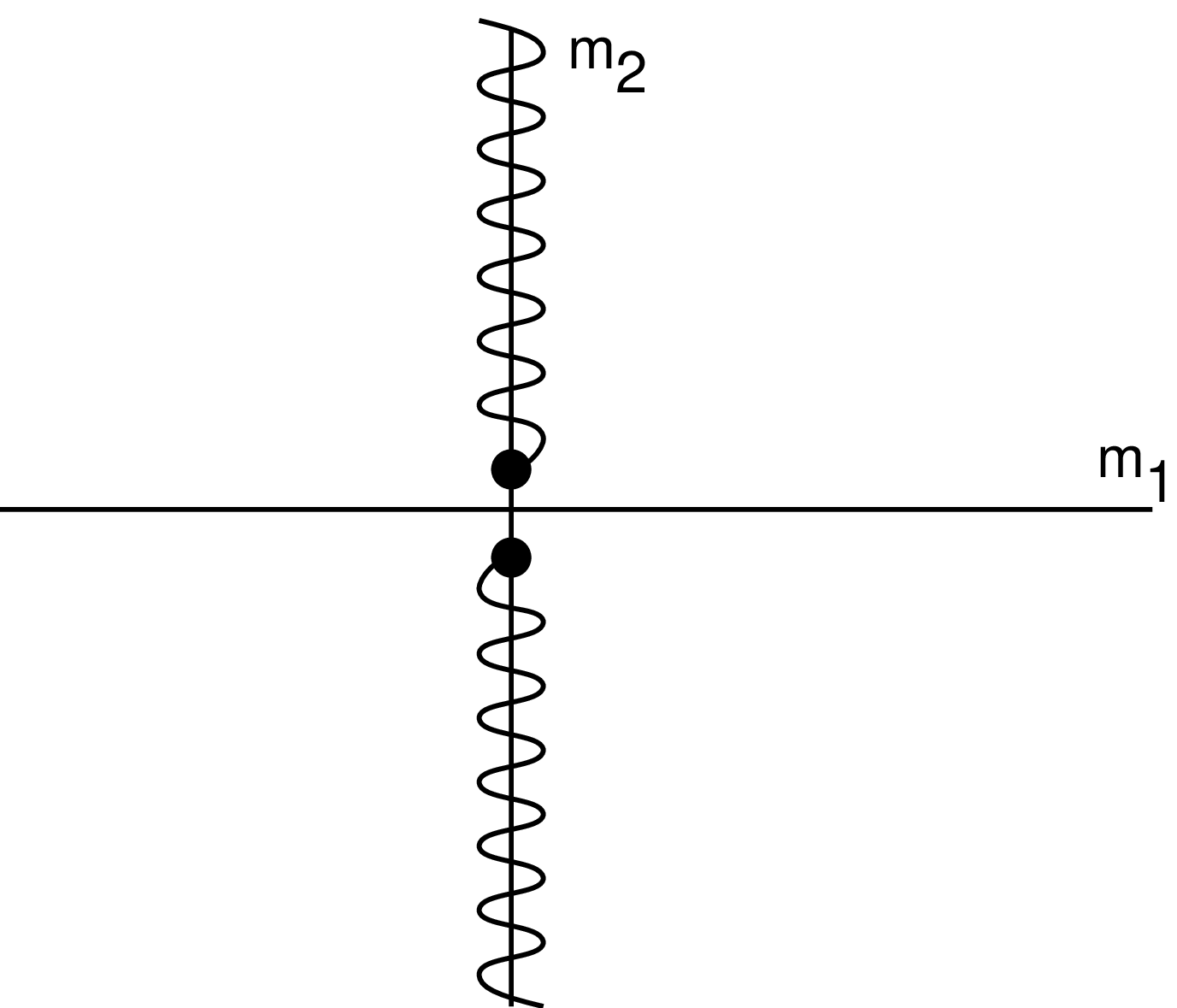}}\bigskip
\narrower {\myfigname\twoflavor The two flavor phase
diagram. First-order lines run up and down the $m_2$ axis.  The second
order endpoints of these lines can be separated by flavor breaking
terms.  The chiral limit is pinched between these endpoints.}
\endinsert

The phase diagram in the $(m_1, m_2)$ plane strongly depends on the
number of Fermion flavors.  In later sections I will be more explicit,
but let me briefly summarize the qualitative picture.  With a single
species, a first-order phase transition line runs down the negative
$m_1$ axis, starting at non-zero $m_1$.  This is sketched in
\oneflavor.  For two flavors there are two first-order phase
transition lines, starting near the origin and running up and down the
$m_2$ axis.  For degenerate quarks these transitions meet at the
chiral limit of vanishing Fermion mass; a small flavor breaking can
separate the endpoints of these first-order lines.  This is sketched
in \twoflavor.  With $N_f>2$ flavors, the $(m_1,m_2)$ plane has $N_f$
first-order phase transition lines pointing at the origin.  The
conventionally normalized parameter $\Theta$ is $N_f$ times the angle
to a point in this plane, and these transition lines are each
equivalent to $\Theta$ being $\pi$.

Whenever the number of flavors is odd, a first-order transition runs
down the negative $m_1$ axis.  Along this line there is a spontaneous
breaking of CP, with the natural order parameter being $\langle i
\overline\psi\gamma_5\psi\rangle$.  This possibility of a spontaneous
breakdown of parity was noted some time ago by Dashen (1971) and has
reappeared at various times in the lattice context (Smit, 1980; Aoki,
1989; Aoki and Gocksch, 1992).  Any valid algorithm for dealing with
an odd number of flavors must be able to distinguish the sign of the
mass.  The following three sections give detailed continuum arguments
for this general structure, dealing successively with the three, one,
and two flavor cases.

\bigskip
{\bf \mysecno Three flavors}
\medskip

To see how the above picture arises naturally in a chiral Lagrangian
approach, start with the three flavor case.  I begin with a lightning
summary of effective theories of Goldstone Bosons.  Suppressing other
indices, consider left and right handed quark fields
$\psi_L^a,\psi_R^a$ with a flavor index $a$ running from one to three.
Spontaneous chiral symmetry breaking appears in a non-vanishing
expectation
$$
\langle \overline \psi^a_L\psi^b_R\rangle=vg^{ab}.
\myeqno
$$
The vacuum is not unique, and is labeled by $g^{ab}$, an element of
$SU(3)$.  The basic $SU(3)_L\otimes SU(3)_R$ chiral symmetry is
realized via the global transformation
$$
g\rightarrow g_L^\dagger g g_R.
\myeqname\chsym
$$
Picking $g=I$ for a standard vacuum, the 8 fields $\pi^\alpha$ which
excite the Goldstone modes are nicely parametrized by
$$
g=e^{i\lambda^\alpha\pi^\alpha/F_\pi}
\myeqno
$$
where the matrices $\lambda^\alpha$ are a set of generators for the
group $SU(3)$ and are normalized ${\rm Tr}\lambda^\alpha
\lambda^\beta=2\delta^{\alpha\beta}$

Effective Lagrangians start from the assumption that in the low energy
limit only the Goldstone excitations are important; one considers an
effective theory depending on $g$ alone.  For low momenta, this theory
is expanded in terms of increasing numbers of derivatives.  To lowest
order the effective Lagrangian is
$$
L={F_\pi^2\over 4} {\rm Tr}(\partial_\mu g^\dagger \partial_\mu g).
\myeqno
$$
This is invariant under the global symmetry of \chsym.  The parameter
$F_\pi$ is related to the pion decay, which occurs through the axial
vector coupling to the intermediate weak Boson.  The phenomenological
value is $F_\pi=93$ MeV.

To give the quarks masses, add an explicit symmetry-breaking term.
For degenerate quarks the simplest possibility adds to $L$ a potential
term
$$
V=-{m\over v}{\rm ReTr}g.
\myeqname\massterm
$$
Minimizing $V$ for positive $m$ selects the standard $g=I$ to
represent the vacuum.  The pions are no longer massless, but acquire a
mass proportional to the square root of the quark mass.  For $SU(2)$,
$-I$ is an element of the group, so changing the sign of the mass
simply induces a rotation to a new vacuum represented by $g=-I$.
However $-I$ is not an element of $SU(3)$, requiring a somewhat more
detailed analysis.

This leads me to a digression on some details of the $SU(N)$ group
manifold.  In particular, where is ${\rm ReTr}(g)$ extremal?  At such
a point, first order changes in $g$ must vanish; in particular
$$
0={d\over d\epsilon^\alpha}{\rm ReTr}
(ge^{i\lambda^\beta\epsilon^\beta})\vert_{\epsilon=0}
=i{\Tr}(g-g^\dagger)\lambda^\alpha. 
\myeqno
$$
The $\lambda$ matrices are almost a complete set, only the identity is
missing.  This indicates that $g-g^\dagger$ is proportional to the
identity; thus I write
$$
g-g^\dagger= i s I
\myeqname\constimag
$$
where the constant $s$ is yet to be determined.

Since ${\rm Tr} g$ is a class function, the extrema being discussed
are by definition classes in the group.  Given some extremal
point $g$, all elements of form $h^\dagger g h$, with $h$ an arbitrary
group element, are in the same equivalence class.  Using this freedom,
define a standard member of the class by diagonalizing $g$ and
ordering the diagonal elements by increasing magnitude of the real
part.  After doing this, the diagonal elements are phases, and by
\constimag all have the same imaginary part.  The
diagonals must all be taken from the two numbers $c+is$ or $-c+is$
with $c=\sqrt{1-s^2}$.

In general there are a variety of such extremal elements, some of
which are maxima, some minima, and some saddle points of ${\rm
ReTr}g$.  To distinguish them, look at the second variation
$$
{d^2\over {d\epsilon}^2}{\rm ReTr}
(ge^{i\lambda\epsilon})\vert_{\epsilon=0}
=-{\Tr}((g+g^\dagger)\lambda^2)
\myeqno
$$
with $\lambda$ a generator of the group.  Using diagonal generators,
it is a straightforward exercise to see that for a maximum (minimum),
the real parts of $g$ must all be positive (negative).  All other
solutions are saddle points.  Thus all the maxima and minima are
elements of the group center.

\midinsert
\epsfxsize .6\hsize
\centerline {\epsfbox{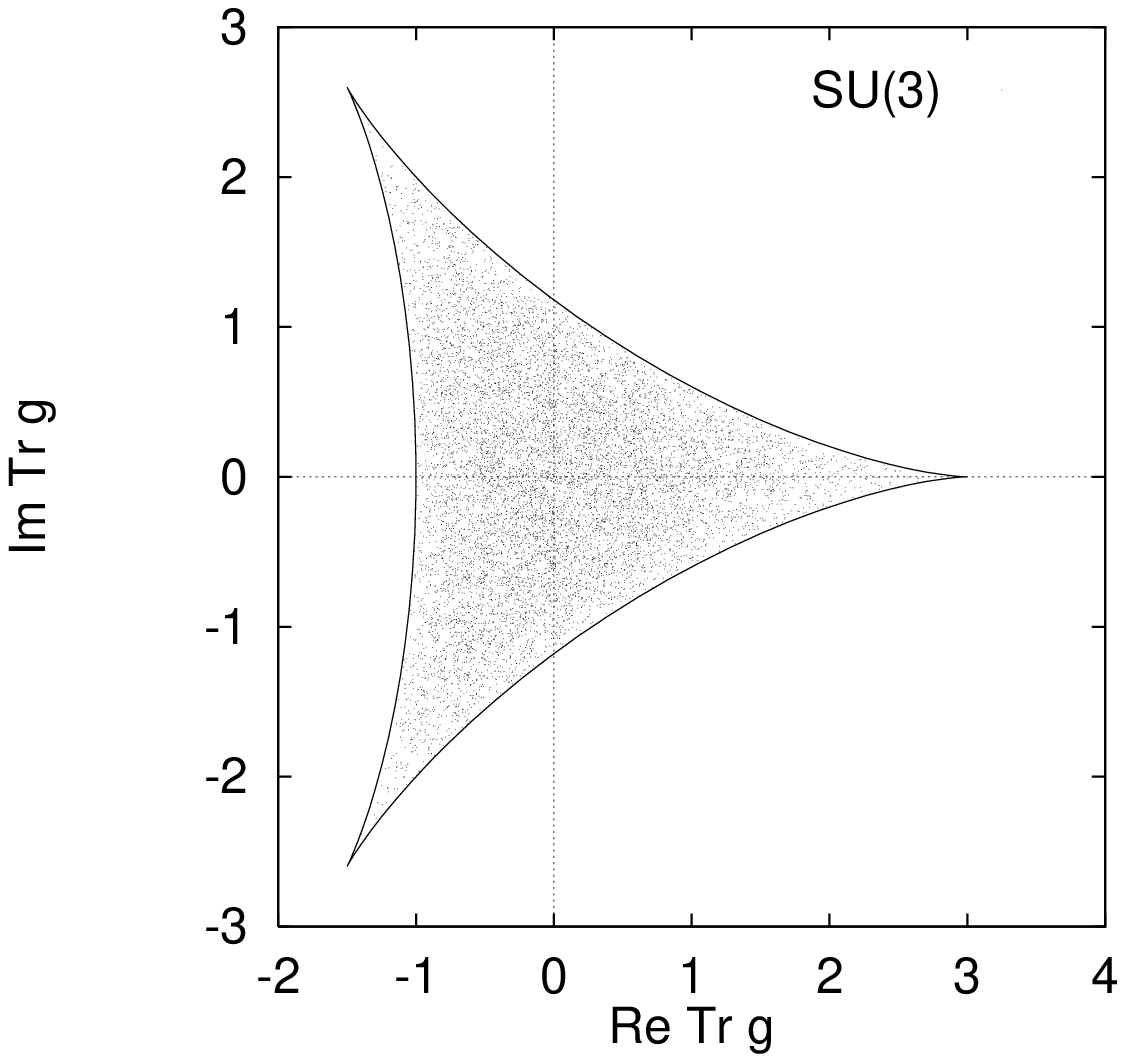}}\bigskip
\narrower {\myfigname\suthree Scatter plot of the traces of 10,000
randomly selected $SU(3)$ matrices.  They are enclosed in the envelope
discussed in the text.  The spikes occur at the center elements of the
group.

}
\endinsert

Specializing to $SU(3)$, there are four extremal classes.  The single
maximum of ${\rm ReTr}g$ occurs at the identity, two degenerate minima
occur at the other two center elements $g=e^{\pm2\pi i/3}$, and a
class of saddle point elements is represented by
$$
g=\pmatrix{ -1 & 0 & 0\cr 0 & -1 &
0\cr 0 & 0 & 1\cr }.
\myeqno
$$
A complex curve that passes through all these extrema and encloses all
values of ${\rm Tr}g$ is given by
$$z={\rm
Tr}\exp(i\theta\lambda)
\myeqno
$$
with 
$$
\lambda=\pmatrix{ 1 & 0 & 0\cr 0 & 1 & 0\cr 0 & 0 & -2\cr }
\myeqno
$$ 
as theta runs from 0 to $2\pi$.  In \suthree I plot this curve along
with the traces of 10,000 randomly chosen $SU(3)$ matrices.

Using this basic structure of the $SU(3)$ group, it is now
straightforward to see
how the picture of first order phase transitions in the
phase of the mass arises from the potential in \massterm.
For $m>0$ the vacuum is the usual one with $g=I$.  For a real but
negative mass, the vacuum should lie at a minimum of ${\rm ReTr}g$,
which occur in a degenerate pair at $g=e^{\pm2\pi i/3}$.  Adding a
small imaginary piece to $m$ breaks this degeneracy, showing the first
order nature of the transition along the negative $m$ axis.

With a diagonal mass term $m=|m|e^{i\Theta/3}$ with a general phase,
the vacuum is always represented by one of these three extremal
classes.  For $\Theta=\pi$ there is a discontinuity, with the system
jumping discontinuously from $g=I$ to $g=Ie^{2\pi i/3}$.  This is
exactly the first order transition alluded to in the previous section.
The three transition lines in the $(m_1,m_2)$ plane are physically
equivalent since a phase change $m\rightarrow m e^{2\pi i/3}$ can be
absorbed by taking $g\rightarrow g e^{-2\pi i/3}$.  Later I will
extend the above discussion to non-degenerate masses, providing a nice
route between various numbers of flavors.

\bigskip
{\bf \mysecno One flavor}
\medskip

The one flavor situation, while not phenomenologically particularly
relevant, is fascinating in its own right.  In this case anomalies
remove all chiral symmetries from the problem.  No massless Goldstone
Bosons are expected, and there is nothing to protect the quark mass
from additive renormalization.  Nevertheless, I have argued previously
that there should be a non-trivial dependence on the phase of the
mass.  A large negative mass should be accompanied by a spontaneous
breakdown of parity, as sketched in
\oneflavor.

A quick but dirty argument gives the expected picture.  With only one
flavor, there is only one light pseudo-scalar meson, which I call the
$\eta$.  Were it not for anomalies, conventional chiral symmetry
arguments suggest the mass squared of this particle would go to zero
linearly with the quark mass,
$$
m_\eta^2 \sim m_q.
\myeqno
$$
But, just as the $\eta\prime$ gets mass from anomalies, a similar
contribution should appear here; assume it is simply an additive
constant
$$
m_\eta^2 \sim m_q+C.
\myeqno
$$
Try to describe this model by an effective potential for the $\eta$
field.  This should include the possibility for these particles to
interact, suggesting something like
$$
V(\eta)= {a m_q+C\over 2} \eta^2 + \lambda \eta^4.
\myeqno
$$ 
At $m_q=-C/a$ the effective mass of the eta goes negative.  This
should give a spontaneous breaking in the canonical manner, with the
field acquiring an expectation value
$$
\langle \eta \rangle \sim \langle \overline\psi \gamma_5\psi\rangle \ne 0.
\myeqno
$$
As this is an odd parity field, parity is spontaneously broken.  In
particular, odd numbers of physical mesons can be created, unlike in
the unbroken theory where the number of mesons is preserved modulo 2.

Note that this transition occurs at a negative quark mass, and nothing
special happens at $m_q=0$.  Of course the bare quark mass is a
divergent quantity in need of renormalization.  Without chiral
symmetry, there is nothing to prevent an additive shift in this
parameter.  Nevertheless, with a cutoff in place, these qualitative
arguments suggest it is only for negative quark mass that this parity
violating phase transition will take place.

The anomaly involves processes that mix left and right handed quarks.
This is exactly the role of a mass term, and thus the quark-antiquark
pseudo-scalar bound state, the analogue of the $\eta^\prime$ meson,
gains a mass.  Now imagine making the quark masses slightly negative.
This can cancel some of the mass generated by the anomaly and should
reduce the mass of the meson.  If the quarks become sufficiently
negative in mass, one might decrease the bare mass of the meson to
negative values.  This gives the classic situation of spontaneous
symmetry breaking and the meson field acquires an expectation value.

This argument is not at all rigorous.  To lend more credence to this
qualitative picture, note that a similar phenomenon occurs in two
dimensional electrodynamics.  The Schwinger model is exactly solvable
at zero bare mass, with the spectrum being a free massive Boson.
However for negative bare mass qualitative semi-classical arguments
indicate the same structure as discussed in the previous paragraph,
with a spontaneous generation of a parity violating background
electric field (Creutz, 1995b).  Under the Bosonizaton process, the
quark mass term corresponds to a sinusoidal term in the effective
potential for the scalar field
$$
m\overline\psi \psi \leftrightarrow m\cos(2\sqrt\pi\eta).
\myeqno
$$
Regularization and normal ordering are required for a proper
definition but are not important here (Coleman, 1976).  Combining this
with the photon mass from the anomaly suggests an effective potential
for the $\eta$ field of form
$$
V(\eta) \sim {e^2\over 2\pi} \eta^2 - m \cos(2\sqrt\pi\eta).
\myeqno
$$
For small positive $m$, the second term shifts the Boson mass and
introduces a four meson coupling, making the theory no longer free.
If the mass is negative and large enough, the cosine term can dominate
the behavior around small $\eta$, making the perturbative vacuum
unstable.  The Bosonization process relates
$\overline\psi\gamma_5\psi$ with $\sin(2\sqrt\pi\eta)$; thus, when
$\eta$ gains an expectation value, so does the the pseudo-scalar
density.  Since the scalar field represents the electric field, this
symmetry breaking represents the spontaneous generation of a
background field.  As discussed by Coleman (1976), this corresponds to
a non-trivial topological term in the action, usually referred to as
$\Theta$.

A third way to understand the one flavor behavior is to consider a
larger number of flavors and give all but one large masses.  In
(Creutz, 1995a) this procedure was done starting with two flavors.
That case required some additional assumptions regarding the spectrum;
these assumptions become less important for larger symmetry groups.
Thus I consider the three flavor case, and give two quarks a larger
mass than the third.

I begin with the effective three flavor Lagrangian of the previous
section.  With two quarks of mass $M$ and one of mass $m$, consider
the potential
$$
V(g)\propto -{\rm ReTr}\left\{ g\pmatrix{
m & 0 & 0\cr
0 & M & 0\cr
0 & 0 & M\cr
}\right\}.
\myeqno
$$
It is convenient to break this into two terms
$$
V(g) \propto -{M+m\over 2}\ {\rm ReTr} (g)
+{M-m\over 2}\ {\rm ReTr}(gh)
\myeqname\competing
$$
where
$$
h=\pmatrix{
1 & 0 & 0\cr
0 & -1 & 0\cr
0 & 0 & -1\cr
}.
\myeqno
$$
The minimum of the first term in \competing was worked out in the
previous section; when $M+m$ is positive this occurs at the identity
element.  For the second term, note that I have written the factor of
$g$ as an $SU(3)$ group element.  Via the previous discussions, the
extrema of this term occur when the product $gh$ is in the group
center.  For the case $M-m$ positive, there is a degenerate pair of
minima occurring at
$$
g=e^{\pm2\pi i/3}h.
\myeqno
$$
\competing has two competing terms, one having a unique minimum and
the other having two degenerate ground states.  For the degenerate
case with $M=m$, only the first term is present and the vacuum is
unique.  However when $m=-M$ only the second term is present with its
corresponding pair of degenerate vacuua.  Between these points there
is a critical value $m_c$ where the situation shifts between a unique
and a doubly degenerate vacuum.

To determine the critical mass, consider matrices of form
$$
g=\exp\left\{i\phi\pmatrix{ -2 & 0 & 0\cr 0 & 1 & 0\cr 0 & 0 & 1\cr }\right\}.
\myeqno
$$
For these the potential is
$$
V(\phi) \propto -m\cos(2\phi)-2M\cos(\phi).
\myeqno
$$
The extremum at $\phi=0$ changes from a minimum to a maximum at
$m=-M/2$, the desired critical point.  As discussed at the beginning
of this section, it occurs at a negative value of $m$.  The only
dimensional scale present is $M$, to which the result must be
proportional.  This analysis immediately generalizes to larger groups:
for $N_f$ flavors $m_c={-1\over N_f-1}$.

This discussion suggests that a similar phenomenon should occur with
one flavor of Wilson Fermion.  Here the bare mass is controlled by the
hopping parameter.  As the hopping parameter increases, the Fermion
mass decreases.  In the plane of the gauge coupling and hopping
parameter, a critical line should mark where the above parity breaking
begins.  In the lattice context the possibility of such a phase was
mentioned briefly by Smit (1980), and extensively discussed by Aoki
(1989) and Aoki and Gocksch (1992).  The latter papers also made some
rather dramatic predictions for the breaking of both parity and flavor
symmetries when more quark species are present.  I will return to this
phenomenon in section XII.

\bigskip
{\bf \mysecno Two flavors}
\medskip

The two flavor case can also be obtained as a limit from three
flavors.  However, since $SU(2)$ is a simple $S_3$ sphere, it is a
somewhat more intuitive manifold.  Here several simplifications make
some features of the physics particularly clear.  In this section for
variety I will switch to a ``linear'' realization of the effective
chiral Lagrangian.  The discussion is closely based on (Creutz,
1995a).

I begin by defining eight fields around which the discussion
revolves
$$
\eqalign
{
&\sigma=c\overline\psi\psi\cr
&\vec\pi=ic\overline\psi\gamma_5\vec\tau\psi\cr
&\eta=ic\overline\psi\gamma_5\psi\cr
&\vec\delta=c\overline\psi\vec\tau\psi.\cr
} \myeqname\fields
$$
The Fermion $\psi$ has two isospin components, for which $\vec\tau$
represents the standard Pauli matrices.  The factor $c$ is inserted to
give the fields their usual dimensions.  Its value is not particularly
relevant to the qualitative discussion that follows, but one
convention is take $c=F_\pi/\vert\langle\overline\psi\psi\rangle\vert$
where $F_\pi$ is the pion decay constant and the condensate is in the
standard {\ae}ther.

Corresponding to each of these fields is a physical spectrum.  In some
cases this is dominated by a known particle.  These include the
familiar triplet of pions around 140 MeV and the eta at 547 MeV.  The
others are not quite so clean, with a candidate for the isoscalar
$\sigma$ being the $f_0(980)$ and for the isovector $\delta$ being the
$a_0(980)$.  I will only use that the lightest particle in the
$\delta$ channel appears to be heavier than the $\eta$.
  
Now consider an effective potential
$V(\sigma,\vec\pi,\eta,\vec\delta)$ for these fields.  I first
consider the theory with vanishing quark masses.  In the continuum
limit, the strong coupling constant is absorbed via the phenomenon of
dimensional transmutation (Coleman and Weinberg, 1973), and all
dimensionless quantities are in principle determined.  In the full
theory with the quark masses restored, the only parameters
should be those masses and $\Theta$.

For the massless theory many of the chiral symmetries become exact.
Because of the anomaly, the chiral transformation of \chiralrot, which
mixes the $\sigma$ and $\eta$ fields, is not a good symmetry.  However
flavored axial rotations should be valid.  For example, the rotation
$$
\psi \longrightarrow e^{i\gamma_5\tau_3\phi/2} \psi             \myeqno
$$
which mixes $\sigma$ with $\pi_3$ 
$$
\eqalign
{
\sigma 
&
\longrightarrow \cos(\phi) \sigma + \sin(\phi)\pi_3\cr 
\pi_3  
&
\longrightarrow \cos(\phi)\pi_3 - \sin(\phi) \sigma \cr
}
\myeqno
$$
should be a good symmetry.
This transformation also mixes $\eta$ with $\delta_3$
$$
\eqalign
{
\eta     
&
\longrightarrow \cos(\phi) \eta + \sin(\phi)\delta_3\cr 
\delta_3 
&
\longrightarrow  \cos(\phi)\delta_3 - \sin(\phi) \eta. \cr 
}
\myeqno
$$
For the massless theory, the effective potential is invariant under
such rotations.  In this two flavor case, the consequences can be
compactly expressed by going to a vector notation.  Define the four
component objects $\Sigma=(\sigma,\vec\pi)$ and
$\Delta=(\eta,\vec\delta)$.  The effective potential is a function
only of invariants constructed from these vectors.  A complete set of
invariants is $\{\Sigma^2,\ \Delta^2,\ \Sigma\cdot\Delta\}$.  This
separation into two sets of fields is special to the two flavor case,
but makes the behavior of the theory particularly transparent.

\midinsert

\epsfxsize .5\hsize
\centerline {\epsfbox{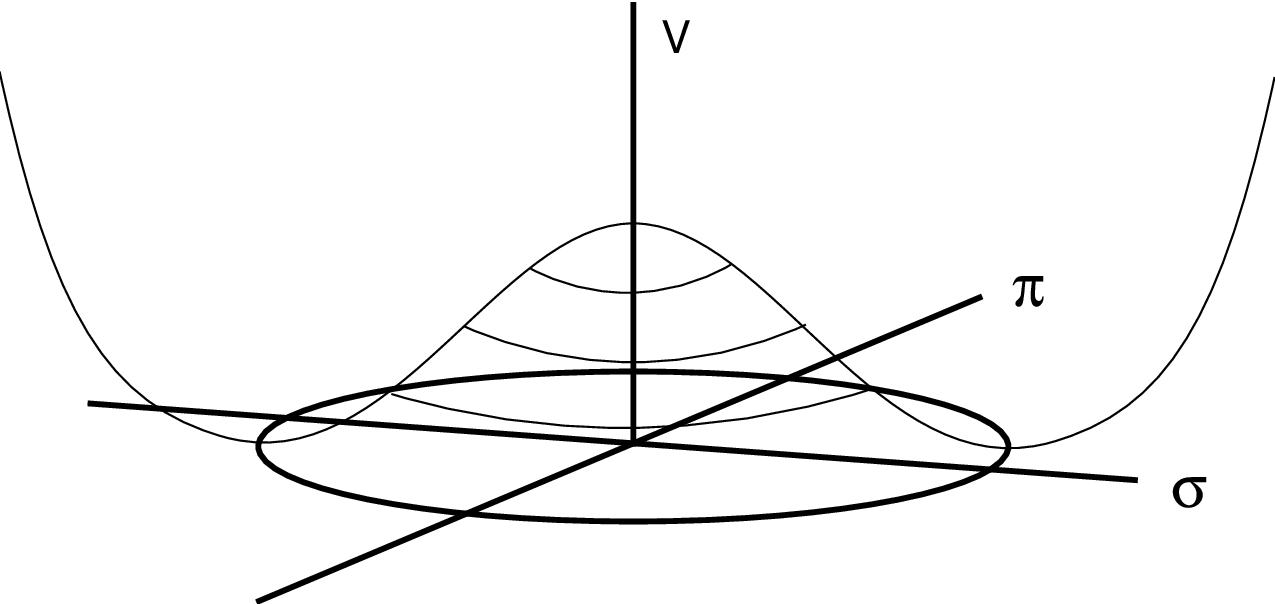}}\bigskip

\narrower {\myfigname\sombrero The ``sombrero'' potential representing
the chiral limit of massless quarks.  Possible distortions of the circle
representing the bottom of this potential are the main subject of this
section.}

\endinsert

I now use the experimental fact that chiral symmetry is spontaneously
broken.  The minimum of the effective potential should not occur for
all fields having vanishing expectation.  Assuming that parity and
flavor are good symmetries of the strong interactions, the expectation
value of the fields can be chosen in the $\sigma$ direction.
Temporarily ignoring the fields $\Delta$, the potential should have
the canonical ``sombrero'' shape, as stereotyped by the form
$$
V=\lambda(\Sigma^2-v^2)^2=\lambda(\sigma^2+\vec\pi^2-v^2)^2. \myeqno 
$$
Here $v$ is the magnitude of the {\ae}ther expectation value for
$\sigma$, and $\lambda$ is a coupling strength related to the $\sigma$
mass.  The normalization convention mentioned below \fields would have
$v=F_\pi/2$.  I sketch the generic structure of the potential in
\sombrero. This gives the standard picture of pions as Goldstone
Bosons associated with fields oscillating along degenerate minima.

Now consider the influence of the fields $\Delta$ on this potential.
Taking small fields, I expand the potential about vanishing $\Delta$
$$
V=\lambda(\Sigma^2-v^2)^2+\alpha \Delta^2 - \beta
(\Sigma\cdot\Delta)^2+\ldots
\myeqname\vexpand
$$
Being odd under parity, $\Sigma\cdot\Delta$ appears
quadratically.

The terms proportional to $\alpha$ and $\beta$ generate masses for the
$\eta$ and $\delta$ particles.  Since $\Delta^2=\eta^2+\vec\delta^2$,
the $\alpha$ term contributes equally to each.  Substituting
$\Sigma\sim(v,\vec 0)$ gives $(\Sigma\cdot\Delta)^2\sim v^2\eta^2$;
thus, the $\beta$ term breaks the $\eta$-$\vec\delta$ degeneracy.
Here is where the observation that the $\eta$ is lighter than the
$\delta$ comes into play; I have written a minus sign in \vexpand, thus
making the expected sign of $\beta$ positive.

Next I turn on the Fermion masses.  I consider small masses, and
assume they appear as a linear perturbation of the effective potential
$$
V\longrightarrow V-(M_1\cdot\Sigma+M_2\cdot\Delta)/c.
\myeqno
$$ 
Here the four-component objects $M_{1,2}$ represent the possible mass
terms.  The normalization constant $c$ appears in \fields.  The zeroth
component of $M_1$ gives a conventional mass term proportional to
$\overline\psi\psi$, contributing equally to both flavors.  The mass
splitting of the up and down quarks appears naturally in the third
component of $M_2$, multiplying $\overline\psi\tau_3\psi$.  The term
$m_2$ of \monemtwo occupies the zeroth component of $M_2$.

The chiral symmetries of the problem require that physics only depends
on invariants.  For these I can take $M_1^2$, $M_2^2$, and $M_1\cdot
M_2$.  That there are three parameters is reassuring; there are the
quark masses $(m_u, m_d)$ and the CP violating parameter $\Theta$.
The mapping between these parameterizations is non-linear, the
conventional definitions giving
$$
\eqalign
{
M_1^2 
&
={1\over 4} (m_u^2+m_d^2)+{1\over 2}m_u m_d\cos(\Theta)\cr
M_2^2 
&
= {1\over 4}(m_u^2+m_d^2)-{1\over 2}m_u m_d\cos(\Theta)\cr
M_1\cdot M_2 
&
={1\over 2} m_u m_d\sin(\Theta).\cr
}
\myeqname\changeparam
$$
If one of the quark masses, say $m_u$, vanishes, then the $\Theta$
dependence drops out.  This is sometimes proposed as a possible way to
remove unwanted CP violation from the strong interactions; however,
having a single quark mass vanish represents a fine tuning which is
not obviously more compelling than simply tuning $\Theta$ to zero.
Also, having $m_u=0$ appears to be phenomenologically untenable
(Donoghue, Holstein, and Wyler, 1992; Leutwyler, 1990; Bijnens,
Prades, de Rafael, 1995).

\midinsert
\epsfxsize .6\hsize
\centerline {\epsfbox{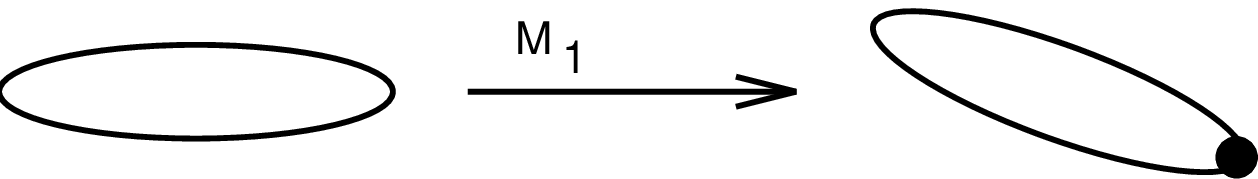}}\bigskip
\narrower {\myfigname\tiltfig 
The effect of $M_1$ is a tilting of the effective potential. The
ellipse in this and the following figures represents the set of minima
for the effective potential from \sombrero.  The dot represents lowest
energy state where the {\ae}ther settles.\par}
\endinsert

I now turn to a physical picture of what the two mass terms $M_1$ and
$M_2$ do to the ``Mexican hat'' structure of the massless potential.
For $M_1$ this is easy; its simply tilts the sombrero.  This is
sketched in \tiltfig.  The symmetry breaking is now driven in a
particular direction, with the tilt selecting the direction for
$\Sigma$ field to acquire its expectation value.  This picture is well
known, giving rise to standard relations such as the square of the
pion mass being linearly proportional to the quark mass (Gell-Mann,
Oakes, and Renner, 1968).

\midinsert
\epsfxsize .6\hsize
\centerline {\epsfbox{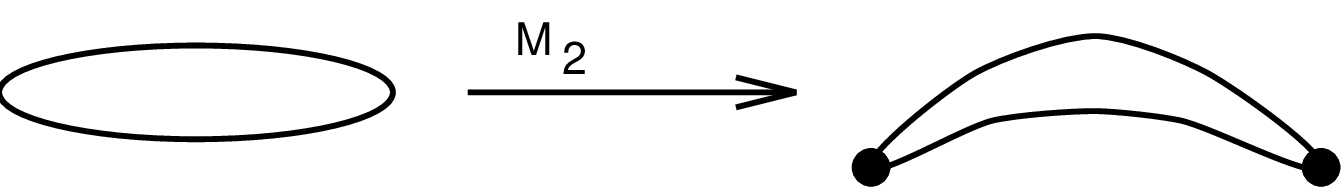}}\bigskip
\narrower {\myfigname\warpfig
The effect of $M_2$ is a warping of the effective potential.  This is
quadratic, with the dots representing the two places where the
{\ae}ther can settle.}
\endinsert

The effect of $M_2$ is more subtle.  This quantity has no direct
coupling to the $\Sigma$ field; so, I must look to higher order.  The
$M_2$ term is a force pulling on the $\Delta$ field; it should give an
expectation value proportional to the strength, $\langle\Delta\rangle
\propto M_2$.  Once $\Delta$ gains an expectation value, it then
affects $\Sigma$ through the $\alpha$ and $\beta$ terms of the
potential in \vexpand.  The $\alpha$ term is a function only of
$\Sigma^2$, and, at least for small $M_2$, should not qualitatively
change the structure of the symmetry breaking.  On the other hand, the
$\beta$ term will warp the shape of the sombrero.  As this term is
quadratic in $\Sigma\cdot\Delta$, this warping is quadratic.  With
$\beta$ positive, as suggested above, this favors an expectation value
of $\Sigma$ lying along the vector $M_2$, but the sign of this
expectation is undetermined.  This effect is sketched in
\warpfig.

To summarize, the effect of $M_1$ is to tilt the Mexican hat, while
the effect of $M_2$ is to install a quadratic warping.  The three
parameters are the amount of tilt, the amount of warping, and,
finally, the relative angle between these effects.  To better
understand the interplay of these various phenomena, I now consider
two specific situations in more detail.

First consider $M_1$ and $M_2$ parallel in the four vector sense.
This is the situation with two mass terms of \monemtwo and no explicit
breaking of flavor symmetry.  Specifically, I take $M_1=(m_1,\vec 0)$
and $M_2=(m_2,\vec 0)$.  In this case the warping and the tilting are
along the same axis.

\midinsert
\epsfxsize .6\hsize
\centerline {\epsfbox{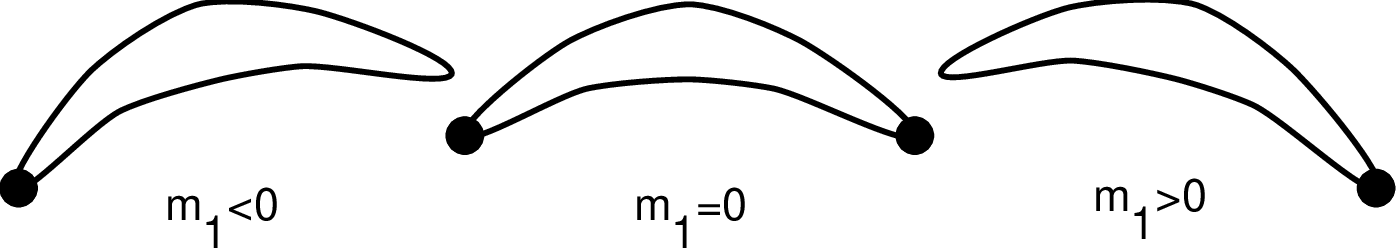}}\bigskip
\narrower {\myfigname\tiltwarpfig 
Varying $m_1$ at fixed $m_2$.  A first-order phase transition is
expected at $m_1=0$.  This corresponds to $\Theta=\pi$.  The dots
represent possible vacuum states.}
\endinsert

Suppose I consider $m_2$ at some non-vanishing fixed value, and study
the state of the {\ae}ther as $m_1$ is varied.  The $m_2$ term has
warped the sombrero, but if $m_1$ is large enough, the potential will
have a unique minimum in the direction of this pull.  As $m_1$ is
reduced in magnitude, the tilt decreases, and eventually the warping
generates a second local minimum in the opposite $\sigma$ direction.
As $m_1$ passes through zero, this second minimum becomes the lower of
the two, and a phase transition occurs exactly at $m_1=0$.  The
transition is first order with the expectation of $\sigma$ jumping
discontinuously.  This situation is sketched in \tiltwarpfig.  From
\changeparam, the transition occurs at $m_u=m_d$ and $\Theta=\pi$.

As $m_2$ decreases, so does the warping, reducing the barrier
between the two minima.  This makes the transition softer.  A small
further perturbation in, say, the $\pi_3$ direction, will tilt the
sombrero a bit to the side.  If the warping is small enough, the field
can then roll around the preferred side of the hat, thus opening a gap
separating the positive $m_2$ phase transition line from that at
negative $m_2$.  In this way sufficient flavor breaking can remove the
phase transition at $\Theta=\pi$ for small masses.

\midinsert
\epsfxsize .6\hsize
\centerline {\epsfbox{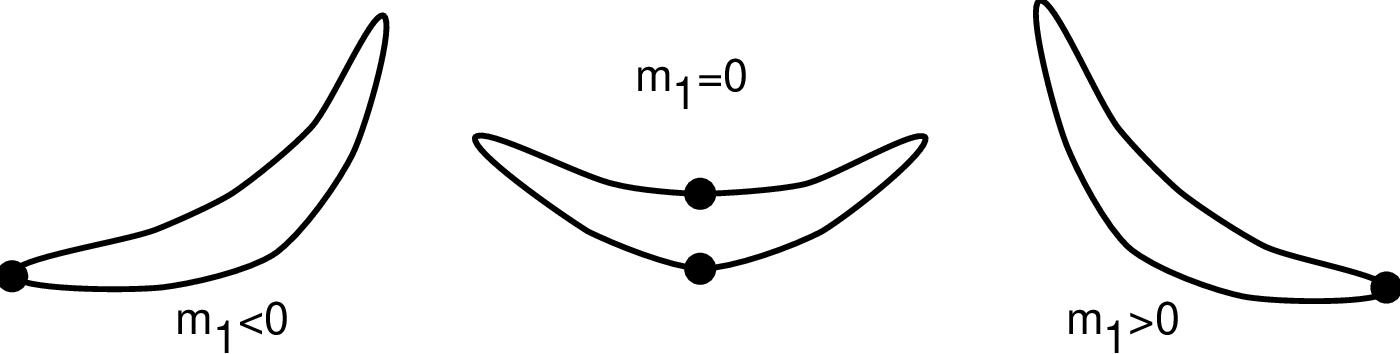}}\bigskip
\narrower {\myfigname\tiltupwarpfig Varying $m_1$ at fixed quark 
mass splitting.  A second order phase transition occurs when the
tilting is reduced sufficiently for a spontaneous expectation of
$\pi_3$ to develop.  The dots represent places where the {\ae}ther can
settle.}
\endinsert

I now turn to a situation where $M_1$ and $M_2$ are orthogonal.  To be
specific, take $M_1=(m_1,\vec 0)$ and $M_2=(0,0,0,\delta m)$, which
physically represents a flavor symmetric mass term $m_1=(m_u+m_d)/2$
combined with a flavor breaking $\delta m=(m_u-m_d)/2$.  Now $M_2$
warps the sombrero downwards in the $\pm \pi_3$ direction.  A large
$m_1$ would overcome this warping, still giving an {\ae}ther with only
$\sigma$ having an expectation value.  However, as $m_1$ decreases in
magnitude with a fixed $\delta m$, there eventually comes a point
where the warping dominates the tilting.  A new symmetry breaking
should occur, with $\pi_3$ acquiring an expectation value.  This is
sketched in \tiltupwarpfig.  As $\pi_3$ is a CP odd operator, this is
a spontaneous breaking of CP.

\midinsert
\epsfxsize .5\hsize
\centerline {\epsfbox{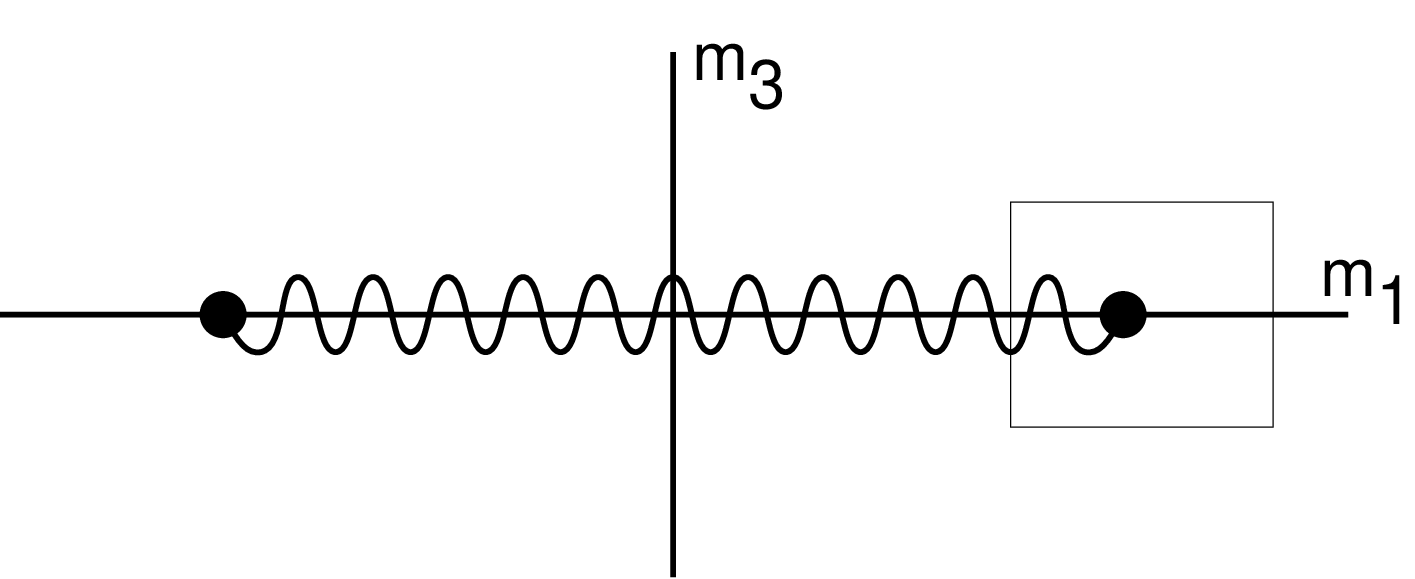}}\bigskip
\narrower {\myfigname\phasediagfig
The $(m_1,m_3)$ phase diagram for unequal mass quarks.  The
wavy line represents a first-order phase transition ending at the
second order dots.  The light box on the right shows how the one
flavor case is extracted.}
\endinsert

To make this into a proper two dimensional phase diagram, I add an
$m_3\pi_3$ piece to the potential.  This effectively twists $M_1$ away
from being exactly perpendicular to $M_2$.  This explicitly breaks CP
and should remove the transition, just as an applied field removes the
phase transition in the Ising model.  We thus have a phase diagram in
the $(m_1,m_3)$ plane with a first-order transition connecting two
symmetrically separated points on the $m_1$ axis.  This is sketched in
\phasediagfig.

Physically, the endpoints of this transition line are associated with
the points where the respective quark masses vanish.  The phase
transition occurs when the two flavors have masses of opposite sign.
Simultaneously flipping the signs of both quark masses can always be
undone by a flavored chiral rotation, say about the $\pi_3$ axis, and
thus is a good symmetry of the theory.

Taking one of the flavors to infinite mass provides another way to
understand the one flavor situation discussed in the previous section.
As sketched in \phasediagfig, this represents looking only at the
vicinity of one endpoint of the transition line.  In terms of the
light species, this transition represents a spontaneous breaking of CP
with a non-vanishing expectation for $i\overline\psi\gamma_5\psi$.

\bigskip
{\bf \mysecno Wilson Fermions for complex mass}
\medskip

I now discuss the suggested implications of these ideas for Wilson's
lattice Fermions.  The picture is rather qualitative, and, as the next
section will show, is not correct in fine details.  Nevertheless, I
hope to convince you that the doublers, although raised to high masses
in the Wilson approach, are responsible for a rather complex phase
structure.

Generalizing \wilsonferm to complex mass leads to the action
$$
\eqalign
{
&
S(K,r,m_1,m_2)=\cr
&
\sum_{j,\mu}
     K\left(\overline\psi_j         ( i\gamma_\mu+r) \psi_{j+e_\mu}
           +\overline\psi_{j+e_\mu} (-i\gamma_\mu+r) \psi_{j}\right) +
           \cr
&
\sum_{j} (m_1 \overline\psi_j\psi_j+i m_2 \overline\psi_j\gamma_5\psi_j). \cr
}
\myeqno
$$
Here $j$ labels the sites of a four dimensional hyper-cubic lattice,
$\mu$ runs over the space-time directions, and $e_\mu$ is the unit
vector in the $\mu$'th direction.  I have scaled out all factors of
the lattice spacing.  The parameter $K$ is called the hopping
parameter, and $r$ controls the strength of the so called ``Wilson
term,'' which separates off the famous doublers.  I have also added
$m_2$ type mass term as in \monemtwo to connect with my earlier
discussion.

Being quadratic with only nearest neighbor couplings, the free
spectrum is easily found by Fourier transformation.  Conventionally, a
massless Fermion appears at $m_1=8Kr$, but there are other values of
the parameters where this original particle is massive while other
doublers from the naive theory become massless.  At $m_1=-8Kr$ one
such species does so, at each of $m_1=\pm 4Kr$ there are four massless
doublers, and at $m_1=0$ I find the remaining $6$ of the total 16
species present in the naive theory.

\midinsert
\epsfxsize .75\hsize
\centerline {\epsfbox{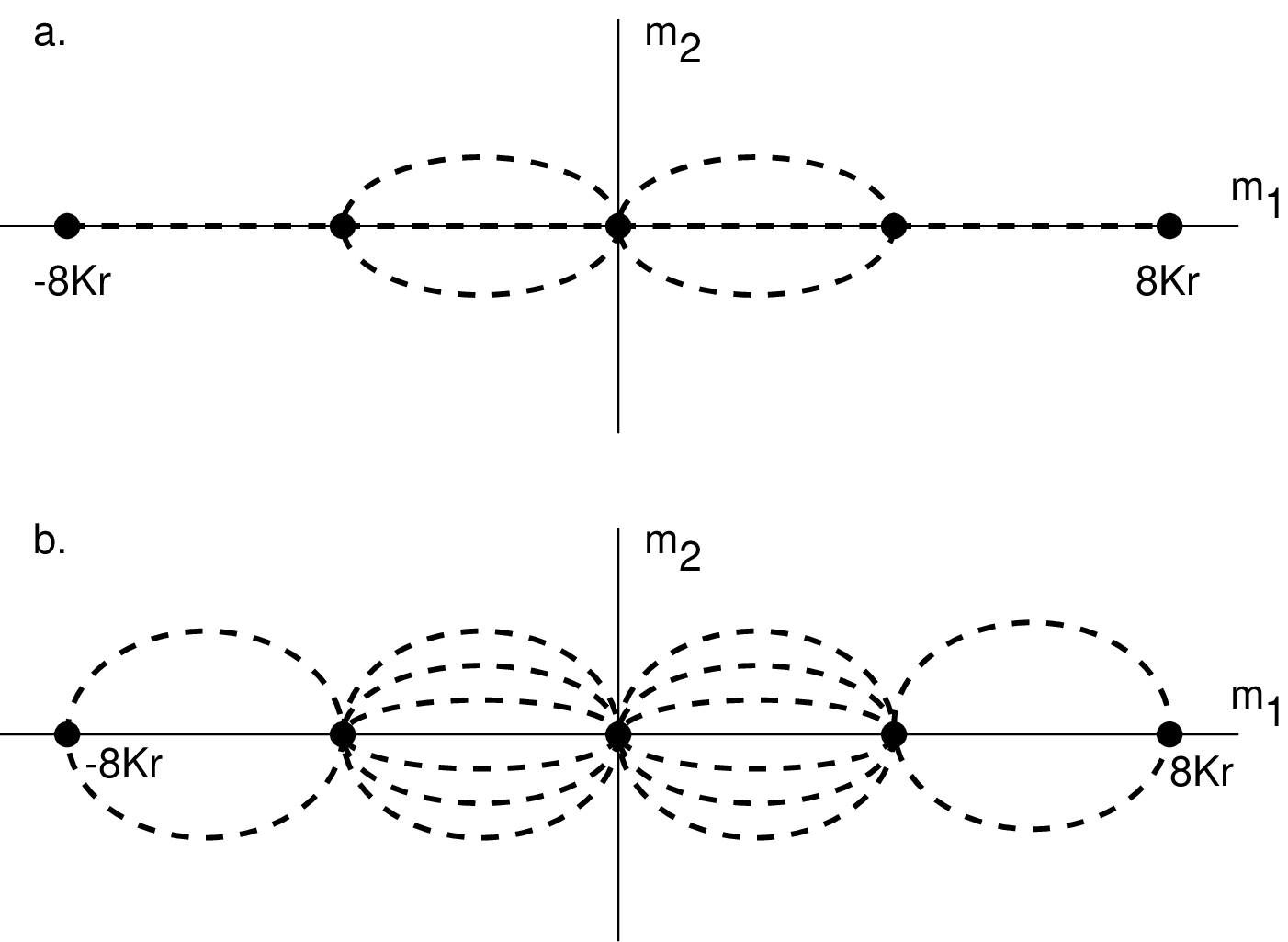}}\bigskip
\narrower {\myfigname\wilsonthetafig 
Possible phase diagrams for lattice gauge theory with Wilson
Fermions.  The dashed lines represent first-order phase transitions
and the dots represent points where massless excitations should exist.
Parts (a) and (b) are for the one and two flavor cases, respectively.}
\endinsert

A natural conjecture is that these various species should be thought
of as flavors.  When the gauge fields are turned on, the full chiral
structure should be a generalization of the discussion in the previous
sections.  Thus somewhere near $m_1=8Kr$ I expect a first-order
transition to end, much as is indicated in \oneflavor.  This may join
with numerous other transitions at the intermediate values of $m_1$,
all of which then finally merge to give a single first-order
transition line ending near $m_1=-8Kr$.  The situation near $0$ and
$\pm 4Kr$ involves larger numbers of light flavors, giving more
transition lines.  One possible way the lines could join up is shown
in part $a$ of \wilsonthetafig.

For two flavors of Wilson Fermions, near to the singularity
at $8Kr$ one should obtain a picture similar to \twoflavor.  However,
further away these lines can curve and eventually end in the structure
at the other doubling points.  One possible picture is sketched in
part $b$ of
\wilsonthetafig.  While this is likely the qualitative
behavior near the continuum limit, at more modest lattice spacings the
situation is likely somewhat more complicated.  Indeed, since the
Wilson term violates chiral symmetry, the second order phase
transition at vanishing mass is probably not robust.  The resulting
lattice artifacts are quite fascinating, and form the topic of the
next section.

\bigskip
{\bf \mysecno The Aoki phase}
\medskip

In this section I probe further into the rather rich phase structure
expected for lattice gauge theory with Wilson Fermions, paying
particular attention to what happens for hopping parameters very near
the critical value.  This behavior was predicted some time ago (Aoki,
1989; Aoki and Gocksch, 1992; Aoki, Ukawa and Umemura, 1996) in
attempts to understand the meaning of the phase transition that occurs
with Wilson Fermions when the pseudo-scalar masses go to zero.  As
there is no chiral symmetry when the cutoff is in place, the
``continuum'' argument that this represents a spontaneous breaking of
chiral symmetry is inadequate.  This picture involved a spontaneous
breaking of parity for one flavor, and a breaking of both flavor and
parity symmetries for more species.  The one flavor prediction is
exactly the spontaneous breaking of parity discussed in the section on
one flavor.  However, for multiple flavors, the situation is a bit
more complicated: the predicted spontaneous breaking of flavor
symmetry gives additional Goldstone Bosons.  In Aoki's picture with
two flavors, at the critical hopping there are three massless
particles, one being associated with the phase transition and the
other two being the Goldstone Bosons of the flavor breaking.

Here I will argue that this picture is also easily understood in terms
of effective Lagrangians.  Indeed, the result appears to be
inevitable, although variant actions could change the transition to
first order.  This discussion is based on Creutz (1997) and a
similar treatment appears in Sharpe and Singleton (1998).  While I
frame the discussion about the Wilson-Fermion approach, the phenomenon
appears to be generic in any approach that introduces chiral-symmetry
breaking terms at finite cutoff.
 
I begin by reiterating that the only parameters of the strong
interactions are the quark masses.  I implicitly include here the
strong CP violating parameter $\Theta$, as this can generally be
rotated into the mass matrix.  The strong coupling is not a parameter,
having been absorbed into the units of measurement via the phenomenon
of dimensional transmutation (Coleman and Weinberg, 1973).

For this section I take degenerate quarks at $\Theta=0$; thus, I
consider only a single mass parameter $m$.  I discuss only the two
flavor case, as this will make some of the chiral symmetry issues
simpler.  In addition, as I am discussing the theory on the lattice, I
introduce the lattice spacing $a$ as a second parameter.

On the lattice with Wilson Fermions, the physical parameters $(m,a)$
are usually replaced with $\beta$, representing the inverse bare
coupling squared, and the Fermion hopping parameter $K$.  I ignore the
Wilson parameter $r$, which can be regarded as being fixed at unity,
and set $M=1$, the coefficient of diagonal piece of the action from
\wilsonferm.  The mapping between $(m,a)$ and $(\beta,K)$ is
non-linear, well known, and not at issue here.  Note that in
considering the structure of the theory in either of these sets of
variables, I am inherently talking about finite lattice spacing $a$.
This entire section is about lattice artifacts.

I wish to develop a qualitative phase diagram for the $(\beta,K)$
plane.  The $\beta$ axis with $K=0$ represents the pure gauge theory
of glue-balls.  This is expected to be confining without any
singularities at finite $\beta$.  The line of varying $K$ with
$\beta=\infty$ represents free Wilson Fermions (Wilson, 1977).  Here,
with conventional normalizations, the point $K={1\over 8}$ is where
the mass gap vanishes and a massless Fermion should appear.  The full
interacting continuum limit should be obtained by approaching this
point from the interior of the $(\beta,K)$ plane.

While receiving the most attention, the point $K={1\over 8}$ is not
the only place where free Wilson Fermions lose their mass gap.  As
discussed earlier, at $K={1\over 4}$ four doubler species become
massless. Also formally at $K=\infty$ six doublers loose their mass.
(Actually, a more natural variable is ${1\over K}$.)  The remaining
doublers become light at negative $K$.

The $K$ axis at vanishing $\beta$ also has a critical point where the
confining spectrum appears to develop massless states.  Strong
coupling arguments as well as numerical experiments place this point
near $K={1\over 4}$, but this is probably not exact.  The conventional
picture connects this point to $(\beta=\infty, K={1\over 8})$ by a
phase transition line representing the lattice version of the chiral
limit.

Now I move ever so slightly inside the $(\beta,K)$ plane from the
point $(\infty, {1\over 8})$.  This should lead from free quarks to
a confining theory, with mesons, baryons, and glue-balls being the
physical states.  Furthermore, massless quarks should give chiral
symmetry.  Considering here the two flavor case, this symmetry is
nicely exemplified in the ``sigma'' model, with three pion fields and
one sigma field rotating amongst themselves.  Here I only need these
four fields from Section X
$$
\eqalign{
&\sigma=c\overline\psi\psi\cr
&\vec\pi=ic\overline\psi\gamma_5\vec\tau\psi.\cr
} \myeqno
$$
As before, I consider constructing an effective potential.  For
massless quarks this is expected to have the canonical sombrero shape
stereotyped by
$$
V\sim\lambda(\sigma^2+\vec\pi^2-v^2)^2 
\myeqno 
$$
and illustrated schematically in \sombrero.  The normal {\ae}ther is
taken with an expectation value for the sigma field
$\langle\sigma\rangle\sim v$.  The physical pions are massless
Goldstone Bosons associated with slow fluctuations of the {\ae}ther
along the degenerate minima of this potential.
 
As I move up or down in $K$ from the massless case near ${1\over 8}$,
this effective potential will tilt in the standard way, with the sign
of $\langle\sigma\rangle$ being appropriately determined.  The role of
the quark mass is played by the distance from the critical hopping,
$m_q\sim K_c-K$ with $K_c\sim {1\over 8}$.  At the chiral point there
occurs a phase transition, with the sign of $\langle\sigma\rangle$
jumping discontinuously.  At the transition point massless Goldstone
pions represent the spontaneous symmetry breaking.  With an even
number of flavors the basic physics on each side of the transition is
the same; the sign of the mass term is a convention reversible via a
non-singlet chiral rotation.  (For an odd number of flavors the sign
of the mass is significant because the required rotation involves the
$U(1)$ anomaly and is not a good symmetry.  This is discussed in some
detail in earlier sections.)

A similar picture should also occur near $K={1\over 4}$, representing
the point where a subset of the Fermion doublers become massless.
Thus another phase transition should enter the diagram at $K={1\over
4}$.  Similar lines will enter at negative $K$ and further complexity
occurs at $K=\infty$.  For simplicity, let me concentrate only on the
lines entering from large $\beta$ at $K={1\over 8}$ and ${1\over 4}$.

Now I delve a bit deeper into the $(\beta,K)$ plane.  The next
observation is that the Wilson term is explicitly not chiral
invariant.  This should damage the beautiful symmetry of the sombrero.
The first effect expected is a general tilting of the potential.  This
represents an additive renormalization of the Fermion mass, and
appears as a beta dependent motion of the critical hopping away from
${1\over 8}$.  Define $K_c(\beta)$ as the first singularity in the
phase diagram for increasing $K$ at given $\beta$.  This gives a curve
which presumably starts near $K={1\over 4}$ at $\beta=0$ and ends up
at ${1\over 8}$ for infinite $\beta$.

Up to this point this is standard lore.  Now I continue to delve yet
further away from the continuum chiral point at
$(\beta,K)=(\infty,{1\over 8})$.  Then I expect the chiral symmetry
breaking of the Wilson term to increase in importance and do more than
simply tilt the Mexican hat.  I'm not sure to what extent a multi-pole
analysis of this breaking makes sense, but let me presume that the
next effect is a quadratic warping of the sombrero, i.e.  a term
something like $\alpha \sigma^2$ appearing in the effective sigma
model potential.  This warping cannot be removed by a simple mass
renormalization.

There are two possibilities.  This warping could be upward or downward
in the $\sigma$ direction.  Indeed, which possibility occurs might
depend on the value of $\beta$.  Consider first the case where the
warping is downward, stabilizing the sigma direction for the
{\ae}ther.  At the first order chiral transition, this distortion
gives the pions a small mass.  The transition then occurs without a
diverging correlation length.  As before, the condensate $\langle
\sigma \rangle$ jumps discontinuously, changing its sign.
The conventional approach of extrapolating the pion mass to zero from
measurements at smaller hopping parameter will no longer yield the
correct critical line.  The effect of this warping on the potential is
illustrated in \tiltwarpfig from section X.  There is a single first
order phase transition without any divergent correlation lengths.

The second possibility is for the warping to be in the opposite
direction, destabilizing the $\sigma$ direction.  In this case we
expect two distinct phase transitions to occur as $K$ passes through
the critical region.  Small hoppings give a tilted potential with
$\sigma$ having a positive expectation.  As $K$ increases, this
tilting will eventually be insufficient to overcome the destabilizing
influence of the warping.  At a critical point, most likely second
order, it will become energetically favorable for the pion field to
acquire an expectation value, such a case being stabilized by the
upward warping in the sigma direction.  As $K$ continues to increase,
a second transition should appear where the tilting of the potential
is again sufficiently strong to give only sigma an expectation, but
now in the negative direction.  The effect of this upward warping on
the effective potential is equivalent to the situation illustrated in
\tiltupwarpfig.

Thus the critical line splits into two, with a rather interesting
phase between.  This phase has a non-vanishing expectation value for
the pion field.  As the latter carries flavor and odd parity, both are
spontaneously broken.  Furthermore, since flavor is still an exact
continuous global symmetry, when it is broken Goldstone Bosons will
appear.  In this two flavor case, there are precisely two such
massless excitations.  If the transitions are indeed second order, a
third massless particle appears on the transition lines, and these
three particles are the remnants of the pions from the continuum
theory.  This ``Aoki'' phase should be ``pinched'' between the two
transitions, and become of less importance as $\beta$ increases and
the continuum limit is approached.  Whether the phase might be
squeezed out at a finite $\beta$ to the first order situation, or
whether it only disappears in the infinite $\beta$ limit is a
dynamical question.

A similar critical line splitting to give a broken flavor phase may
also enter the phase diagram from $(\beta,K)=(\infty,{1\over 4})$,
representing the first set of doublers.  Evidence from toy models
(Aoki, Boetcher, and Gocksch, 1994; Bitar and Vranas, 1994a,b; Aoki,
Ukawa and Umemura, 1996) is that after this line splits, the lower
half joins up with the upper curve from the $(\beta,K)=(\infty,{1\over
8})$ point.  In these models, there appears to be only one broken
parity phase at strong coupling.

\topinsert
{
\epsfxsize .7\hsize
\centerline {\epsfbox{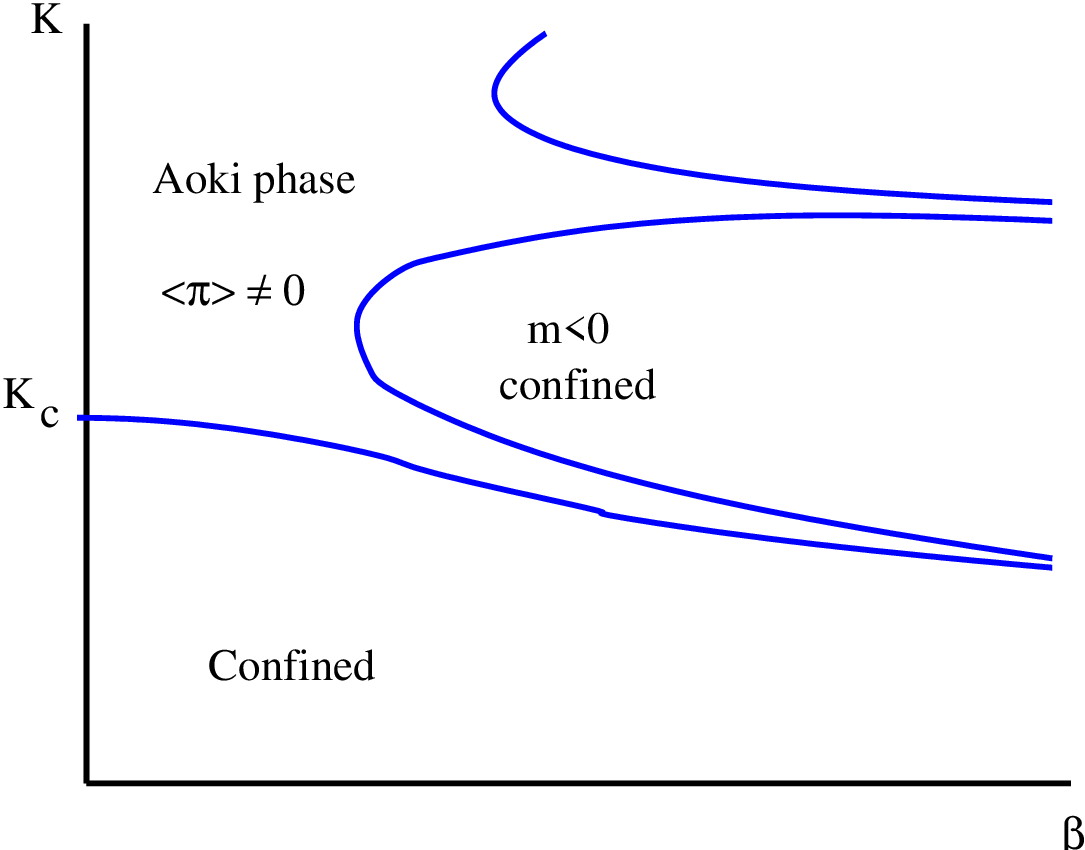}}
\narrower\myfigname\aokiphasefig The conjectured $(\beta,K)$ phase diagram.
In the Aoki phase flavor and parity are spontaneously broken via the
pion field acquiring an expectation value.
\bigskip
}
\endinsert

\aokiphasefig  summarizes the picture.  Small $\beta$ along with
small $K$ gives the massive quark confined phase.  Increasing $K$
brings one to the Aoki phase with spontaneous breaking of flavor and
parity.  As $\beta$ increases, the Aoki phase pinches down into either
a narrow point or a single first order line, leading towards the free
Fermion point at $(\beta,K)=(\infty,{1\over 8})$.

This diagram is wonderfully complex, probably incomplete, and may take
some time to map out.  As a final reminder, this entire discussion is
of lattice artifacts, and other lattice actions, perhaps including
various ``improvements,'' can look dramatically different.  In
particular, other actions may display the alternative option for a
single first order transition instead of the two transitions
surrounding the Aoki phase.

\bigskip
{\bf \mysecno The Wess-Zumino term}
\medskip

The domain-wall Fermion approach, toward which I am leading, uses an
extra space-time dimension.  Earlier sections showed how effective
Lagrangians give considerable insight into the physical consequences
of anomalous processes.  The idea that extra dimensions could be
useful for related phenomena appeared some time ago in the context of
these chiral Lagrangians.  Indeed, with hindsight, the analogy between
one term, called the Wess Zumino term, and domain-wall Fermions is
striking (Creutz and Tytgat, 1996).  In this section I give a
pedagogical discussion of the former as a motivation for looking at
extra dimensions.

In the previous two sections I used what is often called the
``linear'' approach to effective chiral Lagrangians.  Here I return to
the ``non-linear'' approach and repeat some of the equations used in
the three flavor section.  Consider quark fields $\psi^a$ with a
flavor index $a$ running from one to the number of flavors $N_f$.
Since the quark fields are expected to have a chiral condensate, the
vacuum is not unique.  Consider the general matrix valued quantity
$$
{}_g\langle 0 | \overline \psi_L^a \psi_R^b | 0\rangle_g =v g^{ab}.
\myeqname{\vacexp}
$$
Here I pick $v$ to normalize things so that the vacuum is labeled by
$g$, an element of the flavor group $SU(N_f)$.  For the usual vacuum
one would pick the identity element, but flavored chiral rotations
allow turning to alternate degenerate states.  A general chiral
transformation is specified by two $SU(N)$ matrices $\{g_L,g_R\}$ and
takes
$$
\psi_{L,R}^a\rightarrow 
\psi_{L,R}^b
g_{L,R}^{ba}. 
\myeqno
$$
Under this the effective field $g$ from \vacexp transforms as
$$
g\rightarrow g_L^\dagger g g_R.
\myeqname\symmetry
$$

The basic chiral Lagrangian idea is to integrate out all other fields,
generating an effective action for the above matrix $g$.  This is
expanded in powers of derivatives of the group-valued field.  The
lowest term
$$
S_{\rm eff}={F_\pi^2\over 4}\int (dg){\rm Tr} (\partial_\mu g^\dagger
\partial_\mu g)
\myeqno
$$
appeared in earlier sections.  Chiral perturbation theory involves
enumerating and elucidating the higher derivative couplings that can
be added while preserving the chiral symmetries.  As mentioned
earlier, the constant $F_\pi$ has an experimental value 93 MeV.  In
terms of conventional fields, $g=\exp(i\pi\cdot\lambda/F_\pi)$, where
the $n_f^2-1$ traceless matrices $\lambda$ generate $SU(n_f)$ and are
normalized ${\rm Tr}\lambda^\alpha
\lambda^\beta=2\delta^{\alpha\beta}$.
From this action, the equations of motion can be compactly written
$\partial_\mu J_{L,\mu}^\alpha=0$, where the ``left'' current is
$$
J_{L,\mu}^\alpha= {iF_\pi^2\over 4}{\rm Tr}\left(
\lambda^\alpha(\partial_\mu g)g^\dagger
\right).
\myeqno
$$
Equivalently, one can work with ``right'' currents.  There is a vast
literature on possible higher derivative terms (Donoghue, Holstein,
and Wyler, 1992; Leutwyler, 1990; Bijnens, Prades, and de Rafael,
1995).

\midinsert
\epsfxsize .8\hsize
\centerline{\epsfbox{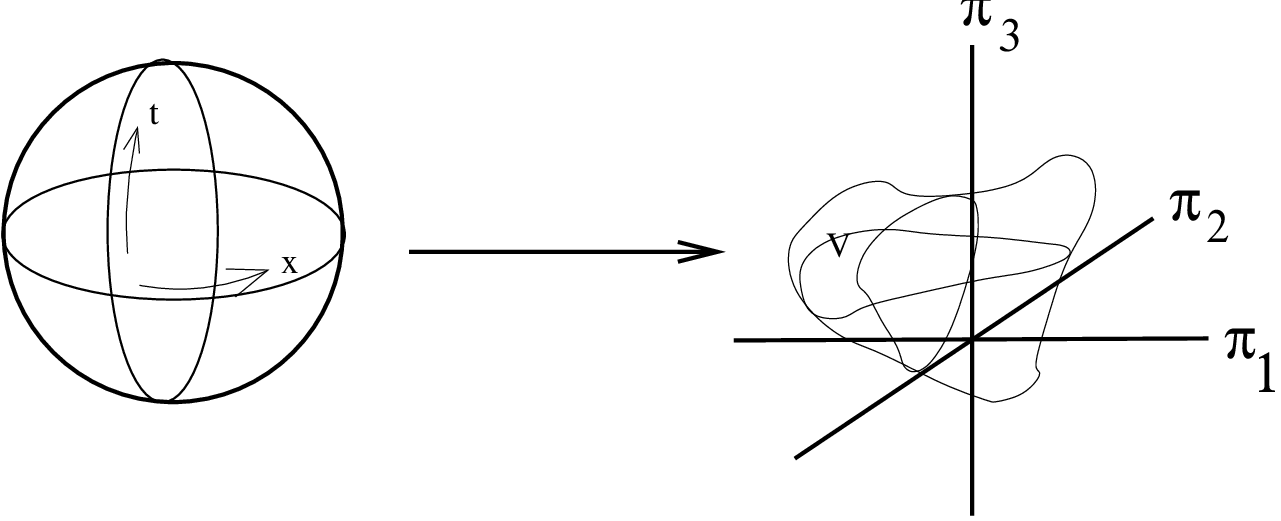}}
{\narrower \myfigname\mappingfig Mapping two dimensional space-time into three
dimensional field space.  The closed surface surrounds a volume which
is invariant under chiral shifts.\par }
\endinsert

For my purposes here, however, I concentrate on one particular term
with a subtle topological significance (Wess and Zumino, 1971; Witten,
1983a,b, 1984).  To understand this pictorially, it is convenient to
work first with the two flavor case in two space-time dimensions.
Then $g\in SU(2)$ represents three massless pion fields.  In the path
integral a field configuration is a mapping of space-time into field
space.  For convenience, imagine I have compactified space-time onto
the surface of a two dimensional sphere.  A general field
configuration is a mapping of this two dimensional sphere into the
three dimensional manifold of fields.  Since the sphere is closed,
this mapping will surround a closed region in field space.  This is
illustrated in \mappingfig An interesting quantity to consider is the
volume of this enclosed region.  Indeed, lattice gauge theorists are
intimately familiar with group invariant integration measures, and,
since $g$ lies in a group, it is natural to use this measure to define
this volume.  Furthermore, the properties of the group measure insure
that the volume is invariant under the chiral symmetry of
\symmetry.  Thus a possible interaction term that satisfies chiral
symmetry is just this volume
$$
S_I =\lambda \int_V dg.
\myeqname\volume
$$

However, the group manifold $SU(2)$ is compact.  This means that it is
not easy to be sure what is ``inside'' the surface mapped out by the
field.  There is an ambiguity in defining this volume up to the volume
of the entire group.  A gradual deformation of the field can take a
small volume and grow it to fill the group such that what was
originally inside the volume now looks like it is outside.  In short,
after normalizing the group measure so that the total volume of the
group is unity, $\int_G dg=1$, the interaction term is ambiguous up to
an integer.  For quantum mechanics this would not be important if the
coupling $\lambda$ was a multiple of $2\pi$.  Then the quantity
relevant to the path integral, $e^{iS}$, is unchanged when the action
changes by $2\pi$ times an integer.  This quantization of a coupling
constant is a field theoretical generalization of the quantization of
magnetic monopole charge in electrodynamics (Witten 1983a,b, 1984).

But note that the expression for this interaction term in
\volume is rather abstract.  Remarkably, this term cannot be
written as a simple space-time integral over the local fields and
their derivatives.  Such an expression would not have the ambiguity
indicated in the previous paragraph.  To obtain an explicit expression
requires introducing an extra dimension, allowing continuation inside
the volume indicated in \mappingfig.  For this let me pick an origin
in the group; this could be the identity element, but I will be more
general and pick some specific element $g_0$.  Then introduce an extra
dimension $s$, which for convenience I let run from 0 to 1.  I extend
my field $g(x,t)$ to a three dimensional one $h(x,t,s)\in SU(2)$.  I
subject the extended field to two constraints
$$
\eqalign{
&h(x,t,0)=g_0\cr
&h(x,t,1)=g(x,t).\cr
}\myeqno
$$
With this extension, the dimension of space-time now matches the
dimension of the field manifold.  This enables a rather compact and
elegant expression for the volume enclosed by the field
$$
\int_V dg = {1\over 24\pi^2}\int_V d^3x\ \epsilon_{\mu\nu\rho}   
\ {\rm Tr}(h^\dagger\partial_\mu h
h^\dagger\partial_\nu h
h^\dagger\partial_\rho h).
\myeqname{\wzone}
$$
To obtain the numerical factor of ${1\over 24\pi^2}$, take a field
configuration that covers the group once and require that the integral
be unity.

This entire discussion extends to higher dimensions with larger
groups.  The dimension of the group must exceed the dimension of space
time, so four dimensions requires at least $SU(3)$.  This allows
mappings into the group space which are ambiguous due to the
possibilities of non-trivial topology (Witten, 1983a,b, 1984).  To
obtain this term from an action requires extending $g(x)$ beyond a
simple mapping of space-time into the group.  As before, introduce an
auxiliary variable $s$ to interpolate between the field $g(x)$ and
some fixed group element $g_0$.  Thus consider $h(x,s)$ satisfying
$h(x,1)=g(x)$ and $h(x,0)=g_0$.  This extension is not unique, but the
equations of motion are independent of the chosen path.  
In analogy with \wzone, write
$$
\eqalign{
S&={F_\pi^2\over 4}\int d^4x\ 
{\rm Tr}(\partial_\mu g\partial_\mu g^\dagger)\cr
&+{n_c\over 240 \pi^2}\int d^4x\int_0^1 
ds\ \epsilon_{\alpha\beta\gamma\delta\rho}
\ {\rm Tr}\ h_\alpha h_\beta h_\gamma h_\delta h_\rho.\cr
}
\myeqname\wzactioneq
$$
Here I define $h_\alpha=i(\partial_\alpha h) h^\dagger$ and regard $s$
as a fifth coordinate for the antisymmetric tensor.

For equations of motion, consider a small variation of $h(x,s)$.  This
changes the final integrand by a total divergence, which integrates to
a surface term.  Working with either spherical or toroidal boundary
conditions in the space-time directions, this surface only involves
the boundaries of the $s$ integration.  When $s=0$, space-time
derivatives acting on the constant matrix $g_0$ vanish.  Writing the
equations of motion in terms of a divergence free current gives
$$
\eqalign{
J_{L,\mu}^\alpha&=
{iF_\pi^2\over 4}{\rm Tr}\left(
\lambda^\alpha(\partial_\mu g)g^\dagger\right)\cr
&+{in_c\over 48\pi^2}\ \epsilon_{\mu\nu\rho\sigma}\ {\rm Tr}
\left(\lambda^\alpha
(\partial_\nu g)g^\dagger
(\partial_\rho g)g^\dagger
(\partial_\sigma g)g^\dagger\right).\cr
}
\myeqname\wztermeq
$$

The last term in \wzactioneq represents a piece cut from the $S_5$
sphere appearing in the structure of $SU(n_f)$ for $n_f\ge 3$.  The
mapping of four dimensional space-time into the group surrounds this
volume.  Chiral rotations shift this region around, leaving its
measure invariant.  As emphasized by Witten (1983a,b, 1984), this term
is ambiguous.  Different extensions into the $s$ coordinate can modify
the above five dimensional integral by an integer multiple of
$480\pi^3$.  To have a well defined quantum theory, the action must be
determined up to a multiple of $2\pi$.  Thus the quantization of $n_c$
to an integer, which Witten has argued represents the number of
``colors.''

Crucial here is the irrelevance of the starting group element $g_0$ at
the lower end of the $s$ integration.  An important issue is the
difficulty of maintaining this condition if the chiral symmetry
becomes local.  As usual, this requires the introduction of gauge
fields.  Under the transformation $g(x)\rightarrow g_L^\dagger(x) g(x)
g_R(x)$, derivatives of $g$ transform as
$$
\partial_\mu g\longrightarrow
g_L^\dagger\left (
\partial_\mu g-\partial_\mu g_L g_L^\dagger g+ g\partial_\mu g_R g_R^\dagger 
\right) g_R.
\myeqno
$$
To compensate, introduce left and right gauge fields transforming as
$$\matrix{
A_{L,\mu} & \longrightarrow
& g_L^\dagger A_{L,\mu} g_L + i g_L^\dagger\partial_\mu g_L \cr
A_{R,\mu} & \longrightarrow 
& g_R^\dagger A_{R,\mu} g_R + i g_R^\dagger\partial_\mu g_R. \cr
}
\myeqno
$$
Then the combination
$$
D_\mu g = \partial_\mu g-iA_{L,\mu} g + ig A_{R,\mu}
\myeqno
$$ 
transforms nicely: $D_\mu g\rightarrow g_L^\dagger D_\mu g g_R$.
The generalized minimal replacement $\partial_\mu g\rightarrow
D_\mu g$ in $S_0$ gives a gauge invariant action.

A problem arises on coupling gauge fields to the Wess-Zumino term.
This requires a prescription for the gauge transformation of the
interpolated group element $h(x,s)$.  Here I anticipate a striking
analogy with the domain-wall approach to chiral Fermions.  There an
extra dimension is also introduced, with the Fermions being surface
modes bound to a four dimensional interface.  The usual method for
adding gauge fields involves, first, not giving the gauge fields a
dependence on the extra coordinate, and, second, forcing the component
of the gauge field pointing in the extra dimension to vanish.  In
terms of a five dimensional gauge field, take $A_\mu(x,s)=A_\mu(x)$
and $A_s=0$ for both the left and right handed parts.  Relaxing either
of these would introduce unwanted degrees of freedom.  The natural
extension of the gauge transformation is to take $h(x,s)\rightarrow
g_L^\dagger(x) h(x,s) g_R(x)$ with $g_{L,R}$ independent of $s$.

I now replace the derivatives in the Wess-Zumino term with covariant
derivatives.  This alone does not give equations of motion independent
of the interpolation into the extra dimension.  However, adding terms
linear and quadratic in the gauge field strengths allows construction
of a five dimensional Wess-Zumino term for which variations are again
a total derivative.  This gives
$$
S_{WZ}=
{n_c\over 240 \pi^2 }\int d^4x \int_0^1 ds\  \Gamma
\myeqno
$$
where
$$\eqalign{
\Gamma&=\Gamma_0+{5i\over2}(i\Gamma_L+i\Gamma_R\cr
&-\Gamma_{LL}-\Gamma_{RR}-\alpha\Gamma_{LR}-(1-\alpha)\Gamma_{RL}),\cr
}
\myeqno
$$
Here $\alpha$ is a free parameter and 
$$\eqalign{
&\Gamma_0=
\epsilon_{\mu\nu\rho\lambda\sigma} 
\Tr D_\mu h h^\dagger D_\nu h h^\dagger D_\rho h h^\dagger 
D_\lambda h h^\dagger D_\sigma h h^\dagger\cr
&\Gamma_L=\epsilon_{\mu\nu\rho\lambda\sigma} 
\Tr D_\mu h h^\dagger D_\nu h h^\dagger D_\rho h h^\dagger
F_{L,\lambda\sigma} \cr
&\Gamma_R=\epsilon_{\mu\nu\rho\lambda\sigma} 
\Tr D_\mu h h^\dagger D_\nu h h^\dagger D_\rho h
F_{R,\lambda\sigma} h^\dagger \cr
&\Gamma_{LL}=
\epsilon_{\mu\nu\rho\lambda\sigma} 
\Tr D_\mu h h^\dagger F_{L,\nu\rho} F_{L,\lambda\sigma} \cr
&\Gamma_{RR}=
\epsilon_{\mu\nu\rho\lambda\sigma} 
\Tr D_\mu h  F_{R,\nu\rho} F_{R,\lambda\sigma} h^\dagger \cr
&\Gamma_{RL}=
\epsilon_{\mu\nu\rho\lambda\sigma} 
\Tr D_\mu h  F_{R,\nu\rho} h^\dagger F_{L,\lambda\sigma} \cr
&\Gamma_{LR}=
\epsilon_{\mu\nu\rho\lambda\sigma} 
\Tr D_\mu h h^\dagger F_{L,\nu\rho} h F_{R,\lambda\sigma} h^\dagger.\cr
}
\myeqno
$$
The covariantly transforming field strengths are
$$
\eqalign
{
&F_{L,\mu\nu}=\partial_\mu A_{L,\nu}-\partial_\nu A_{L,\mu}
-i[A_{L,\mu},A_{L,\nu}]\cr
&F_{R,\mu\nu}=\partial_\mu A_{R,\nu}-\partial_\nu A_{R,\mu}
-i[A_{R,\mu},A_{R,\nu}].\cr
}
\myeqno
$$
For the photon, parity invariance fixes $\alpha=1/2$.  The last four
terms contain the process $\pi\rightarrow 2\gamma$.

I may seem to be drifting from the lattice aspects of the chiral
symmetry problem.  My purpose in this discussion was to point out how
certain field theoretical phenomena require the introduction of extra
dimensions.  These effects are closely tied to anomalies.  The message
is that for chiral symmetry on the lattice, anomalies might lead to
extra dimensions, just as they have for this simple effective
Lagrangian.

\bigskip
{\bf \mysecno Topology and zero modes}
\medskip

The mid 70's saw an explosion of interest in topological structures in
classical field theories.  These were particularly interesting in
their non-perturbative properties.  On quantization, topological
solitons became a new type of particle beyond the simple excitations
of the fundamental fields.  Structures in four dimensions were found
that led to unanticipated tunneling processes, eventually tied to
the anomalies alluded to throughout this document.  Indeed, usual
discussions of the theta parameter start directly with topologically
non-trivial classical gauge field configurations.  My presentation in
terms of chiral Lagrangians relies on using the anomaly to move this
parameter into the mass term.

Shortly after the discovery of these topological structures, it was
realized that Fermions can behave in rather interesting ways in the
presence of non-trivial classical field configurations.  One
provocative point of view (Jackiw and Rebbi, 1976; Goldstone and
Wilczek, 1981) refers to states of ``half-integer Fermion number.''
These modes come about by matching solutions of the Dirac equation in
different backgrounds.  With appropriate boundary conditions, some
solutions only had one state, or chirality, as opposed to the usual
doubling due to spin.  If the basic topological background defect has
spatial extension, such as along a string or membrane, these special
states could move.  This idea eventually evolved into the basis of the
domain-wall formulation.  A four dimensional defect in a five
dimensional underlying theory can support low energy excitations which
mimic free Fermions of a definite chirality.

The connection with anomalies was elegantly presented by Callan and
Harvey (1985) who considered a five dimensional space wherein the mass
of a Fermion changed sign on a four dimensional interface.  Fermionic
zero modes appeared on the interface, and were capable of carrying
momentum in the transverse direction.  At scales well below the basic
Fermion mass, the theory reduces to a four dimensional model of
massless Fermions.  They then discussed how anomalous processes
appeared as a flow of eigenmodes into the fifth dimension.

Kaplan (1992) argued that there was nothing in this mechanism that
precluded working on a lattice.  Indeed, the lattice adds additional
insights.  The continuum model has an ambiguity about which direction
in the fifth dimension the anomalous currents flow.  On the lattice,
this direction is determined by the sign of the Wilson term, the same
quantity that enables the theta parameter to retain its physical
significance.  Also, once the anomaly is known to go only in a single
direction, the interface can be replaced by a surface model, simpler
for calculations (Shamir, 1993b, 1994).

These surface modes were anticipated considerably earlier in the
context of band theory in condensed matter physics.  Considering two
coupled bands, Shockley (1939) showed that when the coupling between
these bands becomes large, the gap between them can close and reopen
leaving behind surface states.  These states are exactly the surface
modes that form the basis of domain-wall Fermions.  The one conceptual
difference is a particle-antiparticle symmetry, which forces the
present states to lie at exactly zero energy.

\bigskip{\bf \mysecno The icetray model}
\medskip

In this section I will approach domain-wall Fermions from a rather
unconventional direction.  Following a recent paper (Creutz, 1999), I
present the subject from a ``chemists'' point of view, in terms of a
chain molecule with special electronic states carrying energies fixed
by symmetries.  For lattice gauge theory, placing one of these
molecules at each space-time site gives excitations of naturally zero
mass.  This is in direct analogy to the role of chiral symmetry in
conventional continuum descriptions.

\midinsert
\epsfxsize .7\hsize
\centerline {\epsfbox{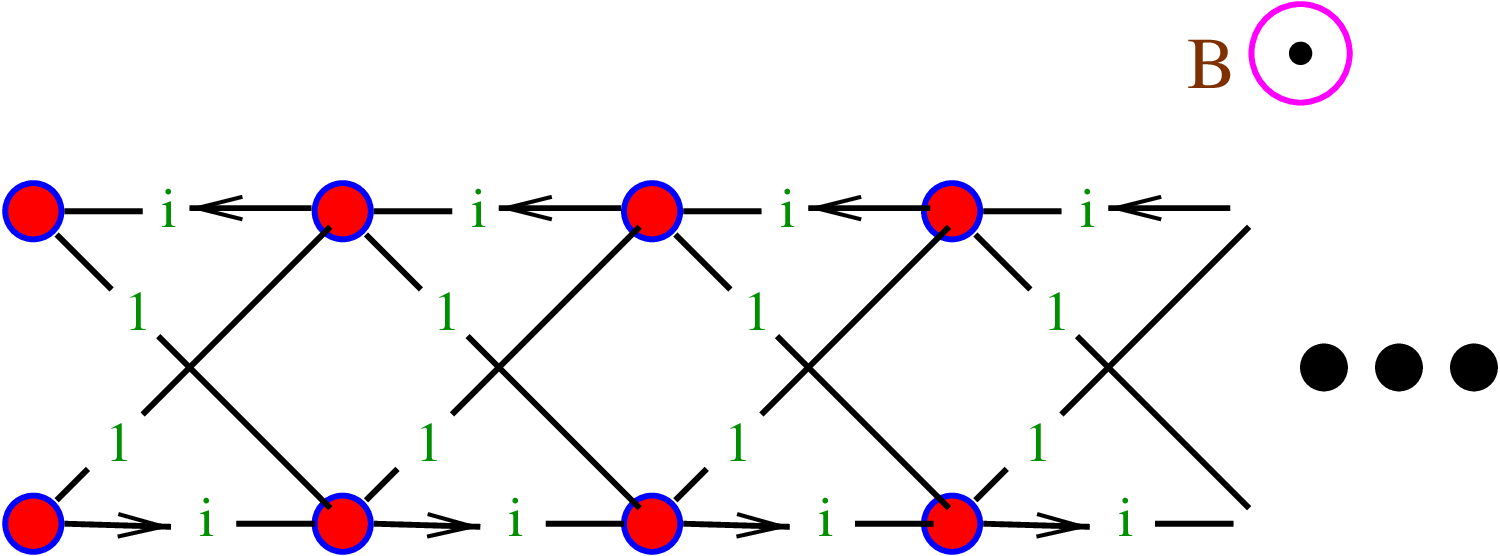}}
\narrower{\myfigname\novert  A ladder molecule in a magnetic field.
The phases represent one unit of magnetic flux per plaquette.\par}
\endinsert

To start, consider two rows of atoms connected by horizontal and
diagonal bonds, as illustrated in \novert.  The bonds represent
hopping terms, wherein an electron moves from one site to another via
a creation-annihilation operator pair in the Hamiltonian.  Later I
will include vertical bonds, but for now consider just the horizontal
and diagonal connections.

Years ago during a course on quantum mechanics, I heard Feynman
present an amusing description of an electron's behavior when inserted
into a lattice.  If you place it initially on a single atom, the wave
function will gradually spread through the lattice, much like water
poured in a single cell of a metal ice cube tray.  With damping, it
settles into the ground state which has equal amplitude on each atom.
To this day I cannot fill an ice cube tray without thinking of this
analogy and pouring all the incoming water into a single chamber.

\midinsert
\epsfxsize .4\hsize
\centerline{\epsfbox{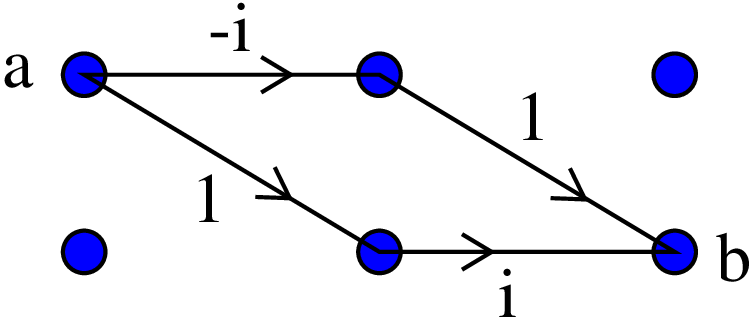}}
\narrower {\myfigname\cancel The two paths from site {\bf a} to site
{\bf b} cancel quantum mechanically when one half flux unit passes
through the loop.\par} 
\endinsert

I now complicate this picture by applying a magnetic field orthogonal
to the plane of the system.  This introduces phases as the electron
hops, causing interesting interference effects.  In particular,
consider a field of one-half flux unit per plaquette.  This means that
when a particle hops around a unit area (in terms of the basic lattice
spacing) the wave function picks up a minus sign.  Just where the
phases appear is a gauge dependent convention; only the total phase
around a closed loop is physical.  One choice for these phases is
indicated by the numbers on the bonds in \novert.  The phase factors
cause cancellations and slow diffusion.  For example, consider the two
shortest paths between the sites {\bf a} and {\bf b} in \cancel.  With
the chosen flux, these paths exactly cancel.  For the full molecule
this cancellation extends to all paths between these sites.  An
electron placed on site {\bf a} can never diffuse to site {\bf b}.
Unlike in the ice tray analogy, the wave function will not spread to
any site beyond the five nearest neighbors.

\midinsert
\epsfxsize .65\hsize
\centerline {\epsfbox{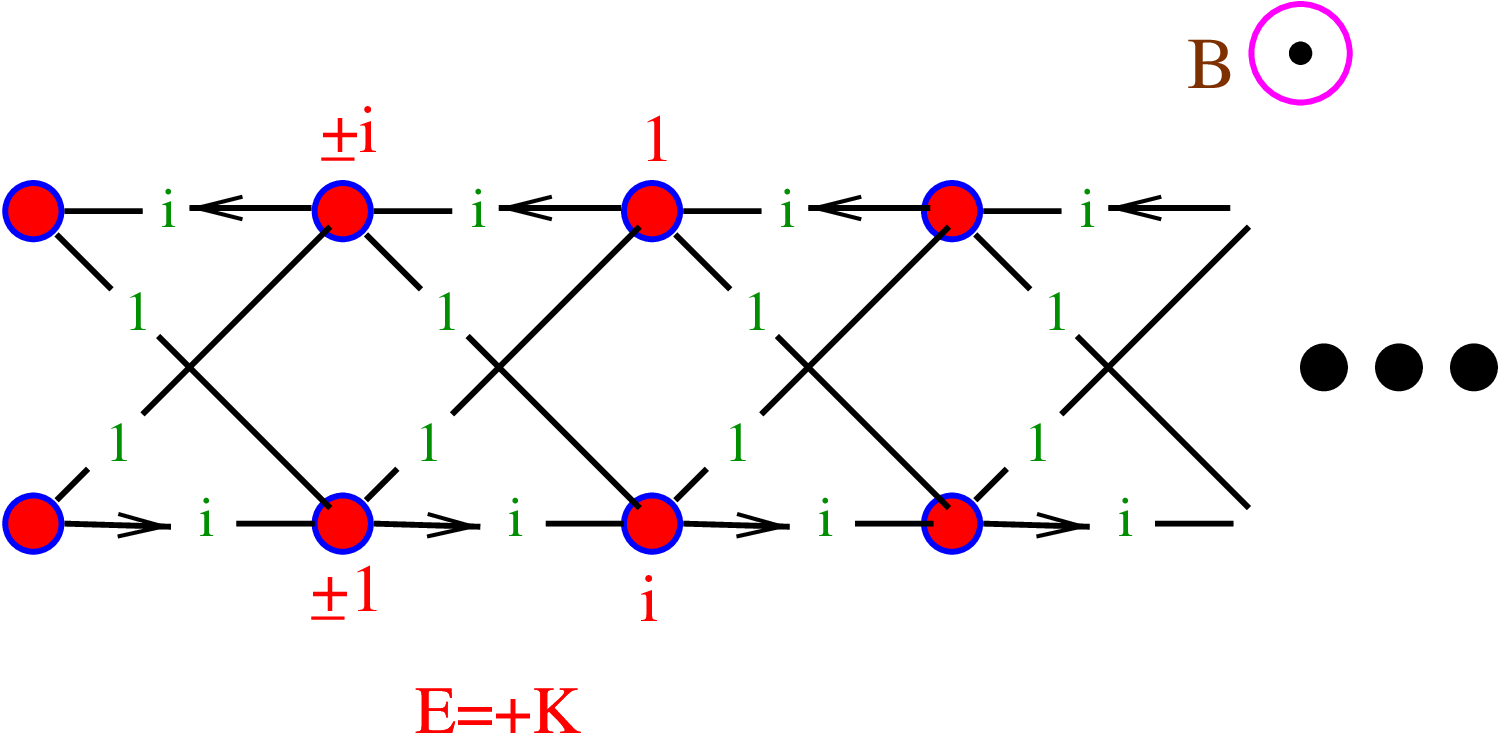}}
\narrower {\myfigname\novertone  There
exist two eigenstates per plaquette that do not spread with time.
The states have opposite signs for the energy.  
\par}
\endinsert

As a consequence, the Hamiltonian has localized eigenstates.
There are two such states per plaquette; one possible
representation for these two states is shown in \novertone.
The states are restricted to the four sites labeled by their relative
wave functions.  Their energies are fixed by the size of the hopping
parameter $K$.

\midinsert
\epsfxsize .65\hsize
\centerline {\epsfbox{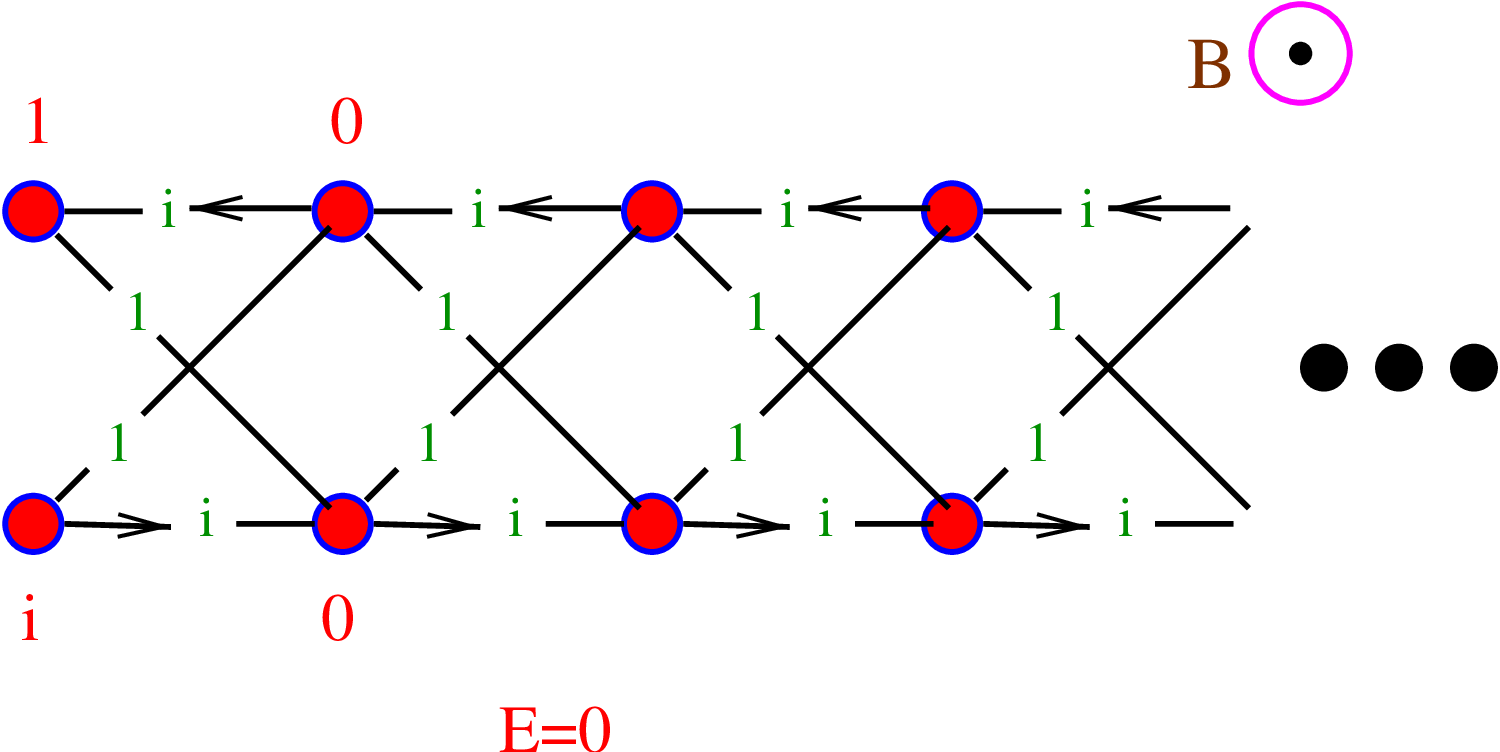}}
\narrower {\myfigname\noverttwo One eigenstate of exactly zero energy
is bound on the end of the chain.  \par}
\endinsert

For a finite chain of length $L$ there are $2L$ atoms, and thus there
should be a total of $2L$ possible states for the electron (ignoring
spin for the moment).  There are $L-1$ plaquettes, and thus $2L-2$ of
the above localized states.  This is almost the entire spectrum of the
Hamiltonian, but two states are left over.  These are zero energy
states bound to the ends of the system.  The wave function for one of
those is shown in \noverttwo.  This is the full spectrum of the
Hamiltonian: $L-1$ degenerate states of positive energy, a similar
number of degenerate negative energy states, and two states of zero
energy bound on the ends.

\midinsert
\epsfxsize .65\hsize
\centerline {\epsfbox{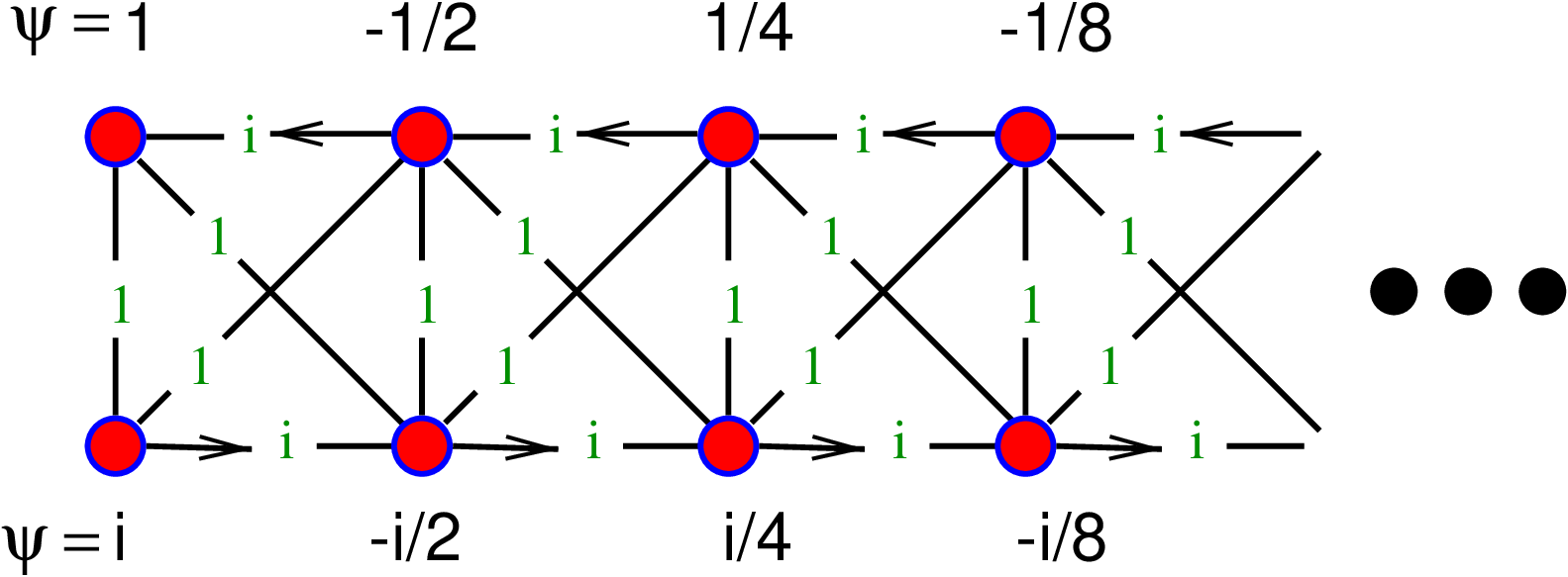}}
\narrower{\myfigname\icetraythree  The zero energy state is robust
under the addition of vertical bonds, although the wave function now
has an exponential decrease into the molecule.\par}
\endinsert

Now consider what happens when vertical bonds are included in this
molecule.  The phase cancellations are no longer complete and the
localized states spread to form two bands, one with positive and one
with negative energy.  However, for the present purposes, the
remarkable result is that the zero modes bound on the ends of the
chain are robust.  The corresponding wave functions are no longer
exactly located on the last atomic pair, but have an exponentially
suppressed penetration into the chain.  \icetraythree shows the wave
function for one of these states when the vertical bond has the same
strength as the others.  There is a corresponding state on the other
end of the molecule.

When the chain is very long, both of the end states are forced to zero
energy by symmetry considerations.  First, since nothing distinguishes
one end of the chain from the other, they must have equal energy,
$E_L=E_R$.  On the other hand, a change in phase conventions,
effectively a gauge change, can change the sign of all the vertical
and diagonal bonds.  Following this with a left-right flip of the
molecule will change the signs of the horizontal bonds.  This takes
the Hamiltonian to its negative, and shows that the states must have
opposite energies, $E_L=-E_R$.  This is also a consequence of a
particle-hole symmetry.  The combination of these results forces the
end states to zero energy, with no fine tuning of parameters.

For a finite chain, the exponentially decreasing penetration of the
end states into the molecule induces a small interaction between them.
They mix slightly to acquire exponentially small energies $E\sim \pm
e^{-\alpha L}$.  As the strength of the vertical bonds increases, so
does the penetration of the end states.  At a critical strength, the
mixing becomes sufficient that the zero modes blend into the positive
and negative energy bands.  In the full model, the mixing depends on
the physical momentum, and, as I will discuss shortly, this
disappearance of the zero modes is the mechanism that removes the
``doublers'' when spatial momentum components are near $\pi$ in
lattice units (Jansen and Schmaltz, 1992; Creutz and Horvath, 1994).

\bigskip{\bf \mysecno Domain-wall Fermions}
\medskip

\midinsert
\epsfxsize .5\hsize
\centerline {\epsfbox{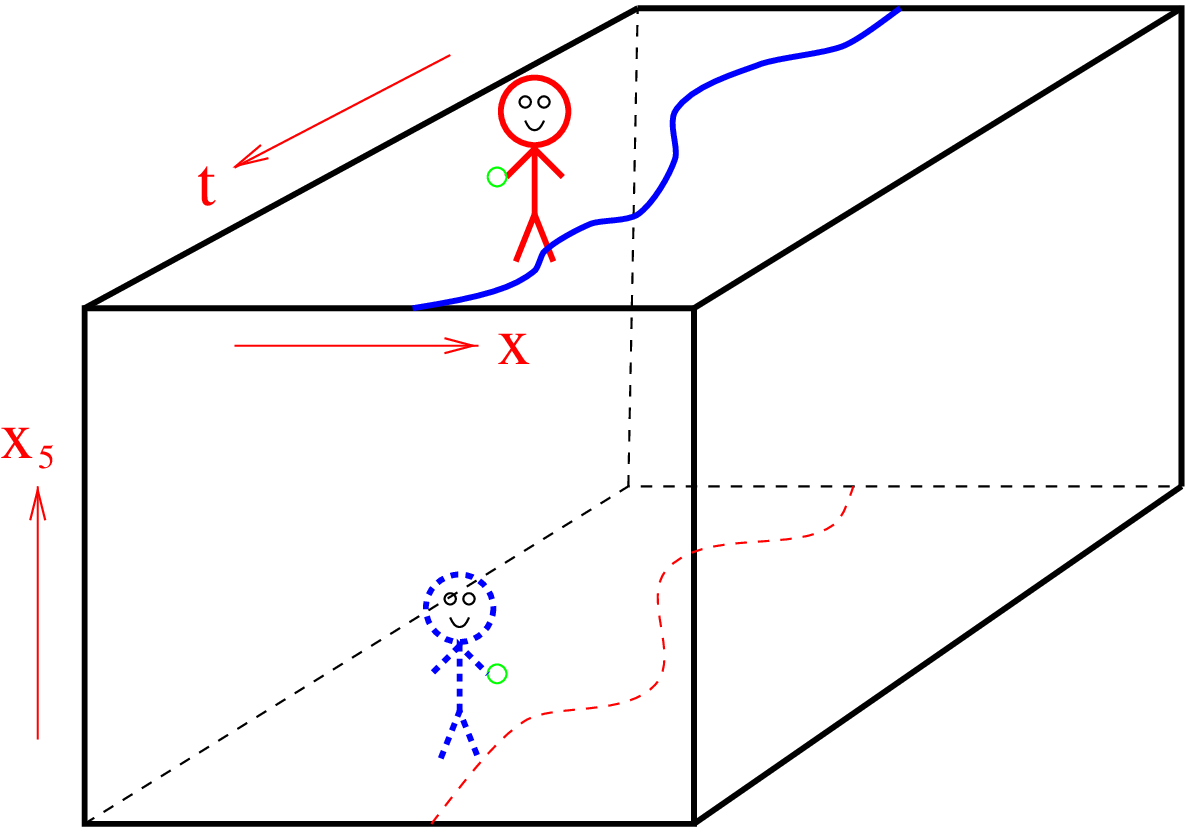}}
\narrower{\myfigname\kaplanfig The basic domain-wall formalism has the
physical particles bound on the surfaces of a five dimensional world.
Left and right handed partners lie on opposing walls.\par}
\endinsert

These energy levels forced to zero by symmetry lie at the core of the
domain wall idea.  On every spatial site of a three dimensional
lattice place one of these chain molecules.  The distance along the
chain is usually referred to as a fictitious ``fifth'' dimension.  The
different spatial sites are coupled, allowing particles in the zero
modes to move around.  These are the physical Fermions.  The
symmetries that protect the zero modes now protect the masses of these
particles.  Their masses receive no additive renormalization, exactly
the consequence of chiral symmetry in the continuum.  The physical
picture is sketched in the cartoon of \kaplanfig.  For this figure I
have rotated the fifth dimension to the vertical.  The world lines of
everyday objects traverse the four dimensional surface of this five
dimensional manifold.  The model is mathematically identical to five
dimensional Wilson Fermions at supercritical hopping.

Actually, the connection with chiral symmetry is much deeper than just
an analogy.  The construction guarantees that the modes are are
automatically chiral.  
To see how this works, first note that the construction of the hopping
in the fifth dimension is consistent with a particular representation
of the Dirac gamma matrices
$$\eqalign{
\gamma_0&=\pmatrix{0 & 1  \cr 1 &  0\cr}\cr
\gamma_5&=\pmatrix{0 & -i \cr i &  0\cr}\cr
\gamma_i&=\pmatrix{\sigma_i & 0  \cr 0 & -\sigma_i\cr}.\cr
}
\myeqno
$$
Note how Pauli spin matrices enter the spatial hoppings.  To see their
effect, it is easiest to go to momentum space in the four physical
dimensions.  Then the spatial hoppings become diagonal and generate
effective vertical bonds which depend on momentum.  Including a
diagonal term proportional to a parameter $M$ as before, the net
vertical bonds are parametrized by the matrix
$$
\pmatrix{0 & M+ K_4(\cos(q_i)+i\sigma_i\sin(q_i))  \cr 
M+K_4(\cos(q_i)-i\sigma_i \sin(q_i))&
  0\cr}.
\myeqname\vertbond
$$
I have explicitly kept the hopping parameter $K_4$ and distinguished
it with the subscript 4 to denote the four physical dimensions.  I
will discuss shortly reasons for distinguishing the hopping in the
physical from the hopping in the fifth direction.
 
Here the terms involving $\sigma\sin(q)$ move the surface modes from
zero energy to a finite value $E^2\sim K_4^2\sum
\sin^2(q_i)$.  The group velocity contains a factor of the Pauli
matrices, indicating that the low energy and low momentum modes are
chiral.  Note that at small momentum this is exactly the naive lattice
Fermion spectrum.  Were that to continue to larger momentum, there
would be doublers when any component of the momentum is near $\pi$.
Removing them involves the terms involving $\cos(q)$, which come from
the Wilson term in the action.  As discussed before, the surface modes
are robust under the addition of a real vertical bond, and thus to
first approximation the $M$ and $\cos(q)$ terms in \vertbond are
unimportant.  However, such terms do influence the penetration of the
modes into the extra dimension, and if they become too large, then
these modes can disappear.  This is useful, since the Wilson terms can
make the surface modes disappear before the momentum reaches the
doubling points.  In other words, $M$ and $K_4$ should take values
such that the surface modes are present at small $q$, but must also be
in a region so that the modes are lost before reaching the doubler
points.

So far I have been treating only free Fermions.  To couple gauge
fields without adding lots of unneeded degrees of freedom, it is
simplest to only have gauge fields in the physical directions.  In
this approach, the extra dimension is perhaps best thought of as a
flavor space (Narayanan and Neuberger, 1993a,b, 1994, 1995).  With a
finite lattice this procedure gives equal couplings of the gauge field
to the Fermion modes on opposing walls in the extra dimension.  The
result is an effective light Dirac Fermion.  In the case of the strong
interactions, this provides an elegant scheme for a natural chiral
symmetry without the tuning inherent in the usual Wilson approach.
The breaking of chiral symmetry arises only through finiteness of the
extra dimension.

In usual discussions of domain-wall Fermions, no distinction is made
between the strength of the hopping in the spatial dimension and that
in the extra direction.  For free Fermions this works well, but Brower
and Svetitsky (2000) have argued that in the strong coupling limit the
doublers can cause difficulties.  While their discussion was in
Hamiltonian language, it is likely to carry through for the Lagrangian
discussion as well.  To resolve this, note that the structure of the
free theory is quite rich if the strength of the spatial hopping,
$K_4$ and that in the extra dimension $K_5$ are varied separately.  In
\myfigname\kfourkfive I sketch the number of independent Fermion
species present in the $(K_4,K_5)$ plane for fixed $M=1$.  The lines
here are places where massless Fermions appear via a vanishing mass
gap for the full five dimensional theory.  When gauge fields are
turned on, all of these lines are expected to be renormalized.  The
Brower and Svetitsky result suggests that these lines shift to larger
$K_4$, pinching out the desired phase with only one zero mode on the
surface.  Whether this pinching is complete at finite gauge coupling
or only in the strong coupling limit is at present unresolved.

\midinsert
\epsfxsize .8\hsize
\centerline {\epsfbox{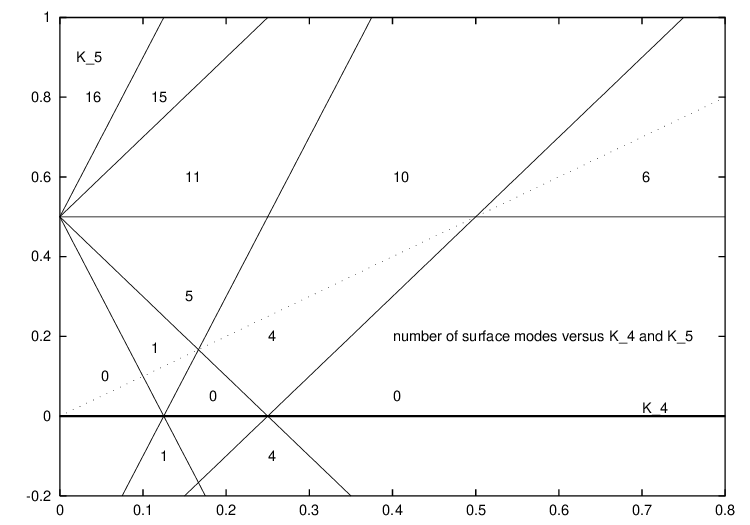}}
\narrower{\myfigname\kcrit A rich phase structure appears when the
four dimensional hopping parameter $K_4$ is allowed to vary separately
from $K_5$ corresponding to the extra dimension.  The lines indicate
the loci of massless five dimensional Fermions.  The numbers labeling
the various regions are the number of independent massless modes
propagating on the 4d surface of a five dimensional lattice.  Usual
domain-wall Fermions utilize the region with only one surface mode.
\par}
\endinsert

When the size of the extra dimension is finite, there is a small
mixing of the modes on the two walls.  Regarded as a theory of light
Fermions plus a set of heavier fields associated with the fifth
dimension, this mixing generates a small residual mass for the low
energy states.  For the strong interactions one wants to include quark
masses anyway, so for practical calculations one might as well work
with finite $L_5$, including this mixing as part of the physical quark
masses.  This point of view is controversial --- integrating out the
heavy fields yields a rather complicated non-local effective theory,
making the interpretation of this mixing as a mass less clear
(Kikukawa, 1999).  Nevertheless, this residual mixing does play a
crucial role towards the physics discussed extensively in the sections
on the phase of quark masses.  The angle $\Theta$ discussed there is
effectively the relative phase a quark acquires in tunneling through
the extra dimension compared to any extrinsic mass added to the
theory.  I will return to this point later.

In the domain-wall scheme there are a large number of heavy fields.
These can cause difficulties for perturbative studies.  The usual way
to eliminate this problem is to add in a comparable number of
auxiliary Bosonic fields (Frolov and Slavnov, 1994; Narayanan and
Neuberger, 1993a; Furman and Shamir, 1995).  These use the same action
as the Fermions, but with periodic rather than open boundary
conditions.  In loops they will cancel the heavy Fermions, eliminating
the need for large perturbative counter-terms.  This is a useful
technical point, but not essential to the basic understanding of the
formulation.

One issue with domain-wall Fermions is that the two helicities on
opposite walls are both present, so this is not a chiral gauge theory
of the type needed to describe the weak interactions.  The main
unresolved question is whether there is some scheme to make the
opposite helicity states on one wall either decouple or move to large
mass.  I return to this topic in later sections.  Before that,
however, I digress and discuss a physical picture of how anomalies
work in this five dimensional model.

\bigskip{\bf \mysecno Anomalies in the domain-wall approach}\medskip

These surface modes form an elegant formulation of a $d$ dimensional
Fermion theory in terms of surface modes in a $d+1$ dimensional
underlying space.  However, it is a bit mysterious how anomalies can
work when the left and right zero modes are so strongly separated.  In
this section I follow the discussion of Creutz and Horvath (1994),
showing how anomalous processes appear as a flow of states in a
Hamiltonian formulation.  The approach is motivated by the earlier
discussion of the anomaly, and displays a mechanism to absorb the
rising eigenvalues without having an infinite band.

The basic picture returns to the continuum context of Callan and
Harvey (1985).  A vector theory with a mass term having a domain-wall
shape in an extra dimension has chiral zero-modes living on the wall.
Anomalous currents associated with this zero-mode must be absorbed in
the underlying $d+1$ dimensional theory since that world is anomaly
free.  What on the interface looks like an anomaly is the flow of
charge into the extra dimension and the role of the heavy modes is to
carry that charge.  Since opposite chirality partners live on opposite
walls, the charge has to be transported through the extra dimension.

For simplicity I concentrate on the one dimensional case with gauge
group $U(1)$.  An external $U(1)$ gauge field is manifested in phase
factors appearing whenever a Fermion hops from one spatial site to the
next.  Calling this factor $U(k)=e^{i\alpha_k}$ for the hopping from
space site $k$ to $k+1$, the Wilson Hamiltonian is
$$
\eqalign{
H=\sum_{k,j} 
\bigg(&K\overline\psi_{k,j+1}
(\gamma_5-1)\psi_{k,j}
-K\overline\psi_{k,j}
(\gamma_5+1)\psi_{k,j+1}\cr
+&K\overline\psi_{k+1,j}U(k)
(\gamma_1-1)\psi_{k,j}
-K\overline\psi_{k,j}U^\dagger(k)
(\gamma_1+1)\psi_{k+1,j}\cr
+&M\overline\psi_{k,j}\psi_{k,j}\bigg).\cr
}\myeqno
$$
Here $k$ represents the spatial coordinate and $j$ the ``fifth''
dimension.  I have taken the Wilson parameter $r=1$ and hopping
parameter $K=K_4=K_5$.  I take antiperiodic boundary spatial
conditions $\psi_{L,j}=-\psi_{0,j}$ for the Fermion field.  Of course,
for domain-wall Fermions the boundaries in the extra dimension are
open.

A gauge transformation by a phase $g$ at site $k$ takes $U(k)$ to
$U(k)g^\dagger$, $U(k+1)$ to $g U(k+1)$, and $\psi_k$ to $g\psi_k$.
The invariance of the Hamiltonian under this symmetry implies that the
spectrum only depends on the product of all the $U(k)$ around the
periodic spatial lattice.  This is the net phase acquired by a Fermion
traveling all the way around the finite periodic system.

A convenient gauge choice is to evenly distribute the phases so that
$U(k)=e^{i\alpha}$ is independent of $k$.  This keeps momentum space
simple, with the Hamiltonian becoming
$$
\eqalign{
H=\sum_{q,j} 
\bigg(&K\overline\psi_{q,j+1}
(\gamma_5-1)\psi_{q,j}
-K\overline\psi_{q,j}
(\gamma_5+1)\psi_{q,j+1}\cr
+&2iK\sin(2\pi q/L-\alpha)\overline\psi_{q,j}\gamma_1\psi_{q,j}\cr
+&(M-2Kr\cos(2\pi q/L-\alpha))\overline\psi_{q,j}\psi_{q,j}\bigg).\cr
}\myeqno
$$
Here $j$ labels the fifth coordinate and the spatial momentum variable
$q$ runs from 0 to L.  The energy eigenstates are functions of the
momentum shifted by $\alpha$.  As $\alpha$ increases from 0 to
$2\pi/L$, sequential momenta rotate into each other. The total net
phase in this case is $2\pi$ and, as expected, physics goes back to
itself.  In natural units, one unit of flux represents $\alpha =
2\pi/L$.

\midinsert \vskip -.25in
\epsfxsize=.65\hsize
\centerline {\epsfbox {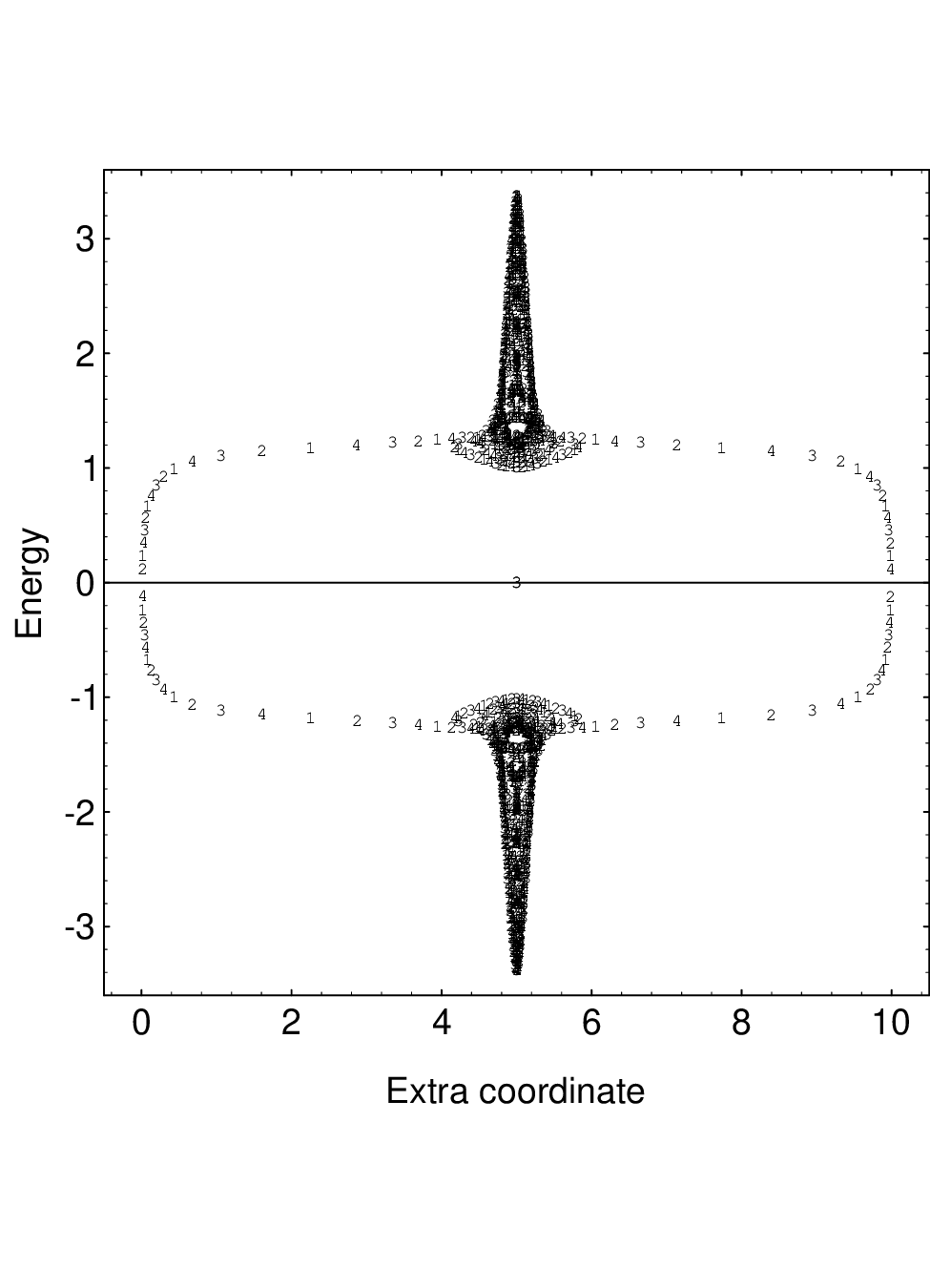}}
\vskip -.3in \noindent \narrower { 
\myfigname\evsxfive The energy spectrum as a function of $<x_5>$ for
$\alpha=0,{1\over4},{1\over2},{3\over4}$ on a $L_5=11$ by $L=16$
lattice.  Successive positions of energy levels are marked by $1,2,3$
and $4$ respectively. The levels rotate in an anti-clockwise
sense.\par}
\endinsert

In the adiabatic limit, time evolution is a continuous flow of one
particle states with $\alpha$.  For the various eigenstates of the
Hamiltonian on a small system, Creutz and Horvath (1994) calculated
the expectation value of the fifth coordinate.  \evsxfive, taken from
that paper, shows the results for four distinct values of
$\alpha=0,{1\over4},{1\over2},{3\over4}$.  States with a low magnitude
for the energy lie at the lattice ends and rise or fall with $\alpha$
without substantially changing their position in the extra dimension.
The same is true for the very high energy states, residing deep in the
lattice interior.  As their energy increases, the surface states
penetrate further into the extra dimension.  When the energy of such a
level gets close to the cutoff, it moves swiftly towards the
middle. At the same time, another level from the middle lowers its
energy and runs towards the opposite wall. This is also true for
corresponding levels with negative energy; they just move in the
opposite direction. The heavy modes near the cutoff are responsible
for carrying the charge on and off the surfaces.

On the lattice there is considerable freedom to define an axial
charge; any definition assigning opposite charges to the two zero-modes
living on the opposite walls should yield a correct continuum limit.
One possibility is to define the axial charge as the Fermion number
weighted by the location in the extra dimension
$$
Q_5={1\over L_5-1}\sum_{q,j}(L_5-1-2j)
\psi^\dagger_{q,j}\psi_{q,j} \myeqno
$$
with $j=0, 1,2,...,L_5-1$.  This assigns to a one particle state $+1$
if it is exactly bound to the left wall with $j=0$ and $-1$ if it is
bound to the right wall. For states smeared uniformly over $j$, we
obtain zero.  Regarding the extra dimension as an internal space, we
see that the axial charge is nothing but a particular combination of
flavor charges.

The vacuum of the theory fills all the negative energy eigenstates.
Physical energies are taken relative to this state.  At zero $\alpha$,
the vacuum has zero axial charge because of the mirror symmetry of the
Hamiltonian with respect to the middle of the extra space
$(j\rightarrow L_5-1-j)$. To see the anomaly, evolve the vacuum in
an adiabatic field, increasing the value of $\alpha$ from $0$ to $1$
unit of flux and look for the change in the total axial charge.  As
the field is turned on, the levels in the Dirac sea start to move
anti-clockwise $(\alpha>0)$, increasing the total axial charge.

When the value of the field is close to one half unit of flux, one of
the filled levels on the right surface is just about to become the
positive energy particle and one empty level on the left is just about
to drop into the sea. However, as long as the extra dimension is
finite, the surface states are not exactly massless.  There is always
a tiny mass $\delta$ present, caused by mixing of the states on the
opposite walls. When the fields are truly adiabatic, with the typical
time $\tau$ for turning on the fields longer than any other time scale
in the problem, e.g.  $\tau\gg 1/\delta$, instead of creation of a
particle-hole pair, the two levels will have enough time to exchange
between the walls.  As a consequence, there is a jump at
$\alpha=1/2$, which allows vacuum to evolve into its original state as
$\alpha$ approaches one unit of flux.  This is the reason for the
appearance of the point labeled with ``3'' in the exact center of
\evsxfive.

\bigskip
{\bf \mysecno Weak interactions and mirror Fermions}
\medskip
With an exact gauge invariance and a finite size for the extra
dimension, the surface models are inherently vector-like.  The
Fermions always appear with both chiralities, albeit separated in the
extra dimension.  However, it is an experimental fact that only left
handed neutrinos couple to the weak Bosons.  In this section I discuss
one way (Creutz and Horvath, 1994) to break the symmetries between
these states, resulting in a theory with only one light gauged chiral
state.  Here I keep the underlying gauge symmetry exact, but do
require that the chiral gauge symmetry be spontaneously broken, just
as observed in the standard model.  The picture also contains heavy
mirror Fermions.  If anomalies are not canceled amongst the light
species, these heavy states must survive in the continuum limit.  It
remains an open question when anomalies are properly canceled whether
it might be possible to drive the heavy mirror states to arbitrarily
large mass.

I start with two separate species $\psi_1$ and $\psi_2$ in the surface
mode picture.  However, I treat these in an unsymmetric way.  For
$\psi_1$ use the previous Hamiltonian.  For $\psi_2$ change the
sign of all terms proportional to $\gamma_5$.  On a given wall, the
surface modes associated with $\psi_1$ and $\psi_2$ will then have
opposite chirality.

Now introduce the gauge fields.  Since I want to eventually couple
only one-handed neutrinos to the vector Bosons, consider gauging
$\psi_1$ but not $\psi_2$.  Indeed, at this stage $\psi_2$ represents
a totally decoupled right handed ``spectator'' Fermion on one wall.
A mirror situation appears on the opposite wall, consisting of a
right handed gauged state and a left handed decoupled Fermion.
 
The next ingredient is to spontaneously break the gauge symmetry, as
in the standard model, by introducing a Higgs field $\phi$ with a
non-vanishing expectation value.  This field can generate
masses as in the standard model by coupling $\psi_1$ and $\psi_2$ with
a term of the form $\bar\psi_1\psi_2\phi$.

The new feature is to allow the coupling to the Higgs field to depend
on the extra coordinate.  In particular, let it be small or vanishing
on one wall and large on the other.  The surface modes are then light
on only the first wall.

\midinsert
\epsfxsize=.6\hsize
\centerline {\epsfbox {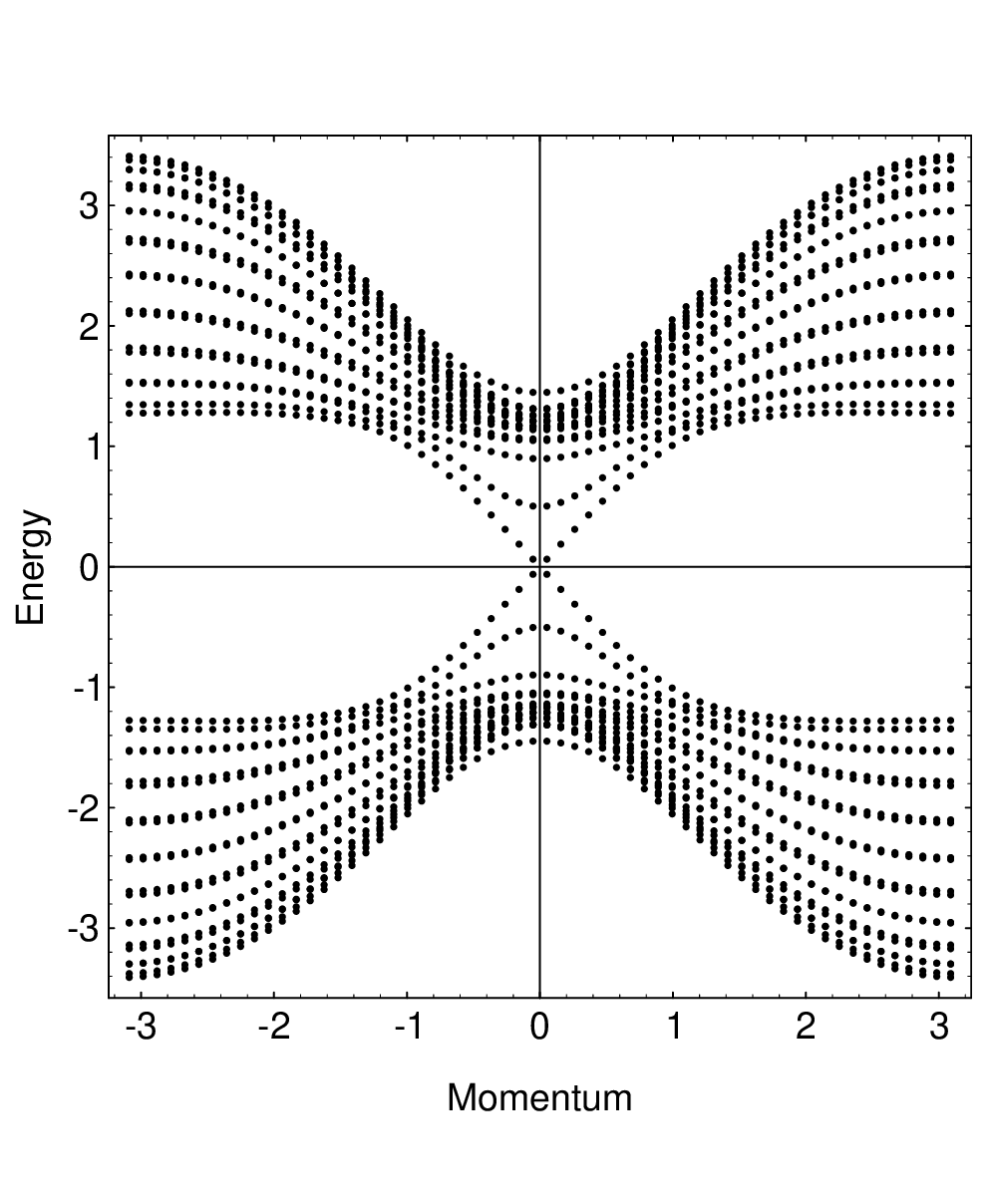}}
\noindent \narrower {\myfigname\mirror 
Energy levels versus momentum for the two species model discussed in
the text. Here the lattice is $L_5=10$ by $L=60$, and the Higgs
coupling is switched off at the middle of the extra dimension. Note
how one species is effectively massless and the other massive.}
\endinsert

\mirror, taken from Creutz and Horvath (1994), considers one space 
dimension and sketches the Fermion spectrum of this model with
vanishing gauge fields and a constant Higgs field.  As in other mirror
Fermion models (Montvay, 1987, 1993), triviality arguments suggest
bounds on the masses of the heavy particles.  This is certainly the
case when the light Fermions alone give an anomalous gauge theory, in
which case the mirror particles cannot become much heavier than the
vector mesons, i.e. the $W$.  Golterman and Shamir (1985) have shown
that naively taking the Higgs-Fermion coupling to infinity on one wall
can introduce a plethora of new low energy bound states.  It is
conceivable that the restrictions on the mirror Fermion masses are
weaker when anomalies cancel amongst the light states.  In this case
there is no perturbative need for the heavy states, and perhaps they
can be driven to infinite mass in the continuum limit.  If so, this
would be a candidate for a lattice discretization of the standard
model.

This model, however, does not lead to baryon number violation (Distler
and Rey, 1993).  The anomaly will involve a tunneling of baryons from
one wall to the opposite, where they become mirror baryons.  Even if
these extra particles are heavy, the decay can only occur through
mixing with the ordinary particle states.  In this sense, the mirror
particles still show their presence in low energy physics.

A speculative proposal is to use the right handed mirror states in
some way as observed particles.  Indeed, the world has left handed
leptons and right handed anti-baryons.  Any simple extension of this
idea to a realistic model must unify these particles.  The next
section presents a possible construction along this line.

\bigskip{\bf \mysecno Leptons and anti-quarks}\medskip

Elementary particle interactions are parity violating, i.e. the theory
must be chiral.  This is implemented in the standard model of
elementary particle interactions via the coupling of the electroweak
bosons to chiral currents.  How to formulate such a theory on the
lattice remains controversial.  Some time ago Eichten and Preskill
(1986) pointed out that a lattice approach must accomodate the 't
Hooft (1976a,b) baryon decay process.  Thus any fully finite theory
must incorporate baryon violating terms in the underlying Lagrangian.
Despite difficulties (Golterman, Petcher, and Rivas, 1993) with their
specific model, the idea of including such terms remains compelling.
In this section I will discuss a specific recent scheme (Creutz,
Tytgat, Rebbi, and Xue, 1997) building on the domain-wall approach.
This approach has not been widely accepted due to a rather complex
action.  Nevertheless, I hope the following interpretation of the
standard model will at least seem provocative.

The standard model is based on the product of three gauge groups,
$SU(3)\times SU(2) \times U(1)_{em}$.  Here the $SU(3)$ represents the
strong interactions of quarks and gluons, the $U(1)_{em}$ corresponds
to electromagnetism, and the $SU(2)$ gives rise to the weak
interactions.  I gloss over the technical details of electroweak
symmetry breaking and the mixing between the $U(1)$ field and the
neutral $SU(2)$ gauge Boson.

The full model is, of course, parity violating, as necessary to
describe observed helicities in beta decay.  This violation is
normally considered to lie in the $SU(2)$ of the weak interactions,
with both the $SU(3)$ and $U(1)_{em}$ being parity conserving.
However, this is actually a convention, adopted primarily because the
weak interactions are small.  I argue below that reassigning degrees
of freedom allows a reinterpretation where the $SU(2)$ gauge
interaction is vector-like.  Since the full model is parity violating,
this process shifts the parity violation into the strong,
electromagnetic, and Higgs interactions.  The resulting theory pairs
the left handed electron with a right handed anti-quark to form a
Dirac Fermion.

With a vector-like weak interaction, the chiral issues which complicate
lattice formulations now move to the other gauge groups.  Requiring
gauge invariance for the re-expressed electromagnetism then clarifies
the mechanism behind a recent proposal for a lattice regularization of
the standard model.

To begin, consider only the first generation, which involves four left
handed doublets.  These correspond to the neutrino/electron lepton
pair plus three colors for the up/down quarks
$$
\pmatrix{\nu \cr e^-\cr}_L,
\ \pmatrix{{u^r} \cr {d^r}\cr}_L,
\ \pmatrix{{u^g} \cr {d^g}\cr}_L,
\ \pmatrix{{u^b} \cr {d^b}\cr}_L.
\myeqno
$$
Here the superscripts from the set $\{r,g,b\}$ represent the internal
$SU(3)$ index of the strong interactions, and the subscript $L$ indicates
left-handed helicities. 

If I ignore the strong and electromagnetic interactions, leaving only
the weak $SU(2)$, each of these four doublets is equivalent and
independent.  I now arbitrarily pick two of them and do a charge
conjugation operation, thus switching to their antiparticles
$$\eqalign{
&\pmatrix{{u^g} \cr {d^g}\cr}_L \longrightarrow 
\pmatrix{\overline{{d^g}} \cr \overline{{u^g}}\cr}_R \cr
&\pmatrix{{u^b} \cr {d^b}\cr}_L \longrightarrow 
\pmatrix{\overline{{d^b}} \cr \overline{{u^b}}\cr}_R. \cr
}
\myeqno
$$
In four dimensions anti-Fermions have the opposite helicity; so, I
label these new doublets with $R$ representing right-handedness.

With two left and two right handed doublets, I can combine them into
Dirac fields (Lee and Schrock, 1988)
$$
\pmatrix{
\pmatrix{\nu \cr e^-\cr}_L\cr
\pmatrix{\overline{{d^g}} \cr \overline{{u^g}}\cr}_R\cr
}
\qquad
\pmatrix{
\pmatrix{{u^r} \cr {d^r}\cr}_L\cr
\pmatrix{\overline{{d^b}} \cr \overline{{u^b}}\cr}_R .\cr
}
\myeqno
$$
Formally in terms of the underlying fields, the construction takes
$$\eqalign{
\psi&=\half (1-\gamma_5)\psi_{(\nu,e^-)}+\half (1+\gamma_5)
\psi_{({\overline{d^g}},{\overline{u^g}})} \cr
\chi&=\half (1-\gamma_5)\psi_{({u^r}, {d^r})}+\half (1+\gamma_5)
\psi_{({\overline{d^b}},{\overline{u^b}})}. \cr
}
\myeqno
$$

From the conventional point of view these fields have rather peculiar
quantum numbers.  For example, the left and right parts have different
electric charges.  Electromagnetism now violates parity.  The left and
right parts also have different strong quantum numbers; the strong
interactions violate parity as well.  Finally, the components have
different masses; parity is violated in the Higgs mechanism.

The different helicities of these fields also have variant baryon
number.  This is directly related to the known baryon violating
processes through weak ``instantons'' and axial anomalies ('t Hooft,
1976a,b).  As discussed earlier, when a topologically non-trivial weak
field is present, the axial anomaly arises from a level flow out of
the Dirac sea (Ambjorn, Greensite, and Peterson 1983; Holstein, 1993).
This generates a spin flip in the fields, {\it i.e.} $e^-_L
\rightarrow ({\overline{u^g}})_R$.  Because of my peculiar particle
identification, this process does not conserve charge, with $\Delta Q=
-{2\over 3} +1={1\over 3}$.  This would be a disaster for
electromagnetism were it not for the fact that simultaneously the
other Dirac doublet also flips, ${d^r}_L
\rightarrow ({\overline{u^b}})_R$, with a compensating $\Delta Q =
-{1\over 3}$.  This is anomaly cancellation, with the total $\Delta Q
= {1\over 3}-{1\over 3}=0$.  Only when both doublets are considered
together is the $U(1)$ symmetry restored.  In this process baryon
number is violated, with $L+Q\rightarrow \overline Q +
\overline Q$.  This is the famous `` `t Hooft vertex.''

This discussion has been in continuum language.  Now I return to the
lattice, and use the Kaplan-Shamir domain-wall approach.  To repeat,
in this picture, the four dimensional world is a ``4-brane'' embedded
in 5-dimensions.  The complete lattice is a five dimensional box with
open boundaries, and the parameters are chosen so the physical quarks
and leptons appear as surface zero modes.  The elegance of this scheme
lies in the natural chirality of these modes as the size of the extra
dimension grows.  With a finite fifth dimension a doubling phenomenon
remains, coming from interfaces appearing as surface/anti-surface
pairs.  It is natural to couple a four dimensional gauge field equally
to both surfaces, giving rise to a vector-like theory.

\midinsert
\epsfxsize .6 \hsize
\centerline{\epsfbox{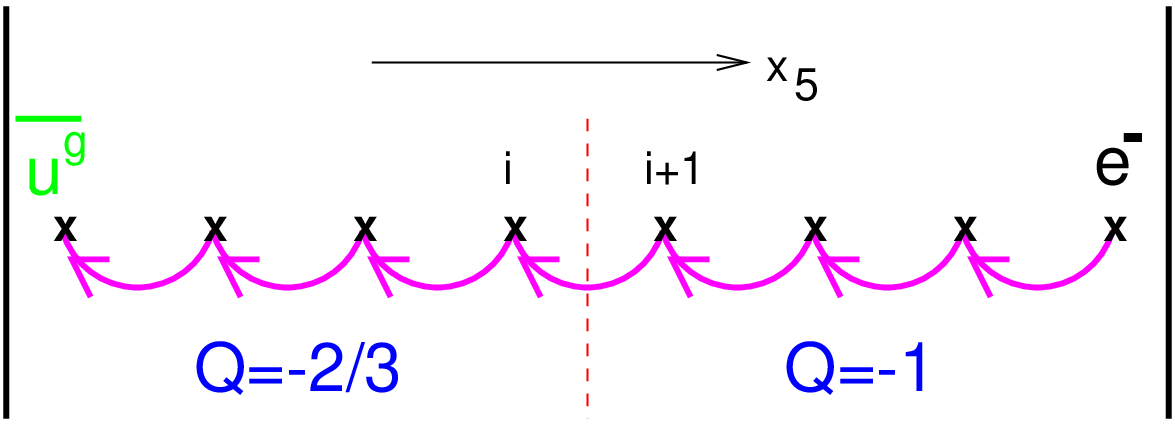}}
\narrower{\myfigname\ehoppingfig Pairing the electron with the 
anti-green-up-quark.\par}
%\label{fig:1}
\endinsert

I now insert the above pairing into this five dimensional scheme.  In
particular, I consider the left handed electron as a zero mode on one
wall and the right handed anti-green-up-quark as the partner zero mode
on the other wall, as sketched in \ehoppingfig.  This provides a
complete lattice regularization for the $SU(2)$ of the weak
interactions.

However, since these two particles have different electric charge,
$U(1)_{EM}$ must be broken in the interior of the extra dimension.  I
now proceed in analogy to the ``waveguide'' picture (Golterman,
Jansen, Petcher, and Vink, 1994) and restrict this charge violation to
$\Delta Q$ to one layer at some interior $x_5=i$.  Then the Fermion
hopping term from $x_5=i$ to $i+1$
$$
\overline\psi_{i}P\psi_{i+1}\qquad{(P=\gamma_5+r)}
\myeqno
$$
is a $Q=1/3$ operator.  At this layer, electric charge is not
conserved.  This is unacceptable and needs to be fixed.

\midinsert
\epsfxsize .6 \hsize
\centerline{\epsfbox{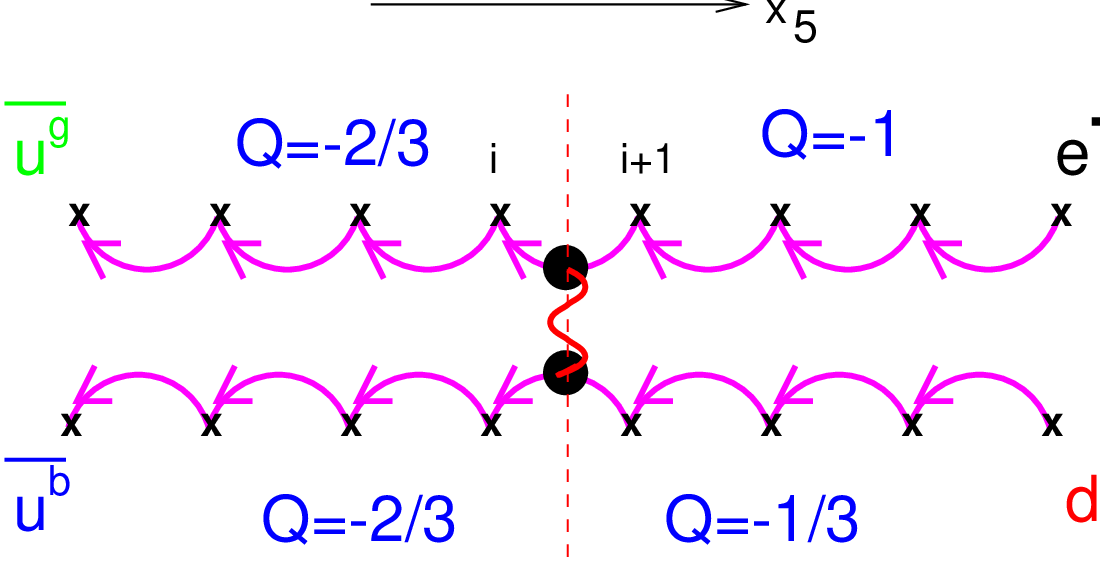}}
\narrower {\myfigname\transferfig Transferring charge between the
doublets.

}
%\label{fig:2}
\endinsert

To restore the $U(1)$ symmetry one must transfer the charge from
$\psi$ to the compensating doublet $\chi$.  For this I replace
the sum of hoppings with a product on the offending layer
$$
\overline\psi_{i}P\psi_{i+1}
{+}\overline\chi_{i}P\chi_{i+1}
\ \longrightarrow
\ \overline\psi_{i}P\psi_{i+1}
{\times}\overline\chi_{i}P\chi_{i+1}.
\myeqno
$$
This introduces an electrically neutral four Fermi operator.  Note
that it is baryon violating, involving a ``lepto-quark/diquark''
exchange, as sketched in \transferfig.  One might think of the
operator as representing a ``filter'' at $x_5=i$ through which only
charge compensating pairs of Fermions can pass.  The need for such an
operator in a fully finite theory was emphasized by Eichten and
Preskill (1986), although the detailed scheme presented there had
technical difficulties (Golterman, Petcher, and Rivas, 1993).

In five dimensions there is no chiral symmetry.  Even for the free
theory, combinations like $\overline\psi_{i}P\psi_{i+1} $ have vacuum
expectation values.  I use such as a ``tadpole,'' with $\chi$
generating an effective hopping for $\psi$ and {\it vice versa}.

Actually the above four Fermion operator is not quite sufficient for
all chiral anomalies, which can also involve right handed
singlet Fermions.  To correct this I need explicitly include the right
handed sector, adding similar four Fermion couplings
(also electrically neutral).

Having fixed the $U(1)$ of electromagnetism, I restore the strong
$SU(3)$ with an antisymmetrization $ {Q^r}{Q^g}{Q^b}{\longrightarrow
\epsilon^{\alpha\beta\gamma}Q^\alpha Q^\beta Q^\gamma}$.  Note that
similar left-right inter-sector couplings are needed to correctly
obtain the effects of topologically non-trivial strong gauge fields.

An alternative view is to fold the lattice about the interior of the
fifth dimension, placing all light modes on one wall and having the
multi-Fermion operator on the other.  This is the model of Creutz,
Tytgat, Rebbi, and Xue (1997), with the additional inter-sector
couplings correcting a technical error (Neuberger, 1997).

Unfortunately the scheme is still non rigorous.  In particular, the
non-trivial four Fermion coupling represents a new defect which should
not give rise to unwanted extra zero modes.  Note, however, that the
five dimensional mass is the same on both sides of defect, removing
topological reasons for such.

A second, and probably the biggest, concern is that the four Fermion
coupling might induce an unwanted spontaneous symmetry breaking of one
of the gauge symmetries.  A paramagnetic phase without spontaneous
symmetry breaking is needed.  Creutz, Tytgat, Rebbi, and Xue (1997)
showed that strongly coupled zero modes preserved the desired
symmetries, but their analysis ignored contributions from heavy modes
in the fifth dimension.

Assuming all works as desired, the model raises several other
interesting questions.  As formulated, I used a right handed neutrino
to provide all quarks with partners.  Is there some variation that
avoids this particle, which completely decouples in the continuum
limit?  Another question concerns possible numerical simulations; is
the effective action positive?  Finally, I have used the details of
the usual standard model, leaving open the question of whether this
model is somehow special.  Can one always use multi-Fermion couplings
to eliminate undesired modes in other anomaly free chiral theories?

\bigskip{\bf \mysecno Chiral identities}\medskip

Returning to the domain-wall lattice formulation of the strong
interactions, it may still be a bit mysterious how it is consistent
for both pions and the $\eta^\prime$ to be made of the same surface
modes and yet have rather different masses.  Some intuition on this
issue can be gained from some exact chiral identities.  There are many
equivalent ways to obtain similar relations; I pick one that appeals
to me.

\midinsert
\epsfxsize .5 \hsize 
\centerline{\epsfbox{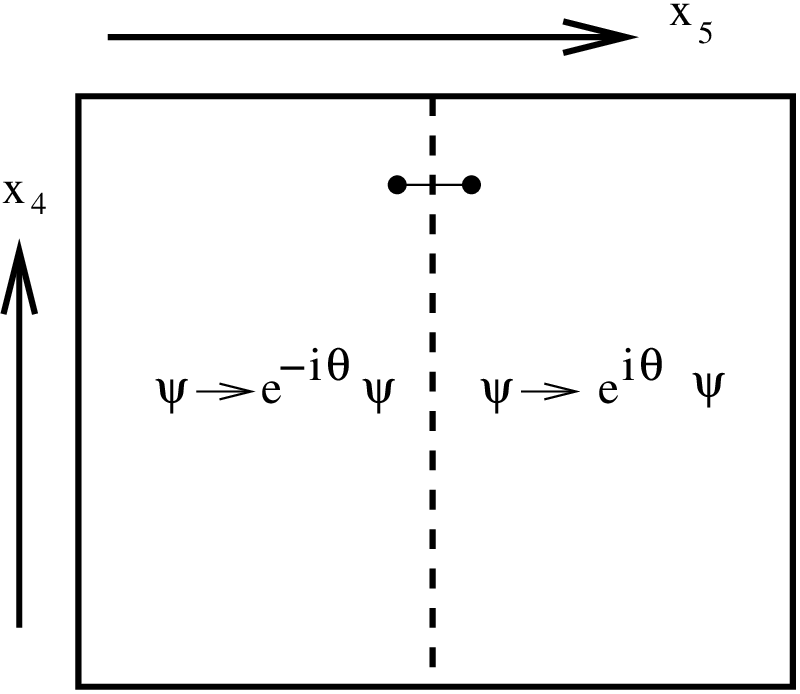}}
\narrower {\myfigname\gmorfig Making opposite phase rotations on opposite
sides of the box sets up an exact chiral symmetry relation.  The bonds
crossing the interface are not invariant under this change of
variables.

}
\endinsert

To explore this, consider taking the simple domain-wall formulation on
a five dimensional lattice and divide the sites in half with a
hyper-plane in the middle of the fifth dimension.  Put this hyper-plane
between two sites in the fifth direction. Take all Fermionic variables
on the ``left'' and multiply by a phase $e^{-i\theta/2}$.  Multiply all
those fields on the ``right'' by $e^{i\theta/2}$.  This is sketched in
\gmorfig.  Every term in the action is invariant except those hopping 
terms that cross the above hyper-plane.  Since the measure is
invariant under this transition, the path integral cannot change.
This is equivalent to the statement that the exponential of the change
in the action must have unit expectation value.  Separating the
exponential factors into terms even and odd in $\theta$, I have
$$
1=\left\langle \exp\left\{\sum_{x_\mu} (A(x_\mu) (\cos(\theta)-1)
+ B(x_\mu) \sin(\theta) )\right\} \right\rangle
\myeqno
$$
where $A$ and $B$ are operators only involving fields adjacent to the
dividing hyper-plane and the sum is over the four dimensions of
space-time.  Explicitly, I define
$$
A=K_5\left(\overline\psi_{x_5+1}(1+\gamma_5)\psi_{x_5}
+\overline\psi_{x_5}(1-\gamma_5)\psi_{x_5+1}\right)
\myeqno
$$
and
$$
B=iK_5\left(\overline\psi_{x_5+1}(1+\gamma_5)\psi_{x_5}
-\overline\psi_{x_5}(1-\gamma_5)\psi_{x_5+1}\right).
\myeqno
$$
Here the hyperplane lies between $x_5$ and $x_5+1$, and $K_5$ is the
hopping parameter in the extra dimension.  Expanding to second order
in $\theta$ and removing an overall spatial volume factor gives an
exact result relating the expectation of $A$ with a vacuum
susceptibility for the operator $B$
$$
\langle A(0)\rangle - \sum_{x_\mu} 
\langle B(0)  B(x_\mu) \rangle=0.
\myeqname\identone
$$
If I average over positions, this is true gauge configuration by gauge
configuration as well as at any volume.  

Now consider two flavors and replace the phase factor by
$e^{i\theta\tau_3}$ where $\tau_3$ is the Pauli matrix of isospin.
Since $\tau_3^2=1$, the $\tau_3$ factor cancels out of the $A$ term.
Corresponding to $B$, another operator appears
$$
B_3=iK_5\left(\overline\psi_{x_5+1}(1+\gamma_5)\tau_3\psi_{x_5}
-\overline\psi_{x_5}(1-\gamma_5)\tau_3\psi_{x_5+1}\right)
\myeqno
$$
giving another exact relation
$$ \langle A(0)\rangle - \sum_{x_\mu} \langle
B_3(0)  B_3(x_\mu) \rangle=0. 
\myeqname\identtwo
$$
Combining \identone with \identtwo shows that a singlet and a
non-singlet susceptibility are exactly equal
$$
\langle B(0)  B(x_\mu) \rangle
=\langle B_3(0)  B_3(x_\mu) \rangle.
\myeqname\identity
$$

For a physical interpretation, consider the particles that can be
created by this operator $B$.  Because of isospin, $B_3$ has the
quantum numbers to create a pion; call the coupling $g_\pi$.  On the
other hand, the operator $B$ can only create the singlet
``$\eta^\prime$'' meson; call this coupling $g_{\eta^\prime}$.  In the
chiral limit, the former becomes light, giving a singularity in the
right hand side of this equation.  Inserting these intermediate states
into \identity gives
$$
g_\pi^2/m_\pi^2 = g_{\eta^\prime}^2/m_{\eta^\prime}^2 +\ldots
\myeqno
$$
The ``$\ldots$'' refers to heavier particles in the spectrum.  As the
chiral limit is approached, these operators deep in the fifth
dimension have vastly different couplings to the physical particles on
the surface
$$
g_{\eta^\prime}^2/g_\pi^2 \sim m_{\eta^\prime}^2/m_\pi^2.
\myeqno
$$
This ratio can be qualitatively understood as a gluon exchange effect.
The gluonic field doesn't know about the fifth dimension, so the
operator $B$ can directly couple to it, which in turn can strongly
couple to the quark modes on the walls.  Since it is a flavor singlet,
both quark flavors contribute equally to this coupling.  On the other
hand, for the flavor non-singlet case, the two flavors couple
oppositely to the gauge field, and cancel out.  In that case, the
coupling to the surface modes must be through the Fermion fields,
which have the exponential damping of the heavy mass in the fifth
dimension.

This argument suggests that the coupling $g_{\eta^\prime}$ remains
finite even in the infinite $L_5$ limit.  In contrast, the expectation
is for $g_\pi$ to go to zero with the pion mass, which should go to
zero exponentially with $L_5$.

Note that the above exact relations can be kept exact when an explicit
mass term is added.  They just need corresponding additions to the
operators $A$ and $B$.  For the non-singlet case these terms represent
a form of the Gell-mann Oakes Renner relation (Gell-Mann, Oakes, and
Renner, 1968).  The operators in the lattice center give exponentially
suppressed corrections due to the discretization and make the relation
exact.

With an explicit mass present it becomes possible in a hopping
parameter expansion for a Fermion to travel first through the extra
dimension and then directly back to the original wall via the mass
term.  An overall phase acquired in such a process gives the $\Theta$
parameter discussed in earlier sections.  Whether the phase is
directly in the mass itself or in the hoppings through the fifth
dimension is a convention; the phase change implemented for the
identity here moves this convention around.  The first order
transition expected at $\Theta=\pi$ shows explicitly that the phase
acquired by a Fermion tunnelling through the extra dimension retains
physical significance even in the infinite $L_5$ limit.

\bigskip
{\bf \mysecno The Ginsparg-Wilson relation}
\medskip

I now drop back to four dimensions and discuss a generic approach to
chiral symmetry that has received considerable recent attention.  This
makes use of a class of lattice Dirac operators which generalize
continuum chiral properties.  Narayanan (1998) showed that this
relation is closely related to the overlap approach, which in turn was
motivated by the domain-wall scheme.  This topic is rapidly evolving
and is not the main thrust of this article.  Nevertheless it is
receiving sufficient attention that I will briefly discuss the central
ideas, although this section will likely soon be out of date.

I begin by considering the Fermionic part of some action as a
quadratic form
$$
S_f=\sum_i \overline\psi  D \psi.
\myeqname\genericferm
$$ 
The usual ``continuum'' Dirac operator $D=\sum\gamma_\mu D_\mu$
naively anti-commutes with $\gamma_5$, i.e. $[\gamma_5, D]_+=0$.
Then the change of variables $\psi \rightarrow e^{i\theta\gamma_5}
\psi$ and $\overline\psi \rightarrow \overline\psi
e^{i\theta\gamma_5}$ would be a symmetry of the action.  This,
however, is inconsistent with the chiral anomalies.  The conventional
continuum discussions map this phenomenon into the Fermionic measure
(Fujikawa, 1979).

On the lattice we work with a finite number of degrees of freedom;
thus, the above variable change is automatically a symmetry of the
measure.  To parallel the continuum discussion, it is necessary to
modify the symmetry transformation so that the measure is no longer
invariant.  Remarkably, it is possible to construct actions exactly
invariant under the altered symmetries.

To be specific, one particular modification (Neuberger, 1998b,c;
Luscher, 1998; Chiu and Zenkin, 1999; Chandrasekharan, 1999) that
leads to interesting consequences starts with the change of variables
$$\eqalign{
&\psi \longrightarrow e^{i\theta\gamma_5(1+aD)} \psi\cr
&\overline\psi \longrightarrow \overline\psi e^{i\theta\gamma_5}\cr
}
\myeqno
$$
where $a$ represents the lattice spacing.  Note the asymmetric way in
which the independent Grassmann variables $\psi$ and $\overline\psi$
are treated.  Inserting this into \genericferm, and requiring the
action to be unchanged gives the relation (Ginsparg and Wilson, 1982;
Hasenfratz, Laliena, and Niedermayer, 1998; Hasenfratz, 1998)
$$
\gamma_5 D + D \gamma_5 +a D\gamma_5 D=0.
\myeqname \gwrel
$$ 
I also assume the Hermeticity condition $\gamma_5 D
\gamma_5 = D^\dagger$.
The ``Ginsparg-Wilson relation'' in \gwrel along with the
Hermititicity condition is equivalent to the unitarity of the
combination $V=1+aD$.

Neuberger (1998b,c) and Chiu and Zenkin (1999) suggested a simple
construction of an operator satisfying this condition.  For this an
appropriate operator $V$ could be found via a unitarization of an
undoubled but chiral symmetry violating Dirac operator, such as the
Wilson operator $D_w$ implicit in \wilsonferm.  This operator should
also satisfy the above Hermeticity condition.  From this build
$$
V=-D_w(D_w^\dagger D_w)^{-1/2}.
\myeqname{\nop}
$$
More precisely, find a unitary operator to diagonalize $D_w^\dagger
D_w$, take the square root of the eigenvalues, and undo this unitary
transformation.

At this point the hopping parameter in $D_w$ is a parameter.  To have
the desired single light Fermion per flavor of the theory, the hopping
parameter should be appropriately adjusted to lie above the critical
value where $D_w$ describes a massless flavor, but not so large that
additional doublers come into play (Neuberger, 1999; Golterman and
Shamir, 2000).  There are actually two parameters to play with, the
hopping parameter of $D_w$, and the lattice spacing.  When the latter
is finite and gauge fields are present, the location of the critical
hopping parameter in $D_w$ is expected to shift from that of the free
Fermion theory.  There is potentially a rather complex phase structure
in the plane of these two parameters, with various numbers of doublers
becoming exactly massless modes.  The relation in \gwrel in and of
itself does not in general determine the number of massless Fermions.
In Section XVI I discussed a similar issue for domain-wall Fermions.

Although the Wilson operator entering this construction is local and
quite sparse, the resulting action is not; it involves direct
couplings between arbitrarily separated sites (Hernandez, Jansen and
Luscher, 1999; Horvath, 1998, 1999).  How rapidly these couplings fall
with distance depends on the gauge fields and is not fully understood.
The five dimensional domain-wall theory is local in the most naive
sense of the word; all terms in the action only couple nearest
neighbor sites.  Were one to integrate out the heavy modes, however,
the resulting low energy effective theory would also involve couplings
with arbitrary range.  Despite these non-localities, recent
encouraging studies (Neuberger 1998c; Edwards, Heller, and Narayanan,
1999; Borici, 1999; Hernandez, Jansen and Lellouch, 2000; Dong, Lee,
Liu, and Zhang, 2000; Gattringer, 2000) show that it may indeed be
practical to implement the inversion implicit in \nop in large scale
numerical simulations.  The overlap operator should have memory
advantages since a large number of fields corresponding to the extra
dimension do not need to be stored.  The extent to which this
outweighs the additional complexity in implementation remains to be
determined.

This approach hides the infinite sea of heavy Fermion states alluded
to above.  It is implicit in the presence of zero modes in the
inversion in \nop.  This is directly related to the connection with
the index theorems in the continuum; for recent reviews see Adams
(2000) and Kerler (2000).  Recent detailed analysis (Luscher, 2000;
Kikukawa and Yamada, 1999) shows that this operator is particularly
well behaved order by order in perturbation theory.  This has led to
hopes that this may lead the way to a rigorous formulation of chiral
models, such as the standard model.

\bigskip{\bf \mysecno Speculations}
\medskip

\midinsert
\epsfxsize .3\hsize
\centerline {\epsfbox{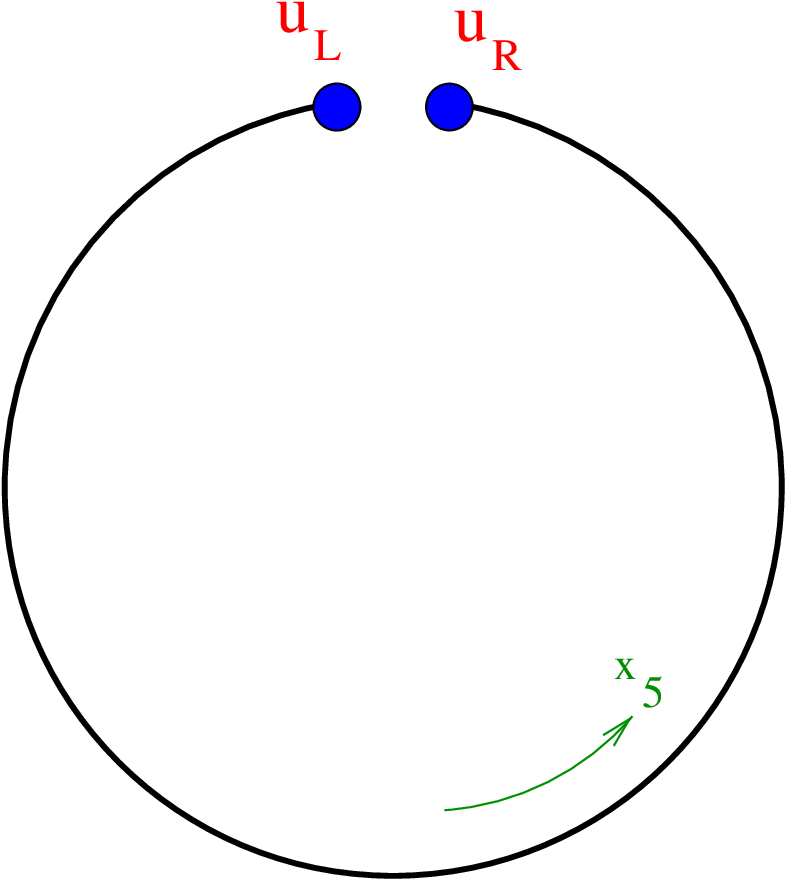}}
\narrower{\myfigname\circleone Cutting a periodic fifth dimension
gives the surfaces necessary for the topological zero modes.\par} 
\endinsert

I now ramble on with some general remarks about the basic domain-wall
scheme.  The existence of the end states relies on using open boundary
conditions in the fifth direction.  If I were to curl the extra
dimension into a circle, they would be lost.  To retrieve them,
consider cutting such a circle, as in
\circleone.  Of course, if the size of the extra dimension is
finite, the modes mix slightly.  This is crucial for the scheme to
accommodate anomalies, as discussed previously.

\midinsert
\epsfxsize .3\hsize
\centerline {\epsfbox{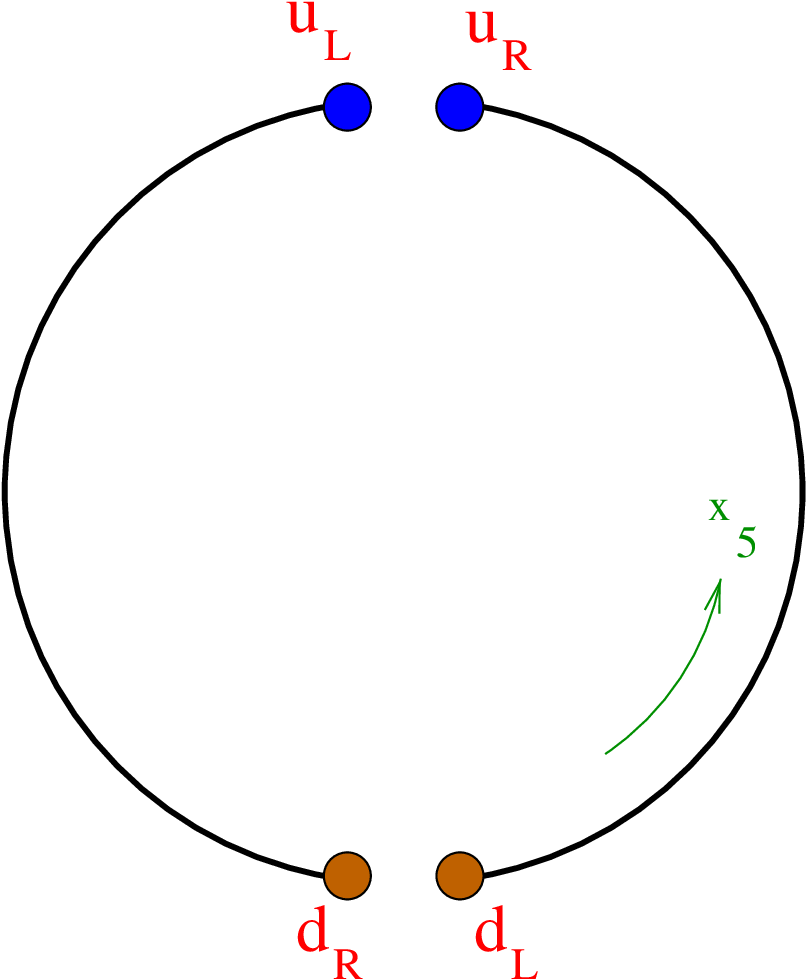}}
\narrower {\myfigname\circletwo Cutting the fifth dimension twice can
give two flavors of quark.\par}
\endinsert

Suppose I want a theory with two flavors of light Fermion, such as the
up and down quarks.  For this one might cut the circle twice, as shown
in \circletwo.  Remarkably, this construction keeps one chiral
symmetry exact, even if the size of the fifth dimension is finite.
Since the cutting divides the system into two completely disconnected
pieces, in the notation of the figure, the number of $u_L+d_R$
particles is absolutely conserved.  Similarly with $u_R+d_L$.
Subtracting gives an exactly conserved axial charge corresponding to
the continuum current
$$
j_{\mu 5}^3 = \overline \psi \gamma_\mu\gamma_5 \tau^3 \psi.
\myeqno
$$
The conservation holds even with finite $L_5$.  There is a small
flavor breaking since the $u_L$ mixes with the $d_R$.  These
symmetries are reminiscent of staggered (Kogut and Susskind, 1975)
Fermions, where a single exact chiral symmetry is accompanied by a
small flavor breaking.  Now, however, the extra dimension gives
additional control over the latter.

Despite this analogy, the situation is physically somewhat different
in the zero applied mass limit.  Staggered Fermions are expected to
give rise to a single zero mass Goldstone pion, with the other pions
acquiring mass through flavor breaking.  In the doubly cut
domain-wall picture, however, the zero mass limit has three
degenerate equal mass particles as the lowest states.  To see how this
works it is simplest to discuss the physics in a chiral Lagrangian
language.  The finite fifth dimension generates an effective mass
term, but it is not in a flavor singlet direction.  Indeed, it is in a
flavor direction orthogonal to the naive applied mass.  In the usual
``sombrero'' picture of the effective Lagrangian, as illustrated in
\sombrero and extensively discussed earlier, the two mass terms
compete and the true vacuum rotates around the Mexican hat from the
conventional ``sigma'' direction to the ``pi'' direction.

\midinsert
\epsfxsize .4\hsize
\centerline {\epsfbox{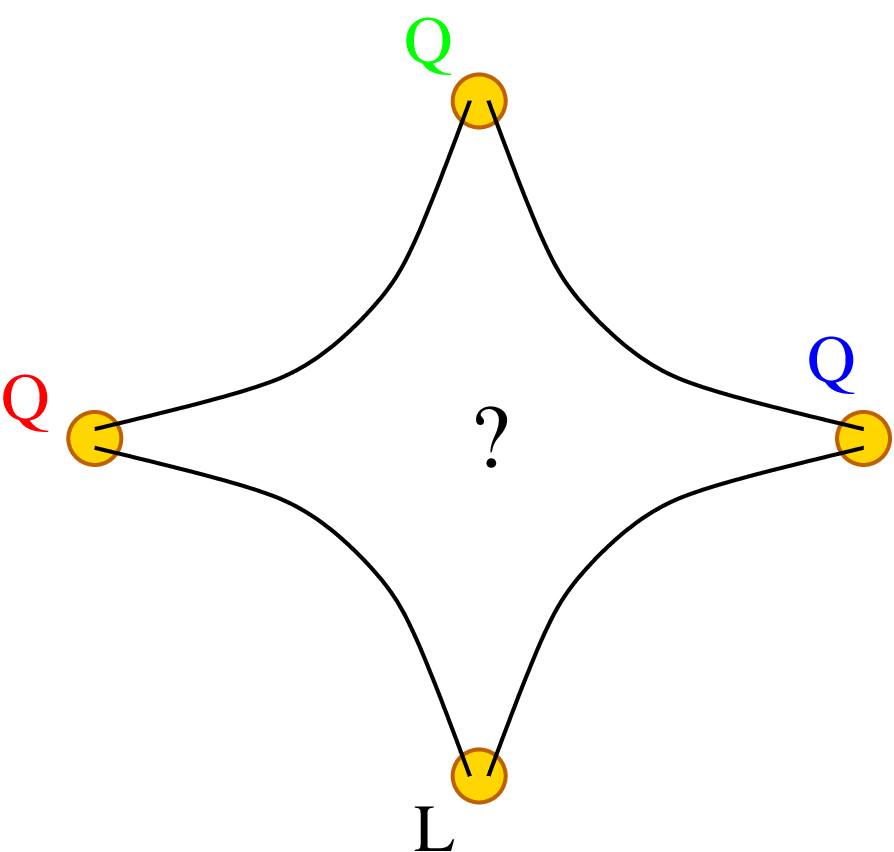}}
\narrower{\myfigname\starfig Perhaps all species of Fermions are
manifestations of a single field at various topological defects in a
higher dimensional space. \par}
\endinsert

Now I become more speculative.  The idea of multiply cutting the fifth
dimension to obtain several species suggests extensions to zero modes
on more complicated manifolds.  Multiple zero modes gives a mechanism
to generate multiple flavors.  Maybe one can have a theory where all
the physical Fermions in four dimensions arise from a single Fermion
field in the underlying higher dimensional theory.  Schematically we
might have something like \starfig where each point represents some
four dimensional surface and the question mark represents structures
in the higher dimension that need specification.  One nice feature
provided by such a scheme is a possible mechanism for the transfer of
various quantum numbers involved in anomalous processes.  For example,
the baryon non-conserving 't Hooft process ('t Hooft, 1976a,b) might
arise from a lepton flavor tunneling into the higher manifold and
reappearing on another surface as a baryon.

I've been rather abstract here.  This generic mechanism is in fact the
basis of the formulation of the standard model on the lattice (Creutz,
Tytgat, Rebbi, and Xue, 1997) presented in Section XIX.  For this model
the question mark in the above figure is the four-Fermi interaction in
the interior of the extra dimension.  As emphasized by Neuberger
(1997), that picture is closely tied with $SO(10)$ grand unified
models.

{\bf \mysecno Concluding remarks}

The last few years have seen remarkable progress towards the
non-perturbative understanding of chiral symmetry in quantum field
theory.  This has largely been driven by the needs of lattice gauge
theory, aided by techniques drawn from effective Lagrangians.  The use
of a tower of auxiliary Fermion fields provides a sink for the flow of
Fermion states in anomalous processes.  Formulation in terms of an
extra dimension provides a useful analogy, and provides hints towards
deeper connections with the higher dimensional theories in vogue
today.  

These approaches lead to greatly improved chiral behavior, such as the
time honored relation $m_Q\sim m_\pi^2$.  Encouraging exploratory
calculations using both the domain-wall (Blum and Soni, 1997a,b; Chen
et al., 2000; Blum et al., 2000).  and the overlap (Neuberger 1998c;
Edwards, Heller, and Narayanan, 1999; Borici, 1999; Hernandez, Jansen
and Lellouch, 2000; Dong, Lee, Liu, and Zhang, 2000; Gattringer, 2000)
formulations are underway.  The new methods may soon dominate
numerical lattice gauge simulations.

The progress has been particularly complete for vector-like theories,
such as the strong interactions.  Here the various methods,
domain-wall/overlap/two cutoff, all appear to have essentially solved
the confusion surrounding the early attempts to formulate chiral
symmetry on the lattice.  The situation is less clear when gauge
fields are coupled to chiral currents, as in the standard model.  Here
formulations based overlap operator with the Ginsparg-Wilson relation
appear promising; however, these schemes still remain somewhat formal.
For the 't Hooft process it would be welcome to have a Hamiltonian
discussion along the lines of section XVII to explicitly trace the
transition of a baryon into a lepton in terms of a flow of states. 

It remains unknown whether there exists a lattice formulation of the
standard model which, in the cutoff form, maintains an exact local
gauge symmetry and a finite number of degrees of freedom per unit
volume.  Models incorporating four Fermion couplings, such as
presented in the section relating leptons and antiquarks, are an
attempt to accomplish this.  Unfortunately, such couplings are
difficult to treat rigorously and these approaches remain
controversial.

Chiral symmetry often plays a central role in extensions of the
standard model.  It has been argued (Narayanan and Neuberger, 1993a,b,
1994, 1995; Nishimura, 1987; Maru and Nishimura, 1998; Kaplan and
Schmaltz, 2000) that the domain-wall approach also may provide a path
to super-symmetry on the lattice.  With adjoint Fermions in the
domain-wall formalism, the states of one handedness can be removed via
a Majorana mass term.  This gives a theory where the low energy
degrees of freedom are the gauge fields and a chiral set of adjoint
fields.  While the heavy modes severely violate super-symmetry, the
low energy states represent exactly the spectrum of super-symmetric
Yang-Mills theory, and do this without fine tuning.  It would be
interesting to see if these ideas can be extended to more general
super-symmetric models.

\bigskip
\hrule
\bigskip

This manuscript has been authored under contract number
DE-AC02-98CH10886 with the U.S.~Department of Energy.  Accordingly,
the U.S. Government retains a non-exclusive, royalty-free license to
publish or reproduce the published form of this contribution, or allow
others to do so, for U.S.~Government purposes.  The work was also
supported by grants No.~95-00077 and No.~98-000302 from the
United-states -- Israel Binational Science foundation (BSF) Jerusalem,
Israel, which enabled visits wherein many of the ideas discussed here
were developed.

\vfill\eject
\noindent{\bf References}
\medskip
\parindent=0pt
\baselineskip=12pt
\parskip=4pt
 
Adams, D., 2000,  e-Print Archive: hep-lat/0001014.

Adler, S. L., 1969, Phys. Rev. 117, 2426.

Alonso, J., Ph.~Boucaud,
J.~Cort\'es, and E.~Rivas, 1991, Phys.~Rev.~D44, 3258.

Ambjorn, J., J.~Greensite, and C.~Peterson,
Nucl.~Phys.~B221 (1983) 381.

Aoki, S., 1989, Nucl. Phys. B314, 79.

Aoki, S., S.~Boetcher, and A.~Gocksch, 1994, Phys.~Lett. B331, 157.

Aoki, S., and A.~Gocksch, 1992, Phys.~Rev. D45, 3845.

Aoki, S., A. Ukawa, and T. Umemura, 1996,
Phys.~Rev.~Lett. 76, 873.

Banks, T., and A. Casher, 1980, Nucl.~Phys. B169, 103. 

Bell, J.S. and R. Jackiw, 1969, Nuovo Cimento 60A, 47.

Bijnens, J., J. Prades, and E. de Rafael, 1995, Phys.Lett. B348 226-238. 

Bitar, K., and P.~Vranas, 1994a, Phys.~Rev. D50, 3406.

Bitar, K., and P.~Vranas, 1994b, Nucl.~Phys. B, Proc.~Suppl. 34, 661 (1994).

Blum, T., P. Chen, N. Christ, C. Cristian, C. Dawson, G. Fleming, A.
  Kaehler, X. Liao, G. Liu, C. Malureanu, R. Mawhinney, S. Ohta, G. Siegert, A.
  Soni, C. Sui, P. Vranas, M. Wingate, L. Wu and Y. Zhestkov, 2000,
   e-Print Archive: hep-lat/0007038.

Blum, T., and A. Soni, 1997a, Phys. Rev. Lett. 79, 3595.

Blum, T., and A. Soni, 1997b, Phys.~Rev.~D56, 174.

Bock, W.,
M. Golterman, and Y. Shamir, 1998a, 
Phys. Rev. Lett. 80, 3444.

Bock, W.,
M. Golterman, and Y. Shamir, 1998b, 
Phys. Rev. D58, 034501 (1998).

Bodwin,G., 1996, Phys.Rev.D54:6497-6520.

Borici A., 1999,  e-Print Archive: hep-lat/9910045.

Borrelli, A.,  L. Maiani, R. Sisto, G.C. Rossi, and
M. Testa, 1990, Nucl.Phys. B333 335.

Brower, R., and B. Svetitsky, 2000, Phys.~Rev.~D61, 114511.

Callan, C., and J.~Harvey, 1985, Nucl.~Phys.~B250, 427. 

Chandrasekharan, S., 1999, Phys.~Rev. D60, 074503.

Chen, P. N. Christ, G. Fleming, A. Kaehler, C. Malureanu,
R. Mawhinney, G. Siegert, C. Sui, L. Wu, Y. Zhestkov, and P. Vranas,
2000,  e-Print Archive: hep-lat/0006010.

Chiu, T. and S. Zenkin, 1999, Phys.~Rev. D59, 074501.

Coleman, S., 1976, Annals Phys.~101, 239.

Coleman, S., and B. Grossman, 1982, Nucl.~Phys. B203, 205.

Coleman, S. and E. Weinberg, 1973,
Phys. Rev. D7, 1888.

Creutz, M., 1983, {\sl Quarks Gluons and Lattices} (Cambridge).

Creutz, M., 1995a, Phys. Rev. D52, 2951 (1995).

Creutz, M., 1995b, Nucl. Phys. B (Proc. Suppl.) 42, 56.

Creutz, M., 1997, e-Print Archive: hep-lat/9608024, in RHIC Summer
Study '96: Theory workshop on relativistic heavy ion collisions,
D. Kahana and Y. Pang, eds., pp. 49-54 (NTIS).

Creutz, M., 1999, Phys. Rev. Letters 83, 2636.

Creutz, M. and I. Horvath, 1994, Phys. Rev. D50, 2297.

Creutz, M., and M. Tytgat, 1996, Phys. Rev. Letters 76, 4671.

Creutz, M., M. Tytgat, C. Rebbi, S.-S. Xue, 1997, Phys. Lett. B402, 341.

Dashen, R., 1971, Phys. Rev. D3, 1879 (1971).

Distler, J., and S. Rey, 1993,  e-Print Archive: hep-lat/9305026.

Dong, S., F. Lee, K. Liu, and J. Zhang, 2000, e-Print Archive:
hep-lat/0006004.

Donoghue, J., B. Holstein, and D. Wyler, 1992, Phys. Rev. Lett. 69, 3444.

Drell, S., M. Weinstein, and S. Yankielowicz, 1976, Phys.~Rev. D19, 3698.

Edwards, R., U. Heller, R. Narayanan, 1999, Phys.~Rev. D59, 094510; 

Eichten, E. and J.~Preskill, 1986,
Nucl.~Phys.~B268, 179.

Evans, N., S. Hsu, A. Nyffeler, and M. Schwetz, 1997, Nucl.~Phys. B494, 200.

Friedberg, R., T.D.~Lee, and Y.~Pang, 1994, J.~Math.~Phys.~35, 5600.

Frolov, S. and A.~Slavnov, 1994, Nucl.~Phys.~B411, 647.

Fujikawa, K., 1979 Phys. Rev. Lett. 42, 1195.

Furman, V., and Y. Shamir, 1995, Nucl.~Phys.~B439, 54.

Gattringer, C., 2000,  e-Print Archive: hep-lat/0003005.

Gell-Mann, M., R. Oakes, and B. Renner, 1968, Phys. Rev. 175, 2195 (1968).

Ginsparg, P. and K. Wilson, 1982, Phys.~Rev.~D25, 2649.

Gockeler, M., A. Kronfeld, G. Schierholz, and U.J. Wiese,
1993, Nucl.~Phys. B404, 839.

Goldstone, J., and F. Wilczek, 1981,
Phys.~Rev.~Lett. 47, 1986.

Golterman, M., K. Jansen, D. Petcher, and J. Vink, 1994, Phys.~Rev.~D49, 1606.

Golterman, M., D. Petcher, and E. Rivas, 1993, Nucl.~Phys.~B395, 596.

Golterman, M., and Y.~Shamir, 1995, Phys.~Rev.~D51, 3026.

Golterman, M., and Y. Shamir, 1999,  e-Print Archive: hep-lat/0007021.

Hasenfratz, P., 1998, Nucl.~Phys. B525, 401.

Hasenfratz, P., V. Laliena, and Ferenc Niedermayer, 1998,
Phys.~Lett. B427, 125. 

Hernandez, P., and P. Boucaud, 1998,
Nucl.~Phys. B513, 593.

Hernandez, P., K. Jansen, and Martin Luscher, 1999, Nucl.~Phys. B552, 363.

Hernandez, P, K. Jansen, and L. Lellouch, 2000, e-Print Archive:
hep-lat/0001008;

Hernandez, P. and R.~Sundrum, 1995, Nucl.~Phys.~B455, 287.

Holstein, B., 1993, Am.~J.~Phys.~61, 142.

't Hooft, G., 1976a, Phys.~Rev.~Lett.~37, 8 (1976). 

't Hooft, G., 1976b, Phys.~Rev.~D14, 3432 (1976).

't Hooft, G., 1986,
Phys.~Rept. 142, 357.

`t Hooft, G., 1995, Phys.~Lett.~B349, 491.

Horvath, I, 1998, Phys.~Rev.~Lett. 81, 4063 (1998); 

Horvath, I, 1999, Phys.~Rev. D60, 034510 (1999).

Hsu, S., 1995, preprint YCTP-P5-95.

Jackiw,R. and C.~Rebbi, 1976, Phys.~Rev.~D13, 3398.

Jansen, K., and M. Schmaltz, 1992, Phys.~Lett. B296, 374.

Kaplan, D., 1992, Phys.~Lett.~B288, 342.

Kaplan, D., and M. Schmaltz, 2000,  e-Print Archive: hep-lat/0002030. 

Karsten, L., and J. Smit, 1981,
Nucl.~Phys. B183, 103.

Kelvin, Lord, 1904, {\sl Baltimore lectures on molecular dynamics and
the wave theory of light} (Clay, London).\footnote*{Kelvin extensively
edited these lectures from the original version reprinted in {\sl
Kelvin's Baltimore lectures and modern theoretical physics} (MIT,
1987).  In the latter the word ``chiral'' is not mentioned, but
Thomson was clearly struggling for a term when on page 186 he writes
``I have objected to the name rotary, because it is not properly
applied, and have taken the name helical because the phenomenon
essentially depends on a screw like form somehow or other.''}

Kerler, W., 2000,  e-Print Archive: hep-lat/0007023.

Kikukawa, Y., 1999,  e-Print Archive: hep-lat/9912056. 

Kikukawa, Y., and A. Yamada, 1999, Phys.~Lett. B448, 265.

Kogut, J. and L. Susskind, 1975, Phys. Rev. D11, 395.

Kronfeld, A., 1995,  e-Print Archive: hep-lat/9504007; 

Lee, I-H., and R. Schrock, 1988, Nucl.~Phys. B305, 305.

Leutwyler, H., 1990, Nucl. Phys. B337 108.

Luscher, M, 1998, Phys.Lett.B428, 342.

Luscher, M., 2000, JHEP 0006, 028.

Maru, N., and  J. Nishimura, 1998, Int.~J.~Mod.~Phys. A13, 2841.

Montvay, I., 1987, Phys.~Lett.~199B, 89.

Montvay, I., 1993, Nucl.~Phys.~B (Proc.~Suppl.) 30, 621.

Narayanan, R., 1998, Phys.~Rev. D58, 097501.

Narayanan, R. and H.~Neuberger, 1993a, 
Phys.~Lett.~B302, 62.

Narayanan, R. and H.~Neuberger, 1993b, 
Phys.~Rev.~Lett.~71, 3251; 

Narayanan, R. and H.~Neuberger, 1994, 
Nucl.~Phys.~B412, 574.

Narayanan, R. and H.~Neuberger, 1995, 
Nucl.~Phys.~B443, 305.

Neuberger, H., 1997, Phys. Lett. B413, 387.

Neuberger, H, 1998a, Phys.~Rev.~Lett. 81, 4060; 

Neuberger, H., 1998b Phys.~Lett. B417, 141.

Neuberger, H., 1998c, Phys.~Lett. B427 353.

Neuberger, H., 1999,  e-Print Archive: hep-lat/9911022.

Nielsen, H., and M. Ninomiya, 1981a, Phys.~Lett.~B105, 219.

Nielsen, H., and M. Ninomiya, 1981b, Nucl.~Phys.~B185, 20,
(Erratum: 1982 ibid.~B195, 541). 

Nielsen, H., and M. Ninomiya, 1981c, Nucl.~Phys. B193, 173.

Nishimura, J., 1997, Phys.~Lett. B406, 215.

Raffelt, G., 1990, Phys. Rept. 198, 1.

Randjbar-Daemi, S., and J.~Strathdee, 1996a,
Nucl.~Phys.~B461, 305.

Randjbar-Daemi, S., and J.~Strathdee, 1996b,
Nucl.~Phys.~B466, 335.

Seiler, E.,  and I.~Stamatescu, 1982, Phys.~Rev. D25, 2177.

Shamir, Y., 1993a, Phys.~Rev.~Lett. 71, 2691.

Shamir, Y., 1993b, Nucl.~Phys.~B406, 90.

Shamir, Y., 1994, Nucl. Phys. B417, 167 (1994). 

Sharpe, S., and R. Singleton, 1998, Phys.~Rev. D58, 074501.

Shockley, W., 1939, Phys.~Rev.~56 317.

Smilga, A., 1999, Phys.~Rev. D59, 114021.

Smit, J., 1980, Nucl.~Phys. B175 307 (1980).

Svetitsky, B., S. Drell, H. Quinn, and M. Weinstein, 1980, Phys.~Rev. D22, 490.

Turner, M., 1990, Phys. Rept. 197, 67.

Tytgat, M., 1999,  e-Print Archive: hep-ph/9909532.

van Baal, P., 1998, Nucl.~Phys.~Proc.~Suppl. 63, 126. 

Verbaarschot,J., 1994, Phys.~Rev.~Lett. 72, 2531. 

Wess, J., and B.~Zumino, 1971, Phys.~Lett.~37B, 95.

Wilson, K., 1977, in {\sl New Phenomena in Subnuclear Physics},
edited by A.~Zichichi (Plenum Press, N.~Y.).

Witten, E., 1983a, Nucl.~Phys.~B223, 422.

Witten, E., 1983b, Nucl.~Phys.~B223, 433.

Witten, E., 1984, Commun.~Math.~Phys.~92, 455.

\bye